\UseRawInputEncoding
\documentclass[11pt]{article}
\pdfoutput=1 
\usepackage{amssymb, amsmath,amsfonts}
\usepackage{graphicx}
\usepackage{xcolor}
\usepackage{color}
\usepackage[pdftex, bookmarks=true,colorlinks,linkcolor=red,urlcolor=blue,citecolor=blue]{hyperref}
\usepackage[normalem]{ulem}
\usepackage{cite}
\usepackage{subcaption}
\usepackage{slashed}

\def\arXiv#1{\href{http://arxiv.org/abs/#1}{arXiv:#1}}
\def\arXiv#1#2{\href{http://arxiv.org/abs/#1}{arXiv:#1}}

\def\doi#1#2{\href{http://doi.org/#1}{#2}}
\def\blue#1{\textcolor{blue}{#1}}

\allowdisplaybreaks
\usepackage{enumerate}
\usepackage[normalem]{ulem}

\newcommand{\bk}{{\mathbf k}}

\makeatletter\@addtoreset{equation}{section}\makeatother

\newcommand{\preprint}[1]{\begin{table}[t]  
             \begin{flushright}               
             {#1}                             
             \end{flushright}                 
             \end{table}}                     
\renewcommand{\title}[1]{\vbox{\center\LARGE{#1}}\vspace{5mm}}
\renewcommand{\author}[1]{\vbox{\center#1}\vspace{5mm}}

\newcommand{\address}[1]{\vbox{\center\em#1}}

\usepackage{bm}
\def\mI{{\bf I}}

\def\be{\begin{eqnarray}}
\def\ee{\end{eqnarray}}
\def\bea{\begin{eqnarray}}
\def\eea{\end{eqnarray}}

\def\Dslash{\,\,{\raise.15ex\hbox{/}\mkern-12mu D}}
\def\Dbarslash{\,\,{\raise.15ex\hbox{/}\mkern-12mu {\bar D}}}
\def\delslash{\,\,{\raise.15ex\hbox{/}\mkern-9mu \partial}}
\def\delbarslash{\,\,{\raise.15ex\hbox{/}\mkern-9mu {\bar\partial}}}
\def\pslash{\,\,{\raise.15ex\hbox{/}\mkern-9mu p}}
\def\calDslash{\,\,{\raise.15ex\hbox{/}\mkern-12mu {\cal D}}}

\def\lae{\mathrel{\mathop{\smash{\lower .5 ex \hbox{$\stackrel<\sim$}}}}}
\def\lae{\mathrel{\mathop{\smash{\lower .5 ex \hbox{$\stackrel>\sim$}}}}}

\begin{document}

\unitlength = .8mm

\begin{titlepage}
\vspace{.5cm}
\preprint{}
\begin{center}
\hfill \\
\hfill \\
\vskip 1cm

\title{\bf Topological phase transitions of semimetal states in effective field theory models}
\vskip 0.5cm

{Xuanting Ji$^{a,b}$}\footnote{Email: {\tt jixuanting@cau.edu.cn
}},{Ya-Wen Sun$^{b,c}$}\footnote{Email: {\tt yawen.sun@ucas.ac.cn}}

\address{${}^a$Department of Applied Physics, College of Science, \\
China Agricultural University, Beijing 100083, China}
\vspace{-10pt}
\address{${}^b$School of Physical Sciences, and CAS Center for Excellence in Topological Quantum Computation, University of Chinese Academy of Sciences, 
Beijing 100049, China}
\vspace{-10pt}
\address{${}^c$Kavli Institute for Theoretical Sciences, \\
University of Chinese Academy of Sciences, 
Beijing 100049, China }
\vspace{-10pt}

\end{center}
\vspace{-10pt}
\abstract{Effective relativistic field theory models capable of realizing various gapless topological states are presented in this work. We study the topological phase transitions of the Weyl and nodal line semimetal states in effective field theories. When the one form field giving rise to the Weyl nodes lie perpendicular to the plane where the nodal ring lives, the nodal ring and Weyl nodes could coexist as the mirror symmetry responsible for the nodal ring is not broken. New phases including a three-node state and a triple degenerate state exist. The nodal ring is immediately destroyed when the one form field lies in the plane of the ring. However, we show that in an eight-component spinor model, Weyl nodes and nodal rings could still coexist even when the one form field is parallel to the plane, due to the expanded symmetry. Topological invariants are calculated which confirm the interesting nontrivial topology of the three-node state and the triple-degenerate node. This work presents potential topological phase transitions in multiphase topological systems, which may be experimentally detected.}

\vfill

\end{titlepage}

\begingroup
\hypersetup{linkcolor=black}
\tableofcontents
\endgroup

\section{Introduction}
\label{sec:1}
The topological properties of certain physical systems are a rapidly developing field in current physics. Among various kinds of topological states, topological semimetals have a lot of interesting properties including but not limited to non-dissipative transports, topological robustness, and realizing particles that cannot be present in the standard model of particle physics. These novel properties have attracted a lot of research interest both in the theory and experimental sides. Topological semimetals include Dirac semimetals\cite{zkliu,Ylchen,kane}, Weyl semimetals\cite{Wan,rmb}, topological nodal line semimetals\cite{burkov1}, and Weyl-$\boldmath{Z}_2$ semimetals\cite{Gorgar1,Gorgar2}, and nodal surface semimetals\cite{liang,zhong}, etc.. 

Weyl semimetal was proposed about ten years ago\cite{Wan}, whose Weyl node is a twofold degenerate point that always appears in pairs due to a no-go theorem\cite{Ninomiya}. Chiral anomalies appear naturally in such systems and can be manifested by a negative longitudinal magneto-resistance and other anomaly transport coefficients\cite{Son:2012wh}. The simplest Weyl semimetal has only one pair of Weyl nodes, namely the ideal Weyl semimetal \cite{pan}. More pairs of Weyl nodes could exist in e.g. $\boldmath{Z}_2$-Weyl semimetal, which has more interesting topological properties\cite{Gorgar1,Gorgar2,Fleury,Ji:2021aan}. More pairs of Weyl nodes also exist in realistic materials such as TaAs family \cite{weng,B.Q.Lv,shuang,SXu}, YbMnBi$_{2}$ \cite{Borisenko} (Type-I Weyl semimetal), Mo$_{x}$W$_{1-x}$Te$_{2}$ \cite{Soluyanov,Deng,HaoZheng}(Type-II Weyl semimetal), and $(TaSe_{4})_{2}$I\cite{yao}(Type-III Weyl semimetal).

The topological semimetals can be described by lattice models in condensed matter. However, as pointed out in\cite{Gooth:2017mbd}, the gravitational anomaly makes significant contributions to the thermoelectric conductance of the Weyl fermions. It is difficult to study such contributions in the lattice model. Furthermore, the existence of strongly coupled topological semimetals has been suggested by the observation of hydrodynamic behavior in Weyl semimetals\cite{Gooth2}. Holography (AdS/CFT correspondence) is an efficient way to study strongly coupled many body systems\cite{Hartnoll:2016apf,Landsteiner:2019kxb}, which provides a useful tool for the study of various properties of strongly coupled topological semimetals systems\cite{Erdmenger:2008rm, Banerjee:2008th, Landsteiner:2011cp, Landsteiner:2011iq, Ji:2019pxx, Gao:2023zbd}. However, originated from string theory, holography relies on a covariant form of formulation, so a first step towards building a holoraphic model is to first obtain the corresponding weakly coupled relativistic field theoretic description of the system, which gives the basis for building the holographic model. Therefore, a different description of the system is necessary for a deeper understanding of these systems.

The above requirements can be satisfied by Lorentz covariant field theory models. A covariant form of the formulation makes it possible to build a hologrphic model. In addition, the contribution of the gravitational anomaly and the chiral anomaly could be included in the effective field theory model in a natural way. People have successfully described various topological semimetals could be described by effective models with intuitive generalizations from the Lorentz breaking relativistic field theory model\footnote{We focus on tree level effective field theories and their spectrums.}. For example, a Weyl semimetal with only one pair of Weyl nodes could be realized just by adding a Lorentz breaking term proportional to the combination of the gamma matrices $\gamma_{\mu}\gamma_5$\cite{Grushin:2012mt, Grushin:2019uuu,Kostelecky:2021bsb}. These effective field theories come from relativistic physics, which could describe many lattice systems in their continuum limit, e.g. the low energy behavior of graphene could be described by a relativistic Dirac fermion, as well as many other systems, including Kitaev's honeycomb model\cite{Kitaev1,Farjami}, \blue{Kitaev chains\cite{Maiellaro1,Maiellaro2},} superconductors\cite{read,Golan,Maiellaro3}. These works demonstrate the power of the effective field theory model for the description of topological semimetals.

As the first realized Weyl semimetal in the laboratory, TaAs shows a Weyl semimetal state when a non-zero spin-orbit coupling is considered. While when spin-orbit coupling becomes zero, TaAs presents a topological nodal line semimetal\cite{weng,shuang}. Topological nodal line semimetal has a nontrivial shape of Fermi surface where Fermi nodal points form a circle under certain symmetries, e.g. mirror reflection symmetry\cite{burkov1,fang1}. The existence of spin-orbit coupling breaks the mirror symmetry and changes a nodal line semimetal state into a Weyl semimetal state. Thus, studying the topological phase transition between the Weyl semimetal state and the topological nodal line state could help understand topological properties clearly and predict new topological materials.

In this paper, we study such a process in relativistic field theory models, the advantage of which has been introduced above. Tuning the parameters in the system, we could realize a nodal line semimetal state which evolves into a Weyl semimetal state along with the change of parameters. In this process, several other states could also exist, including Dirac semimetals, triple-degenerate nodal point semimetals\cite{fang2,fang3,weng2}, and a topological trivial state. 
A novel quantum phase transition point is realized during this phase transition, which is different from the phase transitions occurring in the single Weyl semimetal or the single nodal line semimetal with the Dirac point being the critical point. In this phase transition, the critical phase behaves like a single Dirac point surrounded by nodal walls. This novel phase is quite similar to the state realized in PtGa that breaks the no-go theorem, which brings far-reaching physics meaning\cite{Ma}.


This paper is organized as follows. In Sec. \ref{sec:model} we review the effective field theory models of the Weyl semimetals and the topological nodal line semimetals first. We then build two effective field theory models to realize the phase transition between two semimetal states and show the phase diagram of different models. In Sec. \ref{four}, we build an effective field theory model with a four-component spinor to check the phase transition between Weyl semimetal and nodal line semimetal. We also compute the surface states for a state where the Weyl semimetal and the nodal line semimetal coexist, which served as a potential observable for such a state. We then generalize our results in another effective field theory model with an eight-component spinor in Sec. \ref{eight}. In Sec. \ref{sec:invariant}, we calculate the topological invariants to confirm the phase diagram of the topological phase transition. Sec. \ref{sec:cd} is devoted to conclusions and discussions.

\section{Review of Weyl and nodal line semimetals in effective field theory}
\label{sec:model}

First, we elaborate more on the motivation to study topological semimetals in relativistic effective field theories, which has already been mentioned in the introduction. Topological semimetals are novel electronic band systems that could serve as a bridge between condensed matter and high-energy physics. Weyl semimetal is a Lorentz symmetry-breaking condensed matter system, in which the Weyl fermions are realized by low-energy excitations. The realized Weyl fermions in Weyl semimetals are quasiparticles whose energy scale is much smaller than the rest mass of the electron.  It seems that the propagation of such an electron cannot be described by the Dirac equation, since no relativistic effects need to be taken into account.  However, the existence of the periodic potential in the crystal leads the quasiparticles to a dressed electronic state. Note that an effective low-energy description of such a dressed electron will again resemble the Dirac equation\cite{rmb}. The Dirac equation is the equation of motion derived from quantum electrodynamics in quantum field theory that describes all half-spin massive particles in high energy physics. While for quasiparticles in topological semimetals, one could use an effective field theory model by adding Lorentz breaking terms in the Lagrangian of quantum electrodynamics to describe such a low-energy excitation process. Through these effective field theory models, people could study topological semimetals from another viewpoint, which may open new windows in physics research. Moreover, there are anomaly-induced transports in topological semimetals, which are complicated phenomena that could be more easily studied in relativistic field theories.

From another perspective, how to describe a strongly coupled topological semimetal is an important problem in theoretical research. It is difficult to solve a strongly coupled model under a scheme where perturbative treatments become invalid. Holographic duality has been used in the last decade as a tool to study exactly these kinds of questions\cite{Hartnoll:2016apf, Landsteiner:2019kxb}. The holographic principle ensures that the strong coupling problem in field theory can be solved with weak classical gravity. Anomaly induced transport properties such as the chiral magnetic and chiral vortical effects and their relations to anomalies are also most easily understood with the help of holographic duality\cite{Erdmenger:2008rm, Banerjee:2008th, Landsteiner:2011cp, Landsteiner:2011iq, Ji:2019pxx,Gao:2023zbd}. Therefore, a holographic model of a topological semimetal is a powerful tool. To build a holographic model, we need to follow the holographic dictionary which requires to first get fields and operators in the field theory model. Thus, it is necessary to first construct effective field theory models describing topological semimetals of our interest. Therefore, through effective field theory descriptions of topological semimetals, one could connect the weak and strong coupling problems, as well as high energy physics and condensed matter.

In this section, we will review the effective field theory models of Weyl and nodal line semimetals, which provide the background knowledge for later use.

\subsection{Weyl Semimetal}
\begin{figure}[h!]
  \centering
  \begin{subfigure}[b]{0.32\textwidth}
\includegraphics[width=\textwidth]{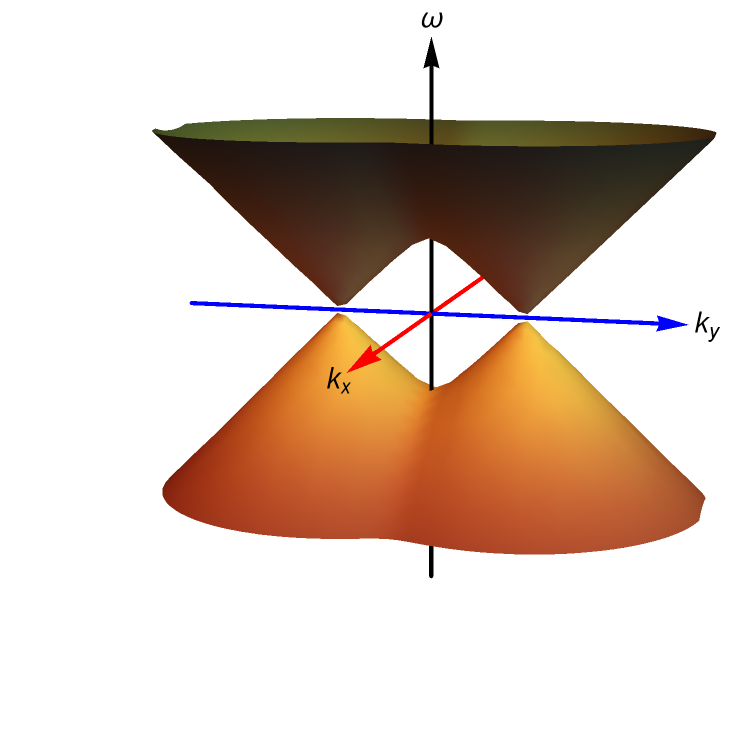}
\caption{\small Weyl semimetal phase}
\end{subfigure}
\begin{subfigure}[b]{0.32\textwidth}
\includegraphics[width=\textwidth]{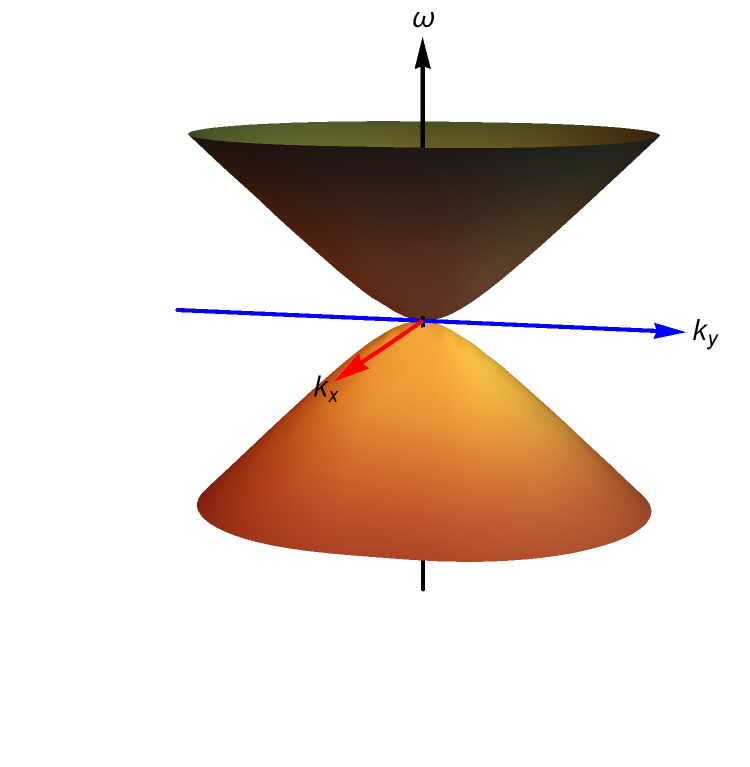}
\caption{\small Critical point}
\end{subfigure}
\begin{subfigure}[b]{0.32\textwidth}
\includegraphics[width=\textwidth]{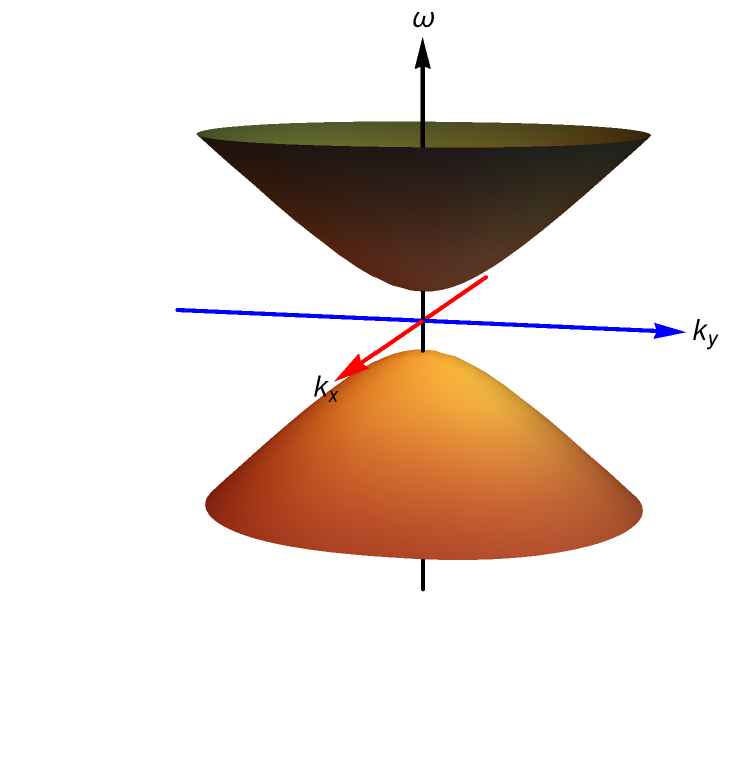}
\caption{\small Trivial phase}
\end{subfigure}
\caption{\small
The figure shows the dispersion relation as a function of momentum for a Weyl semimetal system. The phases of the topological phase transition are shown. From (a) to (c): the system has one pair of Weyl nodes (a), one critical Dirac node (b), and the fully gapped(toplogically trivial phase)(c).}
\label{fig:wsm}
\end{figure}

An effective field theory model for ideal Weyl semimetal could be obtained by adding a Lorentz breaking term into the original Dirac Lagrangian \cite{burkov1, Colladay:1998fq, Grushin:2012mt}
\begin{equation}
\label{eq1}
\mathcal{L}_{wsm} = \bar\psi \left( i \slashed\partial -  e \slashed{A} - \gamma^{\mu}\gamma^5  b_{\mu} + M \right)\psi\,,
\end{equation}
where $\psi$ is a single Dirac spinor, and a time-reversal odd axial gauge field $b_{\mu}$ is introduced. The term $\slashed X = \gamma^\mu X_\mu$, where $\gamma^\mu$ are the Dirac matrices, and $\gamma_5 = i \gamma_0\gamma_1\gamma_2\gamma_3$
allows us to define left- or right-handed spinors via $(1\pm\gamma_5)\psi = \psi_{L,R}$. $\slashed{A}\equiv\gamma^\mu A_\mu$, and $A_\mu$ is the electromagnetic gauge potential. Tuning the ratio between the mass parameter $M$ and the time-reversal symmetry breaking parameter $b$, there exists a topological phase transition from a Weyl semimetal to a trivial semimetal across a critical Dirac semimetal. Without loss of generality, we could choose $\vec{b}$ to be in the $y$ direction so that the two Weyl nodes in the Weyl semimetal phase are separated in the $k_y$ direction of the momentum space. The locations of two Weyl nodes are shown in the left panel of Fig.\ref{fig:wsm}.

When $|b|>|M|$ the spectrum is ungapped. There is a band inversion in the spectrum and at the crossing points the wave function is described by one of the Weyl fermions. The separation of the Weyl points in momentum space is given by $2\sqrt{b^2-M^2}$ along the direction indicated by
the vector $\mathbf{b}$. At low energies it is described by the  effective theory with the Lagrangian of the form (\ref{eq1}) with $M_\mathrm{eff} =0$ and
$\mathbf{b}_{\text{eff}} = \sqrt{b^2-M^2}\mathbf{e}_y$.
For $|b|<|M|$ the system is gapped with gap $2 M_\mathrm{eff} = 2(|M|-|b|)$. The phases of this topological phase transition is shown in Fig.\ref{fig:wsm} and more details could be found in \cite{Landsteiner:2015pdh,Landsteiner:2015lsa}\footnote{Note that here the phases are named based on the value of the corresponding topological invariants, specifically the Chern numbers. In the Weyl semimetal phase, the bands cross at the Weyl point, resulting in non-zero Chern numbers. This differs from gapped topological systems where the band structure only closes at the phase transition points.}.

\subsection{Topological nodal line semimetal}
In three dimensions, two energy bands can intersect either at discrete points or along a closed ring. The crossing points are known as Weyl nodes, which are present in the Weyl semimetal state. The closed ring represents another distinguished state that is named the topological nodal line semimetal. Weyl semimetal states are naturally stable due to strong topological protection. Only the breaking of translation symmetry, which causes the annihilation of the Weyl points, can destroy them. However, to maintain a stable nodal line semimetal state, additional symmetries of the system are required. These symmetries include chiral symmetry, PT symmetry, glide mirror symmetry, and mirror reflection symmetry\cite{fang1}. 

Here, we give an effective theoretic description of a nodal line semimetal with mirror reflection symmetry. The Lagrangian could be written as\cite{Liu:2018bye} 
\be\label{eq3}
\mathcal{L}_{nl}=\bar{\psi}\big(i\gamma^\mu\partial_\mu-m-\gamma^{\mu\nu} b_{\mu\nu}\big)\psi,
\ee
where $\bar{\psi}=\psi^{\dagger}\gamma^{0}$, $\gamma^{\mu\nu}=\frac{i}{2}[\gamma^\mu, \gamma^\nu]\,$ and $b_{\mu\nu}=-b_{\nu\mu}$ is an external antisymmetric two-form field.s We switch on a non-zero constant $b_{xy}$ component of the two-form field and obtain the energy spectrum as:
\be
\label{eqnodalq}
E_\pm^{\left(nodal\right)}=\pm \sqrt{k_z^2+\Big(2b_{xy}\pm\sqrt{m^2+k_x^2+k_y^2}\,\Big)^2}\,.
\ee
\begin{figure}
\begin{center}
\includegraphics[width=\textwidth]{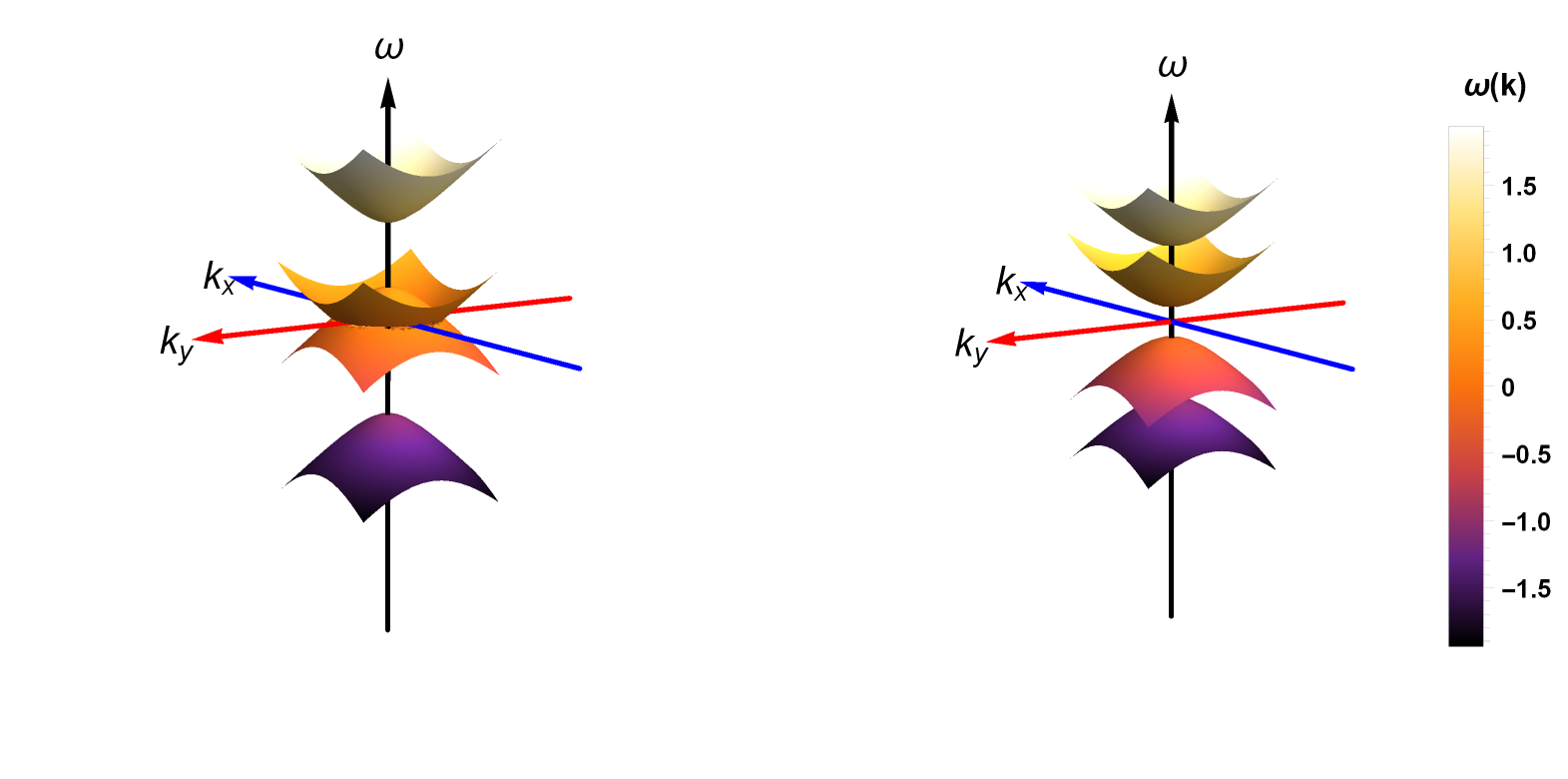}
\end{center}
\vspace{-0.3cm}
\caption{\small The energy spectrum as a function of $k_x, k_y$ at $k_z=0$ for the nodal line semimetal in Eq. (\ref{eqnodalq}). Left: there is a nodal circle at the band crossing when $m^2< 4b_{xy}^2$. Right: for $m^2 > 4b_{xy}^2$ the system is gapped\cite{Liu:2018bye}.}
\label{fig:nodal}
\end{figure}

By adjusting the value of two free parameters $m$ and $b_{xy}$, we obtain three phases in this system, as shown in Fig.\ref{fig:nodal}. The system is a topological nodal semimetal whose Fermi points form a connected circle of radius $\sqrt{4 b_{xy}^2-m^2}$ in momentum space when $m^2< 4b_{xy}^2$. Note that in this parameter regime, a small mass term cannot lead to a gap in the system. The radius becomes zero when $m^2= 4b_{xy}^2$, which is the quantum critical point for the topological phase transition. For $m^2 > 4b_{xy}^2$ the system becomes topological trivial and behaves like an insulator. In the nodal line phase, we see that close to the Fermi line, the dispersion is linear in $\sqrt{k_x^2+k_y^2}-\sqrt{4 b_{xy}^2-m^2}$ with velocity $\sqrt{1-\frac{m^2}{4b_{xy}^2}}$ when $k_z=0$ and linear in $k_z$ with velocity $1$, where we have set $c=1$, if $\sqrt{k_x^2+k_y^2}=\sqrt{4 b_{xy}^2-m^2}$.

\section{Topological phase transitions between Weyl and nodal line semimetals in effective field theory}
\label{four}
As mentioned above, additional symmetries exist to maintain a stable topological nodal line semimetal state. When these symmetries are broken by interactions, the nodal ring is no longer stable and becomes fully gapped or broken into several nodal points. Let us take TaAs as an example. According to the first-principle calculations, there are two stable nodal rings in the energy bands of TaAs under mirror reflection and spin rotation symmetry.  However, when a non-zero spin-orbit coupling appears, these symmetries break, and each nodal ring becomes three pairs of Weyl nodes\cite{weng,shuang}. Thus, it is interesting to study the phase transition between the Weyl semimetal state and the nodal line semimetal state in the effective field theory. In this section we will try to build effective field theory models to check the phase transitions between Weyl semimetal and nodal line semimetal, and study the existence of other possible topological states during this process. 

To achieve this we start from the Lagrangian including both the Weyl semimetal solutions and the nodal line semimetal solutions, i.e. combining Eq.\ref{eq1} and Eq.\ref{eq3}\footnote{We would like to mention here that a complete field theory model for the realisation of the nodal semimetal must take into account the duality relation raised in\cite{Liu:2020ymx}, where a pure imaginary field $b_{\mu\nu}^{5}$ is considered. In this paper, however, we focus only on the combination of the one-form and the real two-form fields. The effects of the pure imaginary field will be considered in a future paper.}
\be
\label{eq4}
\mathcal{L}_{nw}=\bar{\psi}\big( i \slashed\partial -  e \slashed{A}-m-\gamma^{\mu\nu} b_{\mu\nu} - \gamma^{\mu}\gamma^5  b_{\mu}\big)\psi.
\ee
Without loss of generality, we could turn on a non-zero constant $b_{xy}$ component of the two-form field, and then there are two nonequivalent choices of $\vec{b}$. One is to have $\vec{b}$ in the $z$ direction and the other is to put $\vec{b}$ in the $x$(or $y$) direction. Before we proceed, let us look into the physical meanings of the parameters in Eq.\eqref{eq4}. Similar to the the cases of effective field theories for Weyl semimetals and nodal line semimetals, the combinations of the values of $b$, $b_{xy}$ and $m$ determine the radius of the nodal ring, the distance between two Weyl nodes and the energy gap is represented by the effective mass. Adjusting these parameters creates various types of states, including surface states, which may be observed through ARPES experiments in laboratory systems. We will discuss in detail in later sections.

We would be able to see which topological state the system is depending on the relative values of $b_{xy}$ and $b$ and the final result would be a phase diagram in the $b_{xy}$ and $b$ plane. Also, the system behaves differently depending on the direction of $\vec{b}$. This is due to what we emphasized above, that the nodal state is fragile without the protection of mirror symmetry. Since the non-zero $b_{xy}$ preserves the mirror symmetry in the $x-y$ plane, a nodal ring can survive in this plane. If we put the non-zero component of the one-form field in the $z$ direction, the mirror symmetry will not be broken and the nodal line state and the Weyl semimetal state can coexist. On the other hand, as soon as we turn on a non-zero component in the $x$ (or $y$) direction of the one-form field, the mirror symmetry is broken, then the nodal line state vanishes immediately and the Weyl semimetal state and the nodal line state could not coexist in this case. These two cases will be discussed separately below.

\subsection{\texorpdfstring{$\vec{b}$}. in the \texorpdfstring{$z$}. direction}
We start by setting $\vec{b}$ in the $z$ direction and solving the model we could obtain the eigenenergies of the system. Due to the complication of Eq.\eqref{eq4}, the energy spectrum has no explicit analytic expression. Nevertheless, we could plot the energy spectrum numerically. Since the nodal line semimetal stems from the two-form field, thus if we set $b_{xy}=0$ in the obtained energy spectrum, the energy spectrum of Eq.\eqref{eq4} should return to the Weyl semimetal case, i.e. the eigenenergies of Eq.\ref{eq1}. Then we can switch on a non-zero value of $b_{xy}$ to check how the Weyl semimetal state evolves under the contributions from the two-form field. As we know, with different choices of parameters $b$ and $M$, there are three phases for the Lagrangian \eqref{eq1}, including the Weyl semimetal phase, the Dirac semimetal (critical) phase, and the topologically trivial (gapped) phase. By turning on the non-zero value of $b_{xy}$ in each of these phases, the transition between the Weyl semimetal state and the nodal line state could be detected. 

\begin{figure}[h!]
  \centering
\includegraphics[width=\textwidth]{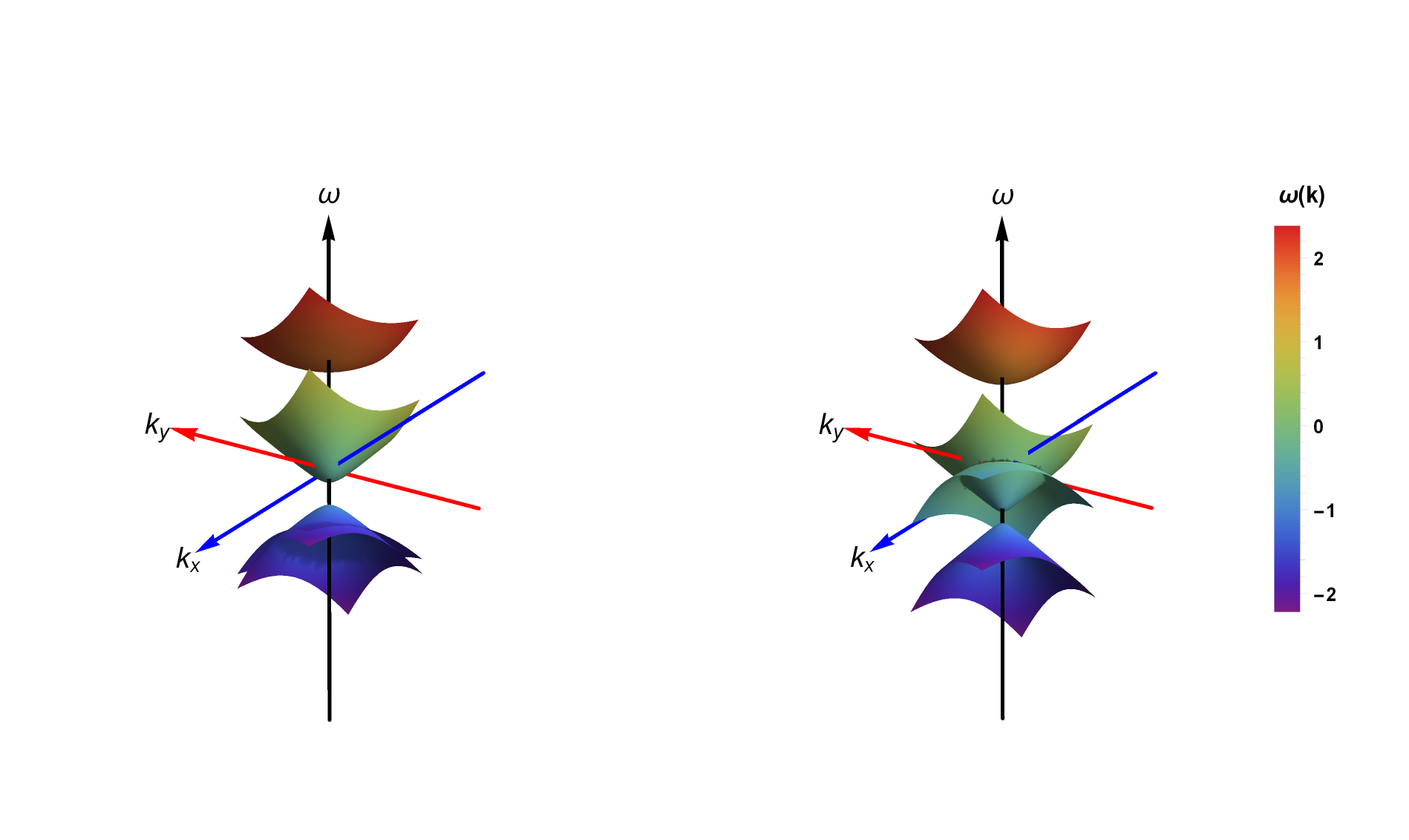}
 \caption{\small The energy spectrum of \eqref{eq4} as a function of $k_x, k_y$ at $k_z=0$ with the given $b$ and $M$. Left: a nodal circle is formed by the lowest and the second lowest energy band when $2b_{xy}<M$(named as nodal ring A). Right: another nodal circle is formed by the second and third lowest energy bands when $2b_{xy}>M$(named as nodal ring B).}
  \label{fig11}
\end{figure}

Since the Weyl nodes locate along the direction indicated by the vector $\vec{b}$, if we set $\vec{b}$ to be in the $z$ direction, we will plot the energy spectrum of \eqref{eq4} as a function of $k_z$ with $k_x=k_y=0$ to check the existence of the Weyl nodes. From this perspective, the potential nodal ring expanding at $k_x-k_y$ plane cannot be observed in the plots. Therefore, before we proceed to check the effect of the interplay of the one-form and two-form fields, we will fix $b$ and $M$ first to check the existence of the nodal line semimetal. The energy spectrum of \eqref{eq4} with fixed $b$ and $M$ as a function of $k_x, k_y$ at $k_z=0$ is shown in Fig.\ref{fig11}.

From Fig.\ref{fig11} we see that with a fixed value of $b$ for the non-zero $z$-component of the one-form field and a fixed value of $M$ for the mass, a nodal ring can form immediately once $b_{xy}$ is non-zero. When $2b_{xy}<M$ the system has a nodal ring formed by the lowest and the second lowest energy bands (left panel in Fig.\ref{fig11}, named as nodal ring A), while when $2b_{xy}>M$ the system forms another nodal ring by the second and the third lowest energy bands (right panel in Fig.\ref{fig11}, named as nodal ring B). The critical point when the radius of the ring becomes zero is at $2b_{xy}=M$. This is different from the case of pure topological nodal line semimetals \eqref{eq3}, where the nodal ring survives only at $2b_{xy}>M$, and no nodal ring can be formed if $2b_{xy}\leq M$. The different behavior indicates that the nodal rings described by \eqref{eq3} and \eqref{eq4} are two different types of nodal rings. These two types of nodal rings can be distinguished by calculating the topological charge. As mentioned in \cite{fang1}, for a nodal ring we could define two independent $\boldmath{Z}_2$ topological charges ($\zeta_1$ and $\zeta_2$). $\zeta_1$ is the topological charge calculated from a circle linking the nodal ring, which describes if each node in the nodal ring is accidental or topologically protected. While $\zeta_2$ is the topological invariant calculated from a sphere enclosing the whole nodal ring, which describes if the nodal ring could be subsequently gapped when it shrinks to a point by tuning the parameters continuously. More precisely, if $\zeta_2=0$ the topological charge of the critical point should also zero, while for $\zeta_2=1$ the topological charge of the critical point should also be 1. From this point of view, the topological charge for the critical point of the nodal ring described in \eqref{eq3} would be zero, while a non-zero topological charge should be obtained for the critical point of the nodal ring described in \eqref{eq4}, we will further confirm this point in section \ref{sec:invariant}. 

Having checked the existence of nodal rings in the system, we now focus mainly on how the three phases of the Weyl semimetal evolve under the non-zero two-form field. The details of parameter region are shown in appendix (\ref{app:a}). A complete phase diagram in $b_{xy}$ and $b$ space is given below in Fig.\ref{fig21}.

\begin{figure}[h!]
  \centering
\includegraphics[width=0.6\textwidth]{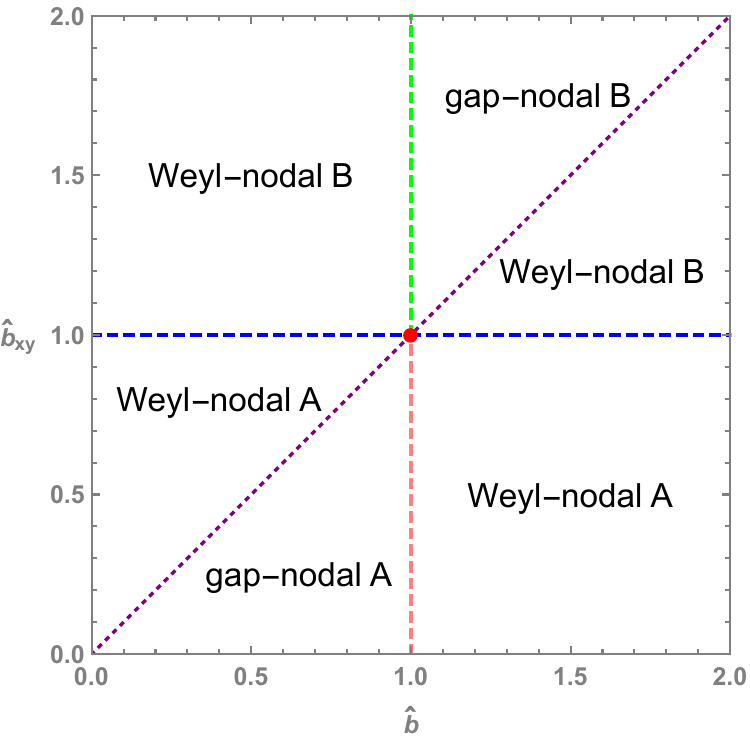}
 \caption{\small The phase diagram of the system \eqref{eq4} with two dimensionless parameters $\hat{b}=b/M$ and $\hat{b}_{xy}=2b_{xy}/M$ when $\vec{b}$ is in the $z$-direction. The red point is the triple degenerate nodal point at which both Weyl nodes and nodal ring become critical (Fig.\ref{fig4}(c)). The horizontal blue dashed line corresponds to the critical three node phase where the radius of the nodal ring becomes zero while a pair of Weyl nodes still exists (Fig.\ref{fig3}(c) in appendix and Fig.\ref{fig5}(e)). The vertical pink dashed line corresponds to the critical phase in which the pair of Weyl nodes annihilates into a critical Dirac point while nodal ring A still survives (Fig.\ref{fig4}(b)). The vertical green dashed line corresponds to the critical phase in which the pair of Weyl nodes annihilates into a critical Dirac point while nodal ring B still survives (Fig.\ref{fig4}(d)). The diagonal purple dotted lines correspond to the phase where the pair of Weyl nodes annihilates into a critical Dirac point while nodal ring A or B still survives (Fig.\ref{fig3}(e) (nodal ring A) and Fig.\ref{fig5}(c) (nodal ring B), the critical Dirac node phase). Note that the $b_{xy}=0$ axis corresponds to states with no nodal rings.}
  \label{fig21}
\end{figure}

With a fixed value of $M$, we can draw the phase diagram with two dimensionless parameters, $\hat{b}=b/M$ and $\hat{b}_{xy}=2b_{xy}/M$. This diagram captures the most essential physics, which are listed below:

\vspace{.25cm}
\noindent {\bf $\bullet$ \em triple degenerate nodal point}

In Fig.\ref{fig21}, the red point is the newly obtained triple degenerate nodal point, where both the Weyl nodes and the nodal ring become the critical point (Fig.\ref{fig4}(c)). 

\vspace{.25cm}
\noindent {\bf $\bullet$ \em Weyl-critical phases}

The horizontal blue dashed line corresponds to the critical phase where the radius of the nodal ring (either nodal ring A or B) becomes zero while a pair of Weyl nodes still exist (Fig.\ref{fig3}(c) and Fig.\ref{fig5}(e), the three nodes phases).

\vspace{.25cm}
\noindent {\bf $\bullet$ \em nodal-critical phases}

The vertical pink dashed line corresponds to the critical phase in which the pair of Weyl nodes annihilate into a critical Dirac point while nodal ring A still survives (Fig.\ref{fig4}(b), the shifted Dirac point phases). 

The vertical green dashed line corresponds to the critical phase in which the pair of Weyl nodes annihilates into a critical Dirac point while nodal ring B still survives (Fig.\ref{fig4}(d), the shifted Dirac point phases). 

The diagonal purple dotted lines correspond to the phase where the pair of Weyl nodes annihilate into a critical Dirac point while nodal ring A or B still survives (Fig.\ref{fig3}(e) (nodal ring A survives) and Fig.\ref{fig5}(c) (nodal ring B survives), the critical Dirac node phase).

\vspace{.25cm}
\noindent {\bf $\bullet$ \em Weyl-nodal phases}

The upper-triangular region of the lower-left and the entire lower-right part of the phase diagram correspond to the phase in Fig.\ref{fig5}(d) and Fig.\ref{fig3}(b) respectively (i.e. the Weyl-nodal A phase).

The entire upper-left part and the lower-triangular region of the upper-right of the phase diagrams corresponds to the phase in Fig.\ref{fig5}(f) and Fig.\ref{fig3}(d) respectively (i.e. the Weyl-nodal B phase).

\vspace{.25cm}
\noindent {\bf $\bullet$ \em gap-nodal phases}

The lower-triangular region of the lower-left of the phase diagrams corresponds to the phase in Fig.\ref{fig5}(b) (i.e. the gap-nodal A phase). 

The upper-triangular region of the upper-right of the phase diagrams corresponds to the phase in Fig.\ref{fig3}(f) (i.e. the gap-nodal B phase). 

In summary, we have coexistence of Weyl nodes and a nodal ring in this case. Note that a non-zero component $b_{tz}$ would have a  contribution as mass terms if we focus the nodal ring formed in the $x-y$ plane. Other components of the two-form field, such as $b_{ty},b_{xz}$, may yield more interesting energy spectrums since they break the mirror reflection symmetry in the $x-y$ plane, and we leave this to future work.

\subsection{\texorpdfstring{$\vec{b}$}. in the \texorpdfstring{$x$}. direction}
\begin{figure}[h!]
\begin{center}
\begin{subfigure}[b]{0.35\textwidth}
\includegraphics[width=\textwidth]{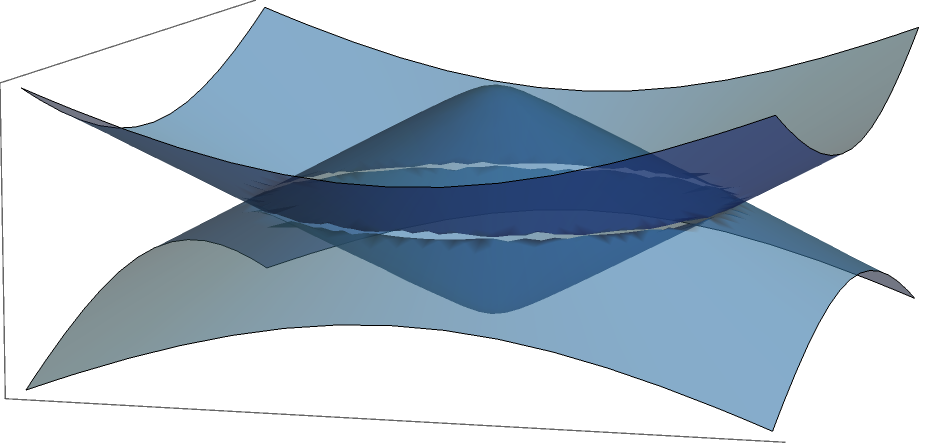}
\caption{\small nodal line semimetal}
\end{subfigure}
\begin{subfigure}[b]{0.35\textwidth}
\includegraphics[width=\textwidth]{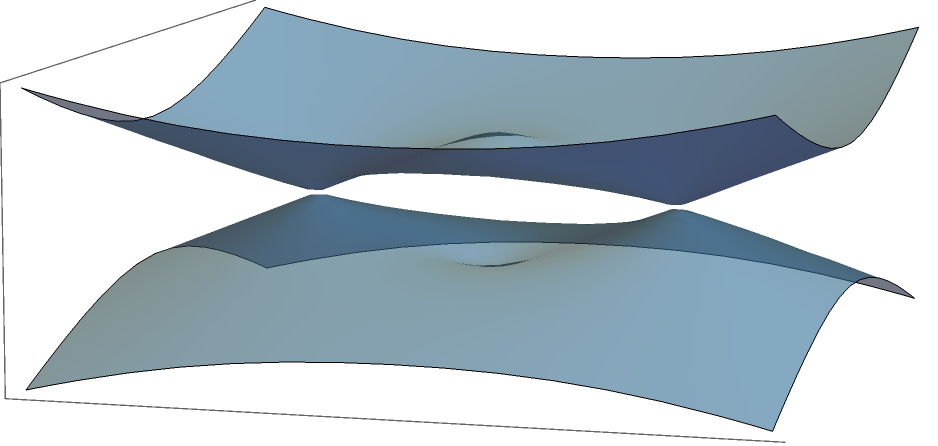}
\caption{\small two Weyl points}
\end{subfigure}
\begin{subfigure}[b]{0.35\textwidth}
\includegraphics[width=\textwidth]{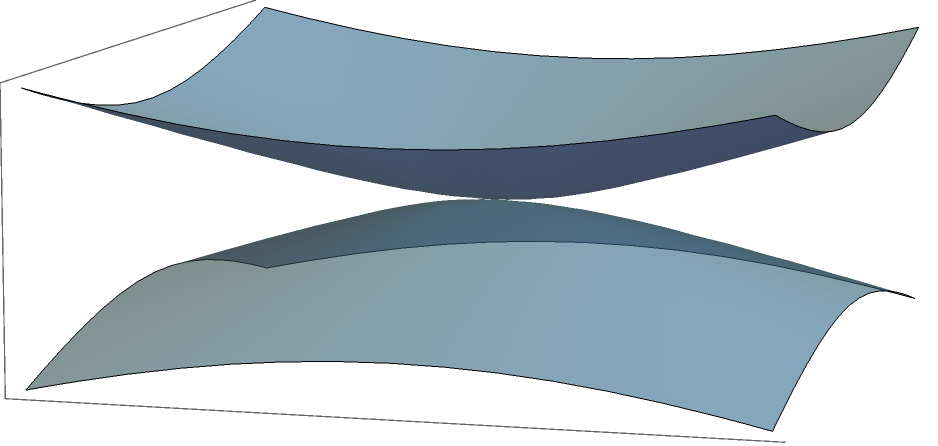}
\caption{\small Dirac point}
\end{subfigure}
\begin{subfigure}[b]{0.35\textwidth}
\includegraphics[width=\textwidth]{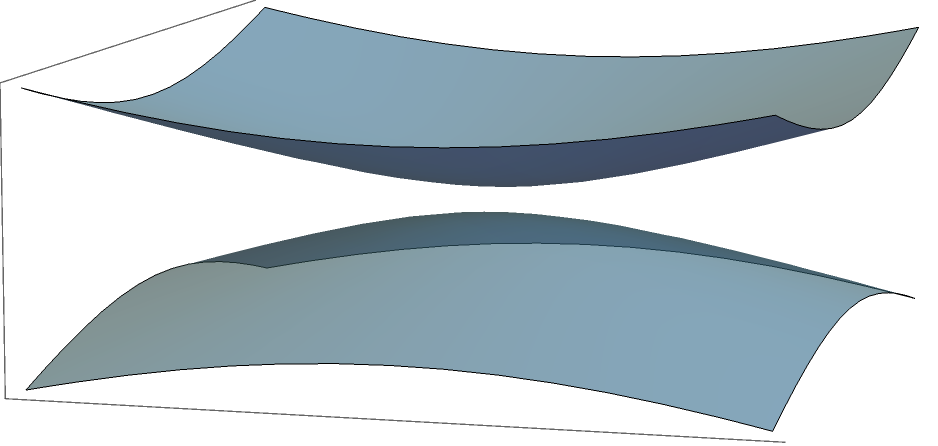}
\caption{\small gapped state}
\end{subfigure}
\end{center}
\vspace{-0.3cm}
\caption{\small The energy spectrum as a function of $k_x, k_y$ for $k_z=0$. From (a) to (d): (a): the system is a topological nodal line semimetal with $M^2< 4b_{xy}^2$ and $b=0$, (b): two Weyl nodes when $b\neq 0$, (c): a critical Dirac node if $M=M_{(c)}=\sqrt{b^2+4b_{xy}^2}$ or $b_{xy}=b_{xy(c)}=\frac{\sqrt{M^2-b^2}}{2}$ is satisfied, and (d): gapped states if $M>M_{(c)}$ or $b_{xy}<b_{xy(c)}$.}
\label{fig6}
\end{figure}

Although the phase structure of the system has been presented from the above prescription, it is not sufficient since we can only study the system starting from the Weyl semimetal state if the location of the Weyl nodes is orthogonal to the plane where the nodal line lives. As mentioned above, if we switch on the $x$ or $y$ component of the one-form field, we may intuitively check how this influences and destroys the nodal line. This will move the location of the Weyl nodes to the $k_z=0$ plane. We try to do this without loss of generality by setting $\vec{b}$ in the $x$ direction, i.e., only the $x$-component of the one-form field is non-zero. The eigen-energies of this case are
\be
\label{eq:eigenstate11}
E_\pm=\pm \sqrt{b^2+4b_{xy}^2+k_x^2+k_y^2+M^2\pm2\sqrt{b^2\left(M^2+k_x^2\right)+4b_{xy}^2\left(M^2+k_x^2+k_y^2\right)}}\,.
\ee
We start from the topological nodal line semimetal state, i.e. the parameters satisfying $M^2< 4b_{xy}^2$ ((a) in Fig.\ref{fig6}). The radius of the nodal ring is $\sqrt{4b_{xy}^2-M^2}$ and this state of the nodal line is the same as the one described in \eqref{eq3}. Then we turn on $b$ while keep the value of $M$ and $b_{xy}$ fixed, and the nodal ring is immediately destroyed with two Weyl nodes formed ((b) in Fig.\ref{fig6}). This is easy to understand from the symmetric analysis since a non-zero x-component of the one-form field breaks the mirror reflection symmetry of the $x-y$ plane. There are two Weyl nodes located along the $k_x$ axis and the distance between them is $2\sqrt{b^2+4b_{xy}^2-M^2}$. For small $b$, its effect is not strong enough, so the two Weyl nodes appear on opposite sides of a diameter of the original ``nodel ring", i.e. close to the points $\left(k_x,k_y,k_z\right)=\left(\pm\sqrt{4b_{xy}^2-M^2},0,0\right)$. The distance between two Weyl nodes changes slowly when $b$ is less than $b_{xy}$. When $b$ increases as it approaches the value of $b_{xy}$, the distance between two Weyl nodes increases much faster than in the case of small $b$. 

The distance of two Weyl nodes could also become small in two ways. For fixed $b$ and $b_{xy}$, an increasing $M$ could descrease the value of $\sqrt{b^2+4b_{xy}^2-M^2}$ to zero (when $M=M_{(c)}=\sqrt{b^2+4b_{xy}^2}$); While for fixed $b$ and $M$, a decreasing $b_{xy}$ could also put the value of $\sqrt{b^2+4b_{xy}^2-M^2}$ to zero (when $b_{xy}=b_{xy(c)}=\frac{\sqrt{M^2-b^2}}{2}$). Two Weyl nodes annihilate to form one Dirac node under each of these two processes ((c) in Fig.\ref{fig6}). Finally, further increasing $M$ and taking $M>M_c$ (or decreasing $b_{xy}$ while $b_{xy}<b_{xy(c)}$) could give a gapped state ((d) in Fig.\ref{fig6}). All of these four phases are shown in Fig.\ref{fig6}.

\begin{figure}[h!]
  \centering
\includegraphics[width=0.6\textwidth]{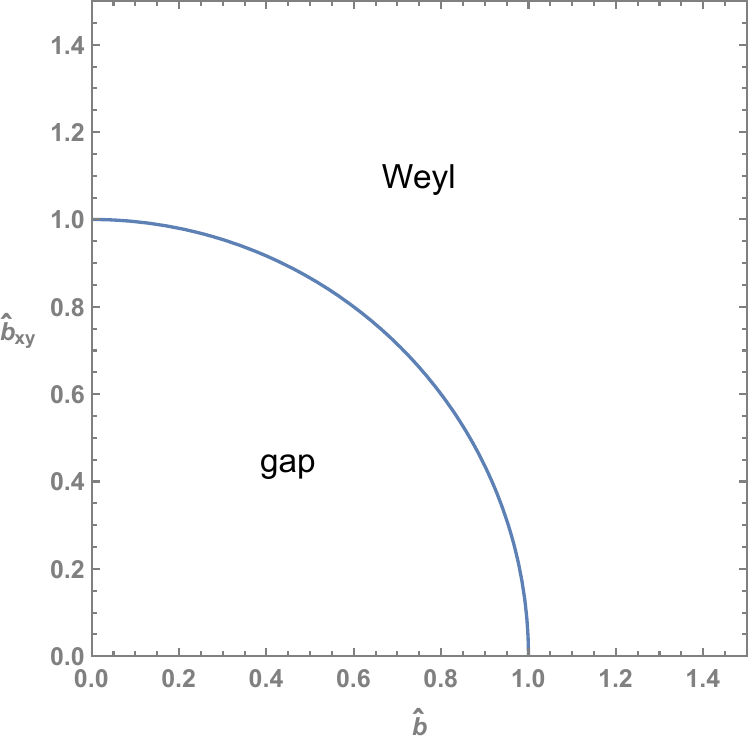}
 \caption{\small The phase diagram of the system \eqref{eq4} with two dimensionless parameters $\hat{b}=b/M$ and $\hat{b}_{xy}=2b_{xy}/M$ when $\vec{b}$ is in the $x$-direction. The quarter-circle satisfying $\sqrt{\hat{b}^2+\hat{b}_{xy}^2}=1$ shows the Dirac point((c) in Fig.\ref{fig6}). Inside and outside the circle is the gapped phase((d) in Fig.\ref{fig6}) and the Weyl semimetal state((b) in Fig.\ref{fig6}), respectively.}
  \label{fig22}
\end{figure}

{Similar to the case in Fig.\ref{fig21}, we can plot a complete phase diagram in Fig.\ref{fig22} depending on $\hat{b}=b/M$ and $\hat{b}_{xy}=2b_{xy}/M$. As mentioned above, the critical value of the mass (or $b_{xy}$) can be obtained when the distance between the two Weyl nodes becomes zero, and we find that it can be described by a quarter-circle $\sqrt{\hat{b}^2+\hat{b}_{xy}^2}=1$. Thus the quarter-circle in Fig.\ref{fig22} represents the critical point in the system ((c) in Fig.\ref{fig6}). Inside and outside the circle is the gapped phase ((d) in Fig.\ref{fig6}) and the Weyl semimetal state((b) in Fig.\ref{fig6}), respectively. This phase diagram is much simpler than the case in Fig.\ref{fig21}, since only the Weyl semimetal survives when the mirror reflection symmetry is broken by the one-form field.}

However, we will not be able to observe the Weyl nodes and the nodal line at the same time if we choose a non-zero component of the one-form field along the $x$ (or $y$) direction. An extension of the Hilbert space might be able to solve the problem. This will be discussed in the next section.

\subsection{Surface states}
These topological phase transitions could also be reflected from the behaviour of the surface states. In the lattice model, the surface state can be obtained naturally at the boundary of the Brillouin zone, e.g. the [001] surface. In the effective theory model there is no such natural boundary, in other words the whole parameter space can be considered as a Brillouin zone. Nevertheless, the surface state can be obtained by adding a fictitious boundary in the component of the field\cite{Goswami, Ammon:2016mwa}. Following this idea, we will first calculate the surface state in the nodal semimetal (Eq.\eqref{eq3}) by setting the $xy$ component of the $b$ field as $b_{xy}(x,y,z)=-b_{yx}(x,y,z)=b_{xy}\theta\left({z}\right)$ as the boundary, where $\theta\left({z}\right)$ is the step function. We then concentrate on the surface states of eq.\eqref{eq4}. Only the case where the one-form field has a non-zero z-component is considered, because under this circumstance a coexistence state with the Weyl semimetal and the topological nodal line state survives, leading to a complicated surface state. We will set $b_z=b \theta\left({x}\right)$ and $b_{xy}=-b_{yx}=b_{xy}\theta\left({x}\right)$ to play the role of a fictitious boundary and check the possible realised surface states.

We start with the Hamiltonian induced from eq.\eqref{eq3} by setting $b_{xy}=-b_{yx}=b_{xy}\theta\left({z}\right)$, which is
\begin{eqnarray}
\label{hnl}
H_{nl}=\left(\begin{array}{cc}
-i\mathbf{\sigma}\cdot\mathbf{\nabla} & m\mathbf{I}+2b_{xy}\theta\left({z}\right)\mathbf{\sigma}_{z}\\
m\mathbf{I}+2b_{xy}\theta\left({z}\right)\mathbf{\sigma}_{z} & i\mathbf{\sigma}\cdot\mathbf{\nabla}
\end{array}\right).
\end{eqnarray}

The eigenvalue equation can be written as $H_{nl}\Upsilon_{nl}\left(x,y,z\right)=E_{nl}\Upsilon_{nl}\left(x,y,z\right)$, where $\Upsilon_{nl}\left(x,y,z\right)$ is the eigenfunction of the eq.\eqref{hnl}. Since we have chosen $b_{xy}=-b_{yx}=b_{xy}\theta\left({z}\right)$, then $\Upsilon_{nl}\left(x,y,z\right)$ can be written as $\Upsilon_{nl}\left(x,y,z\right)=e^{ik_{x}x+ik_{y}y}\varUpsilon_{nl}\left(z\right)$, where $\varUpsilon_{nl}\left(z\right)=\left(\begin{array}{c}
u_{1}\left(z\right)\\
iu_{1}\left(z\right)\\
u_{2}\left(z\right)\\
-iu_{2}\left(z\right)
\end{array}\right)$.

The $u_i\left(z\right),i=1,2$ represent the surface state solutions with $E=+k_{y}$ responsible for generating the drumhead surface states in the ARPES measurements. $u_i\left(z\right),i=1,2$ can be obtained by substituting the form of $\Upsilon_{nl}\left(x,y,z\right)$ and $b_{xy}=-b_{yx}=b_{xy}\theta\left({z}\right)$ into the eigenvalue equation. We obtain 
\begin{eqnarray}
&u_{2}\left(z\right)\left(z<0\right)=A e^{\sqrt{k_{x}^{2}+m^{2}}z},
u_{2}\left(z\right)\left(z>0\right)=A e^{-\sqrt{k_{x}^{2}+m^{2}-4b_{xy}^{2}}z},\nonumber\\
&u_{1}\left(z\right)=B u_{2}\left(z\right).
\end{eqnarray}
Details could be found in appendix \ref{app:b1}.

\begin{figure}[h!]
  \centering
\includegraphics[width=0.8\textwidth]{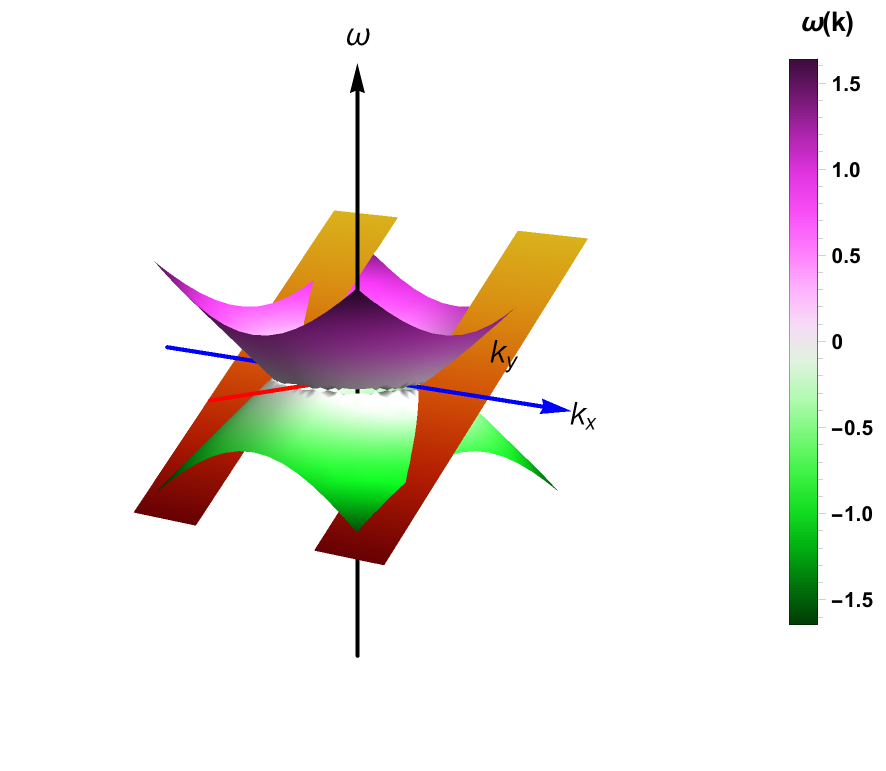}
 \caption{\small The energy spectrum $\omega$ (with fixed values of $b_{xy}$ and $m$) of \eqref{eq3} with respect to $k_x$ and $k_y$. The surface states (with $\omega=k_y$) at $k_x>\sqrt{4b_{xy}^2-m^2}$ and $k_x<-\sqrt{4b_{xy}^2-m^2}$ are also shown in the figure.}
  \label{fig441}
\end{figure}

In Fig.\ref{fig441}, we plot the energy spectrum $\omega$ (with fixed values of $b_{xy}$ and $m$) of \eqref{eq3} with respect to $k_x$ and $k_y$. The surface states (with $\omega=k_y$) at $k_x>\sqrt{4b_{xy}^2-m^2}$ and $k_x<-\sqrt{4b_{xy}^2-m^2}$ are also shown in the figure.The drumhead surface state measured in ARPES can be viewed as a projection of Fig. \ref{fig441} in the plane where $\omega$ is a fixed constant.

Now we will study the Hamiltonian induced from eq.\eqref{eq4}. By setting $b_z=b \theta\left({x}\right)$ or $b_{xy}=-b_{yx}=b_{xy}\theta\left({x}\right)$ we obtain the corresponding Hamiltonian as
\begin{eqnarray}
\label{hnw}
H_{nw}=\left(\begin{array}{cc}
-i\mathbf{\sigma}\cdot\mathbf{\nabla}+\mathbf{\sigma}_{z}\mathbf{b} & m\mathbf{I}+2b_{xy}\mathbf{\sigma}_{z}\\
m\mathbf{I}+2b_{xy}\mathbf{\sigma}_{z} & i\mathbf{\sigma}\cdot\mathbf{\nabla}+\mathbf{\sigma}_{z}\mathbf{b}
\end{array}\right).
\end{eqnarray}

The eigenvalue equation can be written as $H_{nw}\Upsilon_{nw}\left(x,y,z\right)=E_{nw}\Upsilon_{nw}\left(x,y,z\right)$, where $\Upsilon_{nw}\left(x,y,z\right)$ is the eigenfunction of  eq.\eqref{hnw}. Since we have chosen $b_z=b \theta\left({x}\right)$(or $b_{xy}=-b_{yx}=b_{xy}\theta\left({x}\right)$), then $\Upsilon_{nw}\left(x,y,z\right)$ can be written as $\Upsilon_{nw}\left(x,y,z\right)=e^{ik_{y}y+ik_{z}z}\varUpsilon_{nw}\left(x\right)$, where $\varUpsilon_{nw}\left(x\right)=\left(\begin{array}{c}
v_{1}\left(x\right)\\
iv_{1}\left(x\right)\\
v_{2}\left(x\right)\\
-iv_{2}\left(x\right)
\end{array}\right)$. $v_i\left(x\right),i=1,2$ represent the surface state solutions with $E_{nw}=+k_{y}$ responsible for producing the Fermi arc and drumhead surface states in the ARPES measurements. They can be obtained by substituting the form of $\Upsilon_{nw}\left(x,y,z\right)$ and $\mathbf{b}=b \theta\left({x}\right)\vec{e}_{z}$(or $b_{xy}=-b_{yx}=b_{xy}\theta\left({x}\right)$) into the eigenvalue equation and the expressions of $u_{i}\left(x\right),i=1,2$ can be found in the appendix \ref{app:b2}.

\begin{figure}[h!]
  \centering
\includegraphics[width=0.7\textwidth]{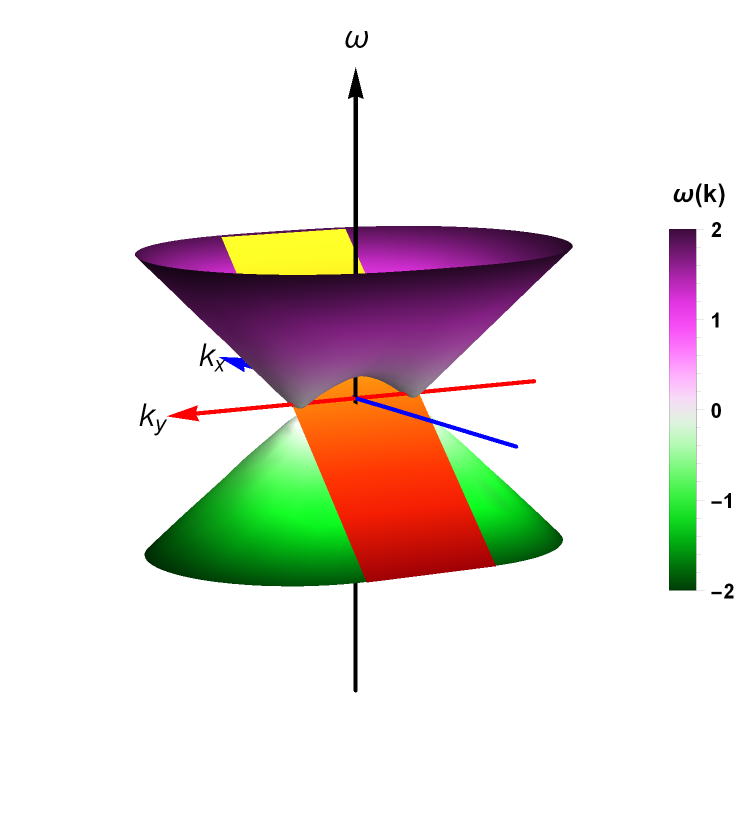}
 \caption{\small The energy spectrum $\omega$ (with fixed values of $b$,$b_{xy}$ and $m$) of \eqref{eq4} with respect to $k_y$ and $k_z$. The surface state (with $\omega=k_y$) at $-\sqrt{b^{2}-m^{2}+4b_{xy}^{2}}<k_z<\sqrt{b^{2}-m^{2}+4b_{xy}^{2}}$ is also shown in the figure in red.}
  \label{fig442}
\end{figure}

In Fig. \ref{fig442}, we plot the energy spectrum $\omega$ (with fixed values of $b$, $b_{xy}$ and $m$) of \eqref{eq4} with respect to $k_y$ and $k_z$. The surface state (with $\omega=k_y$) at $-\sqrt{b^{2}-m^{2}+4b_{xy}^{2}}<k_z<\sqrt{b^{2}-m^{2}+4b_{xy}^{2}}$ is also shown in the figure. The Fermi arc measured in ARPES can be viewed as a projection of Fig. \ref{fig442} in the plane where $\omega$ is a fixed constant. We have not plotted the drumhead surface states because the nodal ring is only present in the $k_z=0$ plane, while the surviving regions of the surface state we obtained are measured by the value of $k_z$. The drumhead surface states survive in the region of $-\sqrt{b^{2}-m^{2}+4b_{xy}^{2}}<k_z<-\sqrt{b^{2}-m^{2}}$ and $\sqrt{b^{2}-m^{2}}<k_z<\sqrt{b^{2}-m^{2}+4b_{xy}^{2}}$ according to the results obtained in appendix \ref{app:b2}. Hence, we may safely conclude that a discontinuous "Fermi arc" observed in a symmetric region is a signature for the coexistence of the Weyl semimetal state and the topological nodal state.

\section{Topological phase transitions of semimetal states in eight-component spinor formalism}
\label{eight}
In section \ref{four}, we showed the phase structure in the presence of a one-form field and a two-form field with two different choices of the non-zero component of the one-form field. Quite different behaviors result from these two choices. Specifically, the nodal state and the Weyl semimetal state can coexist if the non-zero component of the one-form field along the axis orthogonal to the plane where the nodal ring survives. While only the Weyl semimetal state can survive if the non-zero component of the one-form field along the axis inside the the plane where the nodal ring survives since mirror reflection symmetry is broken now. To solve this problem, we can try to extend the original Hilbert space to include more degrees of freedom and symmetries. In this section, we will extend the Hilbert space to realize the Weyl semimetal state and the nodal line semimetal state simultaneously when the non-zero component of the one-form field lies in the plane where the nodal ring survives. The simplest way to enlarge the Hilbert space is just to do the direct product of two small Hilbert spaces. The minimum extension for a Hilbert space that could be described by a four-component spinor is the space that is described by an eight-component spinor. Accordingly, we will define an eight-component spinor to describe the system in the following. This idea has been used in the effective field theory model of the $\boldmath{Z}_2$-Weyl semimetal. A brief review for the $\boldmath{Z}_2$-Weyl semimetal in the eight-component spinor formalism is in the appendix (\ref{appendixz2}) for the reference of readers who are not familiar with eight-component spinors. Here we use an eight-component spinor to study the phase transition between the Weyl semimetal and the topological nodal line semimetal.

Different from the $\boldmath{Z}_2$-Weyl case, we need to introduce an index to describe the mirror symmetry in addition to chirality and particle-hole. Thus another eight-component spinor, different from the case of the $\boldmath{Z}_2$-Weyl semimetal, could be introduced. The new eight-component spinor does not contain the spin degree of freedom anymore. 

We will use the generalized $8\times8$ matrices $\Gamma^{\mu}$ and $\Gamma^{5}$ defined in Eq.\eqref{eq:88marix} to build this effective theory model. We then introduce another eight-component spinor $\Phi$ to enlarge the Hilbert space and include more symmetry. A Lagrangian that could realize Weyl semimetal and nodal line semimetal is written as:
\be\label{eq:1Lagrangian}
\mathcal{L}&=&\mathcal{L}_1+\mathcal{L}_2\nonumber\\
&=&\Phi^{\dagger}\Gamma^{0}\left[\left(i\Gamma^{\mu}\partial_{\mu}-\Gamma^{\mu\nu}b_{\mu\nu}+M_1\right){\mI}_{1}+\left(i\Gamma^{\mu}\partial_{\mu}-\Gamma^{0}\Gamma^{\mu}\Gamma^5 b_{\mu} + M_2\right){\mI}_{2}\right]\Phi,
\ee
where $\Gamma^{\mu\nu}=\frac{i}{2}[\Gamma^\mu, \Gamma^\nu]\,$. $ {\mI}_{1}$ and $ {\mI}_{2}$ are two diagonal matrices with diagonal elements as $ {\mI}_1=\text{diag}\left(1,0,1,0,1,0,1,0\right)$ and $ {\mI}_2=\text{diag}\left(0,1,0,1,0,1,0,1\right)$, respectively. $M_1$ and $M_2$ are two mass terms.

\begin{figure}[h!]
  \centering
  \begin{subfigure}[b]{0.33\textwidth}
\includegraphics[width=\textwidth]{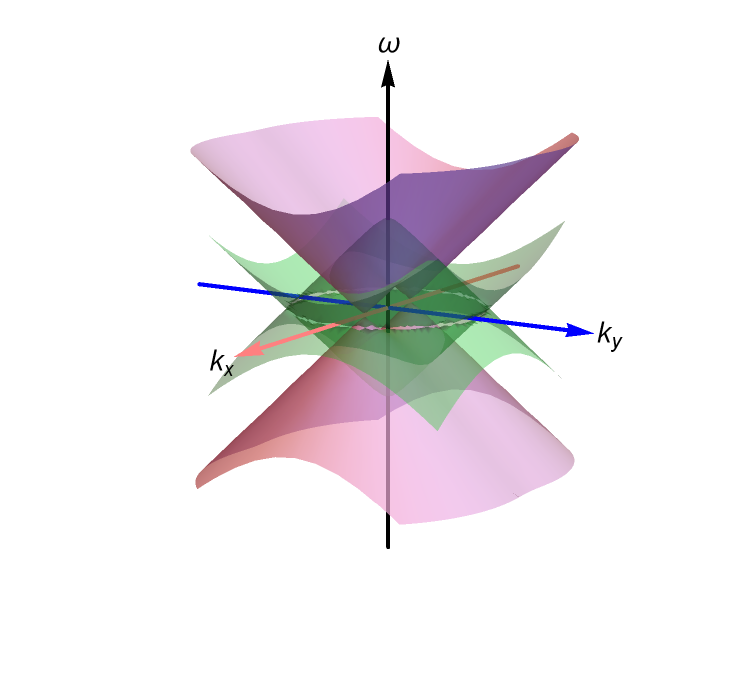}
\caption{\small Weyl-Nodal}
\end{subfigure}
\begin{subfigure}[b]{0.32\textwidth}
\includegraphics[width=\textwidth]{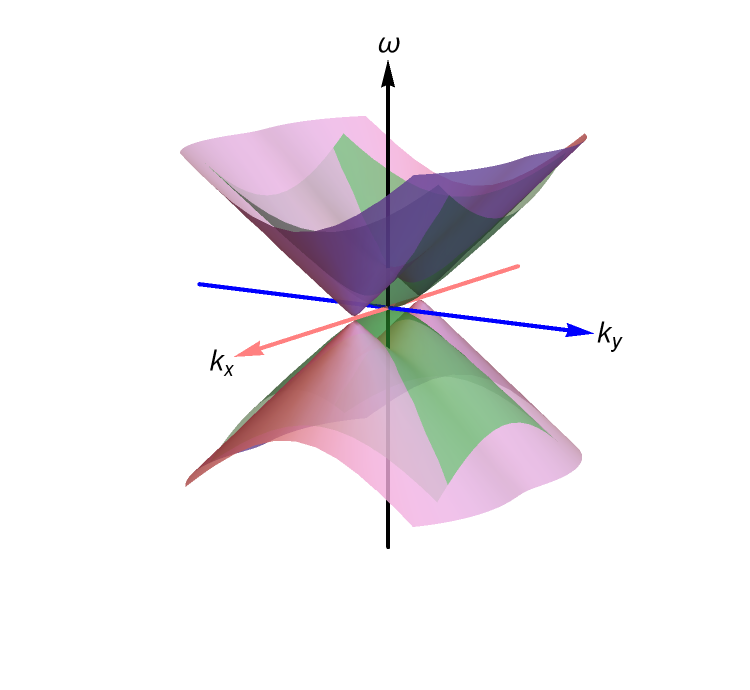}
\caption{\small Weyl-Critical}
\end{subfigure}
\begin{subfigure}[b]{0.33\textwidth}
\includegraphics[width=\textwidth]{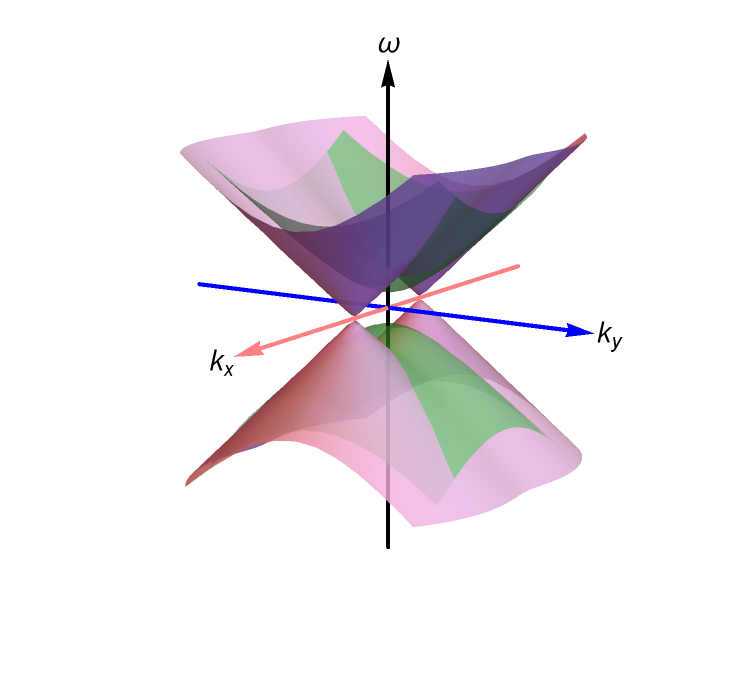}
\caption{\small Weyl-Gap}
\end{subfigure}
\begin{subfigure}[b]{0.33\textwidth}
\includegraphics[width=\textwidth]{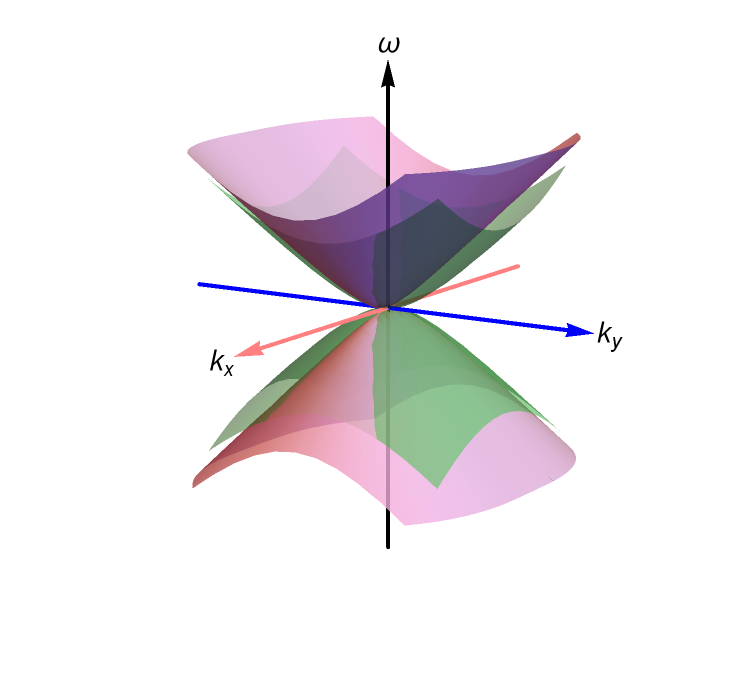}
\caption{\small Dirac Point(Critical)}
\end{subfigure}
\begin{subfigure}[b]{0.32\textwidth}
\includegraphics[width=\textwidth]{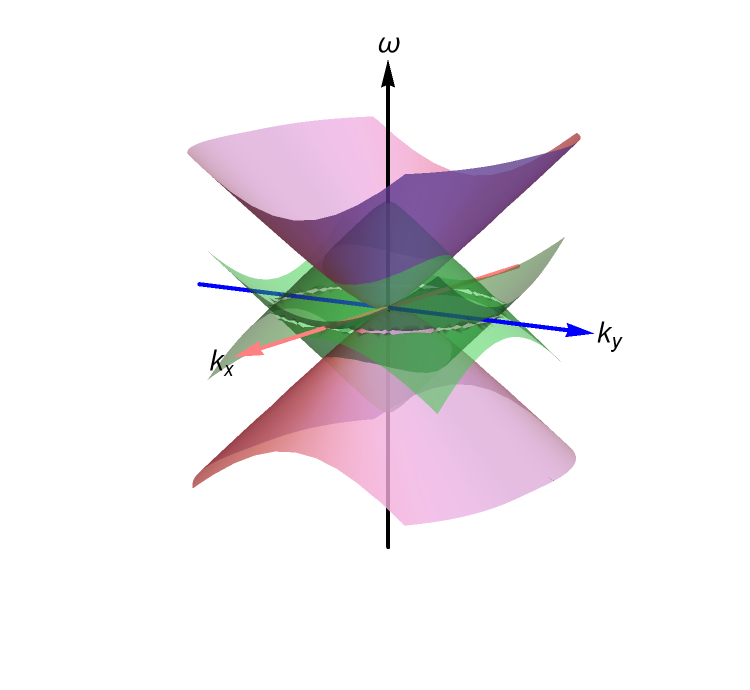}
\caption{\small Critical-Nodal}
 \end{subfigure}
 \begin{subfigure}[b]{0.33\textwidth}
\includegraphics[width=\textwidth]{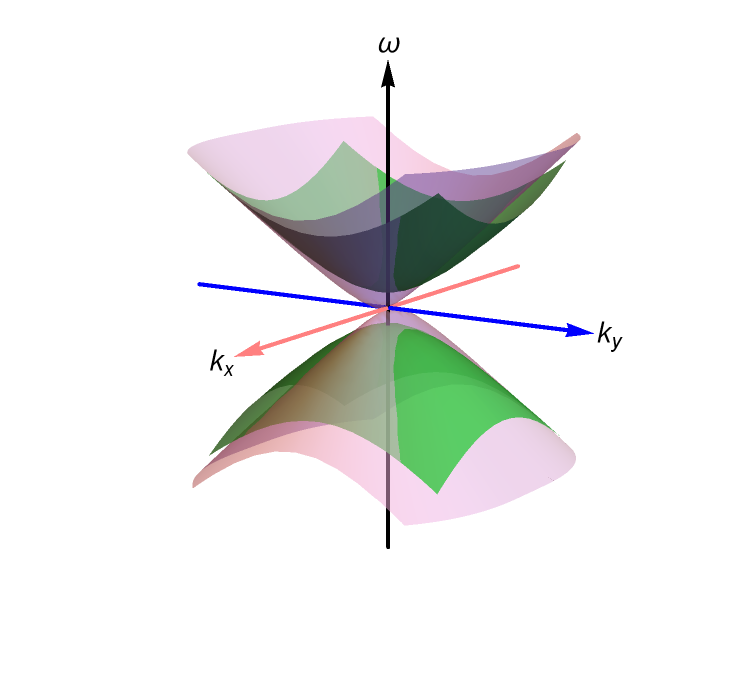}
\caption{\small Critical-Gap}
 \end{subfigure}
 \begin{subfigure}[b]{0.33\textwidth}
\includegraphics[width=\textwidth]{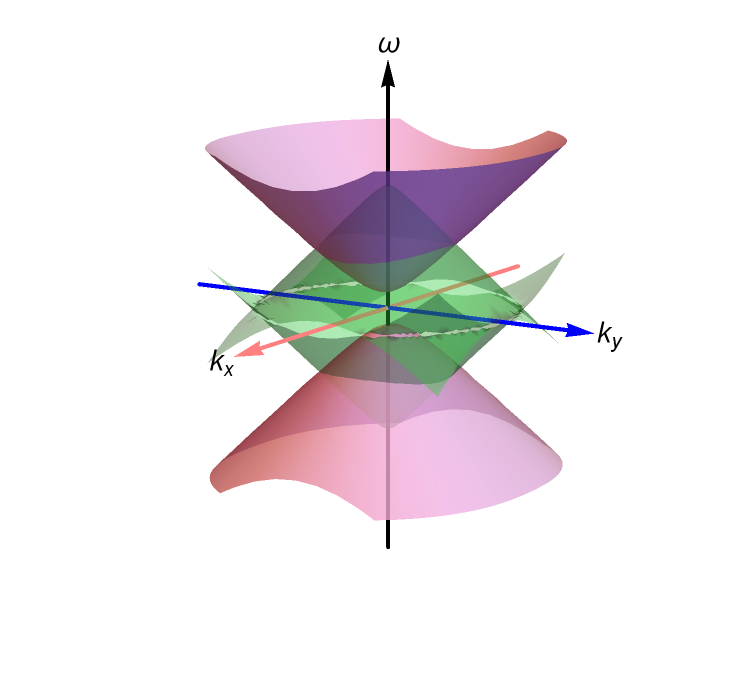}
\caption{\small Gap-Nodal}
 \end{subfigure}
  \begin{subfigure}[b]{0.32\textwidth}
\includegraphics[width=\textwidth]{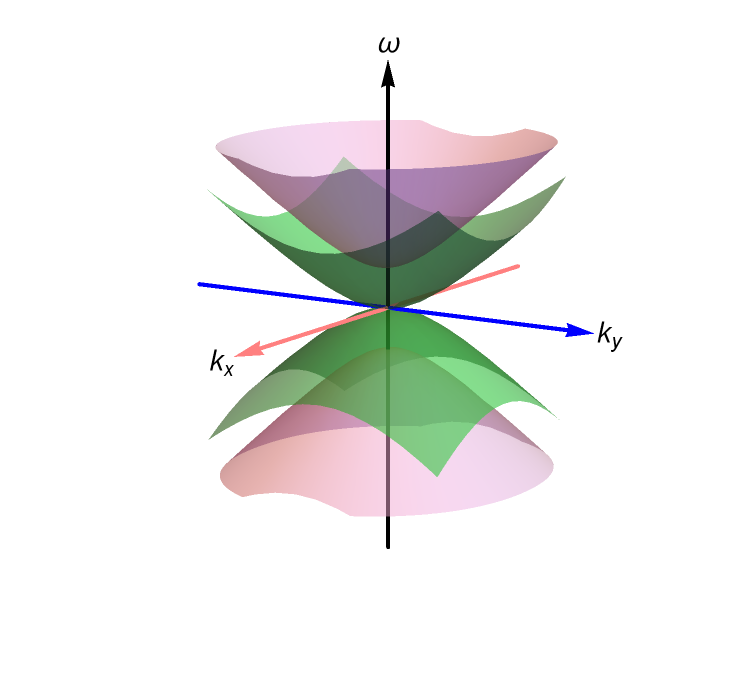}
\caption{\small Gap-Critical}
 \end{subfigure}
  \begin{subfigure}[b]{0.33\textwidth}
\includegraphics[width=\textwidth]{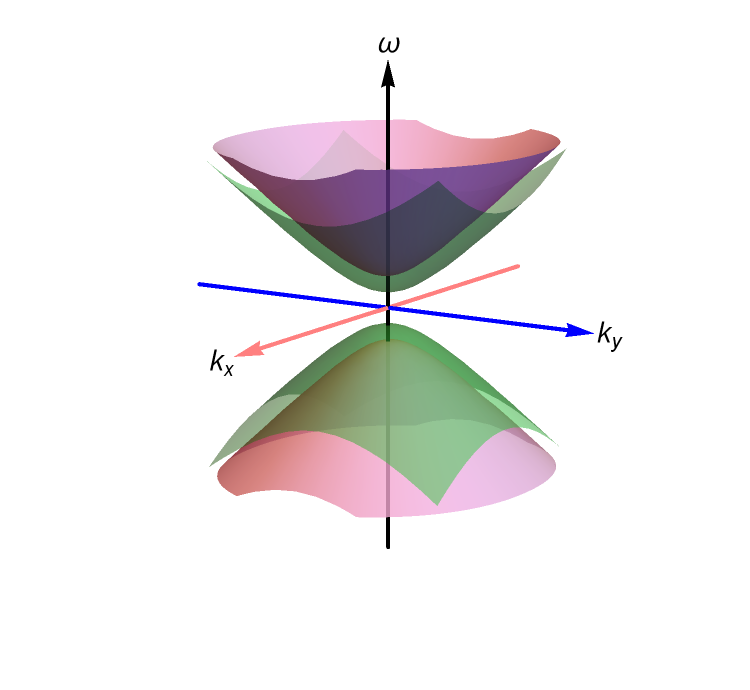}
\caption{\small Gap-Gap}
 \end{subfigure}
 \caption{\small The energy spectrum of \eqref{eq:1Lagrangian} as a function of $k_x$ and $k_y$ with $k_z=0$. From (a) to (i): (a):the system has two Weyl nodes and one nodal line where $M_1<2c$ and $M_2<b$, (b): two Weyl nodes and a critical point where $M_1=2c$ and $M_2<b$, (c): two Weyl nodes and a gap state where $M_1>2c$ and $M_2<b$, (d): a critical Dirac node where $M_1=2c$ and $M_2=b$, (e): a critical Dirac node and one nodal line where $M_1<2c$ and $M_2=b$, (f): a critical Dirac node and a gap state where $M_1>2c$ and $M_2=b$, (g): a gap state and one nodal line where $M_1<2c$ and $M_2>b$, (h): a gap state and one critical point where $M_1=2c$ and $M_2>b$ and (i): fully gap state where $M_1>2c$ and $M_2>b$. }
  \label{fig:eight}
\end{figure}

We turn on a nonzero constant $b_{xy}$ component of the two form field and a nonzero value of $b_{x}$. The eight eigenstates of this system could be obtained at the $k_z=0$ plane:
\be\label{ospectrum}
E_{1n\pm}&=&\pm\sqrt{\left(\sqrt{k_{x}^2+k_{y}^2+M_{1}^2}\mp2b_{xy}\right)^2}\,,\nonumber\\
E_{2w\pm}&=&\pm\sqrt{\left(\sqrt{k_{x}^2+M_{2}^2}\mp b_{x}\right)^2+k_{y}^2}\,.
\ee
With different value of $b,b_{xy},M_1$ and $M_2$, we can obtain nine types of spectrums shown in Fig.\ref{fig:eight}.
\vspace{0cm}

Now we analyze the phase behavior of the energy spectrum (\ref{ospectrum}) obtained above. Similar to the case of $\boldmath{Z}_2$-Weyl semimetal, four free parameters exist in the model \eqref{eq:1Lagrangian}. They are two mass terms $M_1$ and $M_2$, $b_x=b\delta_{\mu}^{x}$ the 1-form field and $b_{xy}=c$ the two-form field. As mentioned above, in the Weyl semimetal case Eq.\eqref{eq1}, changing the value of the dimensionless parameter $M/b$ could give two phases and one critical point. While for the case of $\boldmath{Z}_2$-Weyl semimetal, three dimensionless parameters are employed to present different phases. Based on this, we will fix the parameter $c/b$ as we did in the case of $\boldmath{Z}_2$-Weyl semimetal since $c/b$ only affects the phase structure quantitatively. The system would still depend on two dimensionless parameters $M_1/b$ and $M_2/c$ after $c/b$ is fixed. Thus we would obtain a two-dimensional phase diagram. Phase boundaries would be lines on the phase diagram plane and the phase transition lines would intersect at a particular critical point. We will show that depending on the values of $M_1/b$ and $M_2/c$, there will be four distinct phases and four phase transition lines and one critical point.

The eigen-energies with positive signs within the square root in $E_{1,2}$ of Eq.\eqref{ospectrum} are gapped bands, while the four bands that pick the negative sign within the square root may produce interesting phase structures, thus we will focus on them from now on. For these four bands, depending on the values of $b$, $c$, $M_1$ and $M_2$, the minimum values of $E_{1,2}$ could be either zero or larger than zero, thus the system could have crossing nodes (or lines) or be in a gapped state depending on different parameters.

The behavior of the spectrum as a function of $k_x$ and $k_y$ in the nine different phases is summarized in Fig.~\ref{fig:eight}, where $k_z$ is fixed to be zero. Note that a nonzero $k_z$ would immediately gap the system thus the crossing nodes in the figure are still nodes in the three dimensional momentum space. We explain in detail the four different phases, four phase transition lines and one critical point in the following.

\vspace{.25cm}
\noindent {\bf  $\bullet$ \em The Weyl-nodal phase}

When both $M_1<2c$ and $M_2<b$, we obtain the spectrum in Fig. \ref{fig:eight}(a)
with two Weyl nodes at $(k_x, k_y, k_z)=\left(\pm\sqrt{b^2-M_{2}^2}\,,0\,,0\right)$ and a nodal ring with a radius equal to $\sqrt{4c^2-M_{1}^2}$. For this phase, a small perturbation in the mass terms would not gap the system. This belongs to a topologically nontrivial phase with two nontrivial nodes and a nodal ring.\footnote{We would like to mention here that (a) in Fig.\ref{fig:eight} is a schematic representation where the diameter of the nodal ring is greater than the distance between the two Weyl nodes. However, due to the relative numerical value of $\sqrt{b^2-M_{2}^2}$ and $\sqrt{4c^2-M_{1}^2}$, the distance between the two Weyl nodes can be equal to (where the two Weyl nodes are on two sides of the diameter of the nodal ring) or greater than the diameter of the nodal ring. However, the topological properties of the system do not change significantly. Therefore, we do not show them here.}

Note that the nodal ring formation here is different from the case in eq.\eqref{eq4}. In eq.\eqref{eq4} the Weyl semimetal phase does not arise independently of the nodal line semimetal phase, whereas in the eight-component case, these two phases can form independently.

\vspace{.25cm}
\noindent {\bf $\bullet$ \em The Weyl-critical phases}

Without loss of generality, we fix $b, c$ and tune $M_1, M_2$ to obtain different phases. When we increase $M_1$ from (a) in Fig.~\ref{fig:eight}, the radius of the nodal ring would decrease to zero and form a critical Dirac node while the Weyl nodes keep their original form. We would like to mention that this is the only way to obtain the phase with a pair of Weyl nodes and a critical Dirac point in this system, i.e. the nodal ring in this system could not turn into a pair of Weyl nodes by tuning the parameters.

\vspace{.25cm}
\noindent {\bf $\bullet$ \em The Weyl-gap phases}

Continue to increase $M_1$ from the Weyl-critical phase transition line in (b) of Fig.~\ref{fig:eight}, the critical Dirac node becomes a trivial gap. This corresponds to Weyl-gap phases, as shown in case (c) of Fig.~\ref{fig:eight}.

\vspace{.25cm}
\noindent {\bf $\bullet$ \em The double critical point}

Again starting from the Weyl-critical phase transition line in (b) of Fig.~\ref{fig:eight} and this time we increase the other parameter $M_2$, the two Weyl nodes will also reach a critical point at which two nodes merge into one Dirac node. The system at this special set of parameters $M_1/b$ and $M_2/c$ corresponds to a double critical point on the phase diagram, which is the case (d) in Fig.~\ref{fig:eight}.

\vspace{.25cm}
\noindent {\bf $\bullet$ \em The critical-nodal phases}

A new phase forms if we increase the mass parameters $M_2$ from the Weyl-nodal phase in Fig. \ref{fig:eight}(a). The two Weyl points annihilate to a critical Dirac point which locates at the center of the nodal ring(i.e.$(k_x, k_y, k_z)=\left(0\,,0\,,0\right)$). This phase corresponds to the phase transition line between the Weyl-nodal phase and the gap-nodal phase. This spectrum is shown in (e) of Fig.~\ref{fig:eight}.

\vspace{.25cm}
\noindent {\bf $\bullet$ \em The critical-gap/gap-critical phases}

Continue to increase $M_2$ or $M_1$ from the double critical point in (d) of Fig.~\ref{fig:eight}, the fourfold-degenerate critical point would split into a pair of gapped bands and one twofold-degenerate critical point. This phase corresponds to a phase transition line between the Weyl-gap(or gap-nodal) phase and the gap-gap phase, as shown in the case (f) or (h) in Fig.~\ref{fig:eight}. Note that the gap can form by both the state responsible for Weyl semimetal formation (i.e., the fermion system described by $b$ and $M_2$) and the state responsible for nodal semimetal formation (i.e., the fermion system described by $c$ and $M_1$), so we need to be careful to distinguish which two energy bands form the gap (or critical point).

\vspace{.25cm}
\noindent {\bf $\bullet$ \em The gap-nodal phases}

Continue to increase the mass parameters $M_1$ from the critical-nodal phase in Fig. \ref{fig:eight}(e), the Dirac point becomes a gap. This spectrum is shown in (g) of Fig.~\ref{fig:eight}.

\vspace{.25cm}
\noindent {\bf $\bullet$ \em The gap-gap phase}

Starting from any of the two gap-critical phases transition lines above, and increase the other mass parameter, the system would become fully gapped. This corresponds to (i) in Fig.~\ref{fig:eight}.

The behavior of the energy spectrum with different parameters is summarized in Fig.~\ref{fig:eight}. There are nine different phases (including critical points or phase transition lines) which could be summarized into eight types of spectrums: including the Weyl-nodal,  Weyl-critical, critical-nodal, nodal-gap, Weyl-gap, critical-critical, critical-gap, and gap-gap phases.

\begin{figure}[h!]
  \centering
\includegraphics[width=0.6\textwidth]{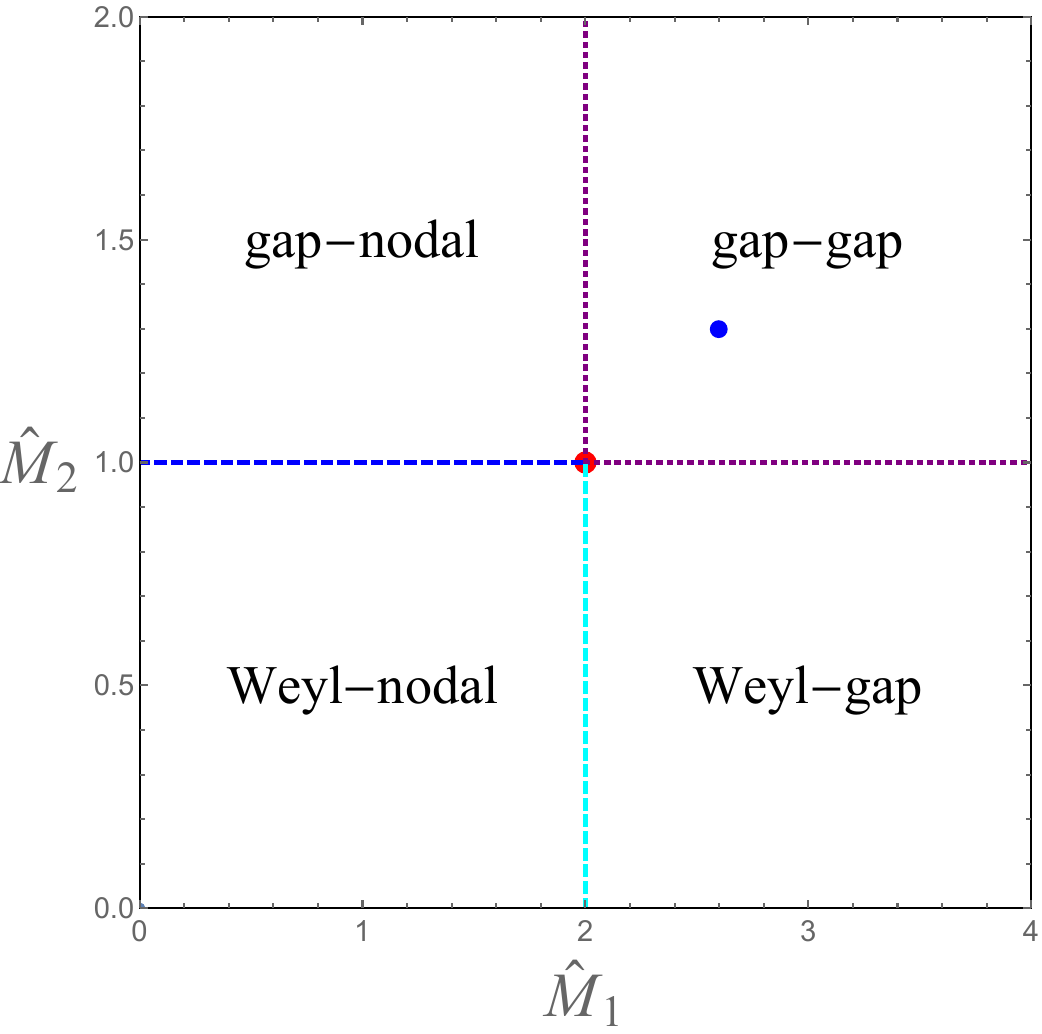}~~~~~
\vspace{-0.3cm}
  \caption{\small The phase diagram of the system \eqref{eq:1Lagrangian} with three dimensionless parameters $\hat{M}_1=M_1/c$, $\hat{M}_2=M_2/b$ and $c/b=1$. The red point is the double critical point at which both Weyl nodes and nodal ring become critical (Fig. \ref{fig:eight}(d)). The horizontal blue dashed line corresponds to the critical phase in which the Weyl nodes annihilate into a critical Dirac node while a nodal ring still exists (Fig. \ref{fig:eight}(e), the nodal-critical phases). The vertical cyan dashed line corresponds to the critical phase in which the radius of the nodal ring becomes zero while a pair of Weyl nodes still exists (Fig. \ref{fig:eight}(b), the Weyl-critical phases). The purple dotted lines correspond to the phase where the pair of Weyl nodes annihilates into a critical Dirac point(or the radius of the nodal ring becomes zero) while the nodal line (or the pair of Weyl nodes) becomes gapped (Fig. \ref{fig:eight}(f)(or (h)), the critical-gap phase). 
  The blue point in the left figure is to show the location of the double critical point where $c/b=1.3$.}
 \label{fig:02}
\end{figure}

The phase diagram can be plotted with three dimensionless parameters $\hat{M}_1=M_1/c$,$\hat{M}_2=M_2/b$ and $c/b$. A full phase diagram of three dimensional can be obtained. The phase diagram with different $\hat{M}_1$ and $\hat{M}_2$, for $c/b=1$ is shown in Fig.~\ref{fig:02}. The red point is the critical point at which both Weyl nodes and nodal ring become critical in Fig.~\ref{fig:eight}(d). The blue dashed line corresponds to the phase transition line where the Weyl nodes annihilates into a critical Dirac node while a nodal ring still exists in Fig.~\ref{fig:eight}(e). The cyan dashed line corresponds to the phase transition line where the radius of the nodal ring becomes zero while a pair of Weyl nodes still exists in Fig.~\ref{fig:eight}(b). The purple dotted lines correspond to another type of phase transition lines where the pair of Weyl nodes annihilates into a critical Dirac point(or the radius of the nodal ring becomes zero) while the nodal line (or the pair of Weyl nodes) becomes gapped in Fig.~\ref{fig:eight}(f)(or (h)). The down-left portion of the phase diagram corresponds to the phase in Fig.~\ref{fig:eight}(a). The up-left portion of the phase diagrams corresponds to the phase in Fig.~\ref{fig:eight}(g). The down-right portion of the phase diagrams corresponds to the phase in Fig.~\ref{fig:eight}(c). The up-right portion of the phase diagram corresponds to the phase Fig.~\ref{fig:eight}(i). In comparison to the location of the double critical point (red point in Fig.~\ref{fig:02}) for $c/b=1$, we also plot the location of the double critical point for $c/b=1.3$ by the blue point in Fig.~\ref{fig:02}.

\vspace{0cm}

As could be seen from the spectrum, when we increase $c/b$, the locations of the critical points $(\frac{M_1}{c})_c$ and $(\frac{M_2}{b})_c$ increase. {Compared to the phase diagram shown in Fig.\ref{fig22}, because more symmetries can be tolerated in this extended Hilbert space, the nodal line state can survive if there is a non-zero component of the one-form field along the axis lying in the plane where the nodal ring lives. The phase transitions between the Weyl semimetal state and the nodal line state are also different from the four component case.}

\section{Topological invariants}
\label{sec:invariant}
Topological states of matter are a new type of quantum states of matter that cannot be described by the Landau-Ginzburg paradigm and do not possess a local order parameter\cite{Witten:2015aoa}. They are otherwise characterized by nontrivial topological structures in their quantum wave functions and possess novel nontrivial properties that are stable under small perturbations. Topological invariants can be employed to classify different kinds of topological states. For weakly coupled topological systems, topological invariants could be defined from the Bloch states, i.e. the eigenstates of the weakly coupled Hamiltonians. A simple example is the nontrivial Berry phase associated with a closed loop in the momentum space of many topological systems, which is calculated from the Berry connection of the eigenstates of the Hamiltonian. To distinguish which topological phase is realized and to verify the occurrence of the topological phase transition, we will compute the corresponding topological invariants to achieve this goal in this section. The definition of topological invariants for various nodes will be reviewed first, then we will calculate the topological invariants both for the four and eight components Hamiltonian.

\subsection{Definition of topological invariants}
For weakly coupled topological systems, a simple example of a topological invariant is the Berry phase with value $0$ or $\pi$, which is the phase accumulated along a closed loop $\gamma$ in the momentum space for the Bloch states, i.e. eigenstates of the Hamiltonian $|n_{\bf k}\rangle$. The formula for Berry phase \cite{berry} is
\be\label{eq:berryphase}
\phi=\oint_\gamma \mathcal{A}_{\bf k}\cdot d {\bf k}\,,
\ee
where the Berry connection is defined by eigenstates
$|n_{\bf k}\rangle$
\be
\mathcal{A}_{\bf k}=i \sum_{j}\langle n_{\bf k}|\partial_{\bf k}|n_{\bf k}\rangle\,,
\ee  where $j$ runs over all occupied bands and $|n_{\bf k}\rangle$ is the eigenvector of the momentum space Hamiltonian.
Berry phase could be defined in general dimensions and here we focus on $3+1$ dimensions for our purpose.
We could also write (\ref{eq:berryphase}) using the Berry curvature as
\be
\phi=\int_S {\bf  \Omega}\cdot d{\bf S}\,,
\ee where
\be \label{eq:berrycur}
\Omega_i=\epsilon_{ijl } \big( \partial_{k_j}\mathcal{A}_{k_l}-\partial_{k_l}\mathcal{A}_{k_j}\big)
\ee
and $d{\bf S}$ is the surface element of $S$ which is a surface surrounded by the closed loop $\gamma$, i.e. $\gamma=\partial S$.

For a closed, orientable two-dimensional surface S in reciprocal space, we can now define the Chern invariant as
\be \label{eq:Chernnumber}
C=\frac{1}{2\pi}\int_S {\bf  \Omega}\cdot d{\bf S}\,.
\ee

The topological charge for the Weyl node could be calculated with the definition Eq.\eqref{eq:Chernnumber}. As we know, Weyl nodes can be viewed as the monopole of Berry curvature in the momentum space. Thus, the value of the Chern number should be $+1$, $-1$, and zero for the ideal Weyl semimetal, which depends on how the closed surface $S$ encloses the Weyl node\cite{rmb}.

\begin{figure}[h!]
  \centering
\includegraphics[width=0.65\textwidth]{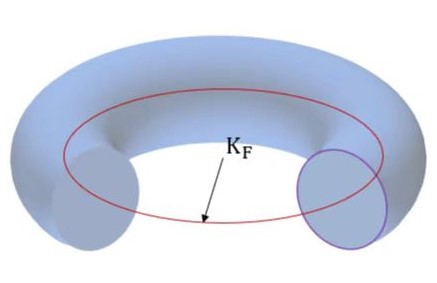}
\vspace{-0.3cm}
  \caption{\small The illustration of the closed path (purple curve) that encloses a Weyl node along the nodal line (red curve).
  }
 \label{fig:loop}
\end{figure}

The calculation of the topological invariant of nodal line semimetal is similar. Follow the idea mentioned in \cite{fang1}, a parameter ${\zeta}$ associated with the topological invariant could be defined with the Berry phase Eq.\eqref{eq:berryphase}, i.e.
\be
(-1)^{\zeta}=\oint{dk}\mathcal{A}(\bk)\cdot{d}\bk\,.
\ee
The Berry phase associated with any loop must be quantized to either $0$ or $\pi$ if the Hamiltonian $H(\bk)$ is real. Thus, a different value of $\zeta$ corresponds to a different topological state. If $\zeta=0$, it means the Berry phase associated with a loop has been quantized to $0$. This is because we can smoothly shrink this loop to a single point then the line crossing is purely accidental and can be removed by an arbitrarily small perturbation. While if $\zeta=1$, i.e. the Berry phase associated with a loop has been quantized to $\pi$, the loop cannot shrink to a point, as an infinitesimal loop necessarily has zero Berry phase. That means there must be a point inside the loop where the Berry phase cannot be defined, where the conduction and the valence bands cross. These crossing parts are topological protected and cannot be removed by an arbitrarily small perturbation. Hence, we can calculate the value of the Berry phase to check the topological properties for topological nodal line semimetal. From the definition of the Berry phase, we need to introduce a loop that encloses the nodal ring. However, it is difficult to define the Berry connection for such kind of energy band structure, because the crossing parts of the energy bands are not isolated from each other. Nevertheless, we can calculate the Berry phase following the cyclotomic method. Considering that a node ring can be regarded as a collection of nodes, we can divide the ring into several discrete nodes. It's easy to choose a circle that encloses a single node(see Fig.\ref{fig:loop}\cite{Liu:2020ymx}) and define a discrete Berry phase as
\be
\label{eq:bpdis}
e^{-i\phi_{i_1i_2}}=\frac{\langle n_{i_1}|n_{i_2}\rangle}{|\langle n_{i_1}|n_{i_2}\rangle|}\,,
\ee

\vspace{0cm}
$| n_{i_1}\rangle$ and $| n_{i_2}\rangle$ are eigenstates at two adjacent discrete points $i_1$ and $i_2$ along the path. The total Berry phase along the closed path is then the summation of all the discrete phases. If the result for the summation is $\pi$, it means that we have a topological nontrivial state.

\subsection{Calculations of topological invariants}
In this subsection we show the calculations and the results of topological invariants in the effective field theory models that we studied above.

\subsubsection{System with one-form field along the $z$ axis.}

{We begin to calculate the topological invariant of \eqref{eq4} when the non-zero component of the one-form field is along the $z$ axis. Besides the three-node state and the triple degenerate nodal point, several other different types of phases shown in Fig.\ref{fig3} to Fig.\ref{fig5} can be divided into three classes. They are the Weyl type including two Weyl nodes and two shifted Weyl nodes state, the Dirac type including the Dirac node and shifted Dirac node, and the gap type including gapped and shifted gapped state, respectively. The topological invariant for these three types are obtained in\cite{Liu:2018djq}, showing that the Chern numbers for the Weyl nodes are $\pm1$, while the Dirac point and the gapped state possess zero Chern numbers. We have calculated the Chern numbers for the three classes of phases mentioned above and obtained the same results, whose details will not be shown here. In the following we will show the calculations and results for the topological invariants of the three-node state and the triple degenerate nodal point.}

\vspace{.25cm}
\noindent {\bf $\bullet$ \em The three nodes state}

{We start by computing the topological invariant for the three nodes state ((c) in Fig.\ref{fig3} and (e) in Fig.\ref{fig5}) when $\vec{b}$ is in the $z$-direction and $\vec{b}_z=b$ in Eq.\ref{eq4}. Without loss of generality, we can set $M=0.4,b_{xy}=0.2,b=0.8,k_x=k_y=0$ to obtain the three nodes state. To compute the corresponding Chern number, we have to substitute the corresponding eigenvectors into the definition of the Berry connection and the Berry curvature. A sphere ${\bf S}: k_{x}^2+k_{y}^2+\left(k_z\pm a_0\right)^2=k_{0}^{2}$ enclosing the node needs to be introduced according to Eq.\eqref{eq:Chernnumber}, where $k_0$ is a small constant representing the radius of the sphere and the value of $a_0$ is the location of the node. There are no analytic eigenenergies and eigenvectors for the three nodes state, so we compute the corresponding Chern number numerically. Choosing three suitable small spheres enclosing the corresponding three nodes, we find that all these three Chern numbers are non-zero. Note that the node located at $k_x=k_y=k_z=0$ can be viewed as a critical point of the nodal line semimetal state, where the radius of the nodal ring is zero. Thus, we obtain a non-zero topological charge \footnote{Since chiral charges always occur in pairs, as in the case of Weyl nodes, this topological charge cannot be called a chiral charge.} for the critical point of the nodal rings described in \eqref{eq4}. As we have explained in previous sections, this indicates that the nodal ring could not first shrink to a point and then become gapped and on the contrary, the ring will first shrink to a point and then grown into a ring again when tuning the parameters slightly. We have also checked the topological charge for the critical point of the nodal rings described in Eq.\ref{eq3} and found it to be zero. The difference of the topological charge for the critical point indicates the difference between two types of nodal rings in \eqref{eq3} and \eqref{eq4}.}

\vspace{.25cm}
\noindent {\bf $\bullet$ \em The triple degenerate nodal point}

{The topological charge for the triple degenerate nodal point ((c) in Fig.\ref{fig4}) has a similar calculation procedure. Without loss of generality, we set $M=b=0.4,b_{xy}=0.2,k_x=k_y=0$ to obtain the triple degenerate nodal point. Solving Eq.\ref{eq4} under these parameters we obtain the eigenenergies as
\be
\label{tdnp}
E_{1,2}=-\frac{2}{5}\pm k_z, ~~~
E_{3,4}=\frac{1}{5}\left(2\pm \sqrt{16+25k_{z}^{2}}\right).
\ee}

{From Eq.\ref{tdnp} we see that when $k_z=0$, $E_{1}$, $E_{2}$ and $E_{3}$ are equal to each other, so they form the triple degenerate nodal point. From the definition of Eq.\ref{eq:Chernnumber}, we need to find the corresponding eigenvectors. They are
\be
\label{evtdnp}
|n_1\rangle&=&\left(\frac{-k_{z}+\sqrt{k_{x}^{2}+k_{y}^{2}+k_{z}^{2}}}{k_x+i k_y},\frac{k_{x}^{2}+k_{y}^{2}+2k_{z}\left(k_z-\sqrt{k_{x}^{2}+k_{y}^{2}+k_{z}^{2}}\right)}{k_{x}^2+k_{y}^2},\frac{k_{z}-\sqrt{k_{x}^{2}+k_{y}^{2}+k_{z}^{2}}}{k_x+i k_y},1\right)^{T},\nonumber\\
|n_2\rangle&=&\left(-\frac{k_{z}+\sqrt{k_{x}^{2}+k_{y}^{2}+k_{z}^{2}}}{k_x+i k_y},\frac{k_{x}^{2}+k_{y}^{2}+2k_{z}\left(k_z+\sqrt{k_{x}^{2}+k_{y}^{2}+k_{z}^{2}}\right)}{k_{x}^2+k_{y}^2},\frac{k_{z}+\sqrt{k_{x}^{2}+k_{y}^{2}+k_{z}^{2}}}{k_x+i k_y},1\right)^{T},\nonumber\\
|n_3\rangle&=&\left(-\frac{-4+5k_{z}+\sqrt{16+25\left(k_{x}^{2}+k_{y}^{2}+k_{z}^{2}\right)}}{5\left(k_x+i k_y\right)},-1,\frac{4+5k_{z}-\sqrt{16+25\left(k_{x}^{2}+k_{y}^{2}+k_{z}^{2}\right)}}{5\left(k_x+i k_y\right)},1\right)^{T}.\nonumber
\ee}

By direct but somewhat complicated calculation we find that we obtain zero Chern invariant if we calculate from $|n_1\rangle$ and $|n_2\rangle$. However, if we calculate the corresponding Chern invariant from $|n_3\rangle$ (the orange band in Fig.5) we obtain a non-zero topology charge. These results are consistent with the topology charge obtained in the $k\cdot p$ model of the triple degenerate nodel point\cite{ayang}. There are several types of topological charges for a triple degenerate node. We denote the three crossing bands as bands 1, 2, 3, and the several types of topological charges describe whether any band in 1, 2, and 3 stay crossed with either one of the other band when we tune the parameter in both directions slightly, or whether all three bands become gapped from each other. For the latter case, the topological charge would be trivial while for the first cases, if band 1 (or 2/3) is the one that stays crossed, then the topological charge calculated from the eigen vector of band 1 should be nontrivial while the ones calculated from the other two bands should be trivial. Thus here the fact that topological charge is only nontrivial when calculated for the orange band in Fig.5 is totally consistent with the behavior of the node in that figure.

Modifying the above parameter slightly, we can obtain (b) or (d) in Fig.\ref{fig4}. By careful calculation of the corresponding Chern number, we find that both of these states have a Chern number equal to zero. Since the topology charge changes as the system evolves from (b) to (d), we can say that the Dirac node and the triple degenerate model point are two different topological states, and a topological phase transition occurs when these two states evolve toward each other.

\subsubsection{System with the one-form field along the $x$ axis.}

In the following, we will check the topological properties of Eq.\eqref{eq:eigenstate11} when the non-zero component of the one-form field is along the $x$ axis .

\vspace{.25cm}
\noindent {\bf $\bullet$ \em nodal line semimetal}

We set $M=0.2,b_{xy}=0.3,k_y=b=0$ in Eq.\eqref{eq4} without loss of generality. At these parameters, we obtain a nodal line semimetal phase ((a) in Fig.\ref{fig6}). From the definition Eq.\eqref{eq:bpdis}, we need to pick up suitable eigenstates. Carefully checking the spectrum we find that the bands with eigenenergies
\be
\label{eqnodal}
E_{1,2}=\pm \sqrt{b^2+4b_{xy}^2+k_x^2+k_y^2+M^2-2\sqrt{b^2\left(M^2+k_x^2\right)+4b_{xy}^2\left(M^2+k_x^2+k_y^2\right)}}
\ee
form the nodal ring. The eigenvectors for Eq.\eqref{eqnodal} have no explicit analytic form, then we solve the discrete Berry phase Eq.\eqref{eq:bpdis} numerically. The closed path can be chosen as
\be
\label{eq:discrete}
\left(k_x,~k_y,~k_z\right)=\left(k_F+l~\text{sin}~\theta,~0,~l~\text{cos}~\theta\right)\,,~~~~\text{with} ~~\theta \in\left[0,~2\pi\right)\,.
\ee
For the calculation of the discrete Berry phase, we have $\theta$ in (\ref{eq:discrete}) with $\theta_i=\frac{2\pi i}{N}$ with $i\in \{1,..., N\}$.
With summation from $1$ to $N$ we obtain the total Berry phase for the system to be $\pi$. The result indicates the nontrivial topological properties of the nodal line semimetal phase.

\vspace{.25cm}
\noindent {\bf $\bullet$ \em Weyl semimetal}

Now we calculate the topological charge for the Weyl semimetal phase, i.e. (b) in Fig.\ref{fig6}. As mentioned above, a non-zero $b$ destroys the nodal ring which cannot be restored. We set $M=0.2,b_{xy}=0.1,b=0.5$ in the Eq.\eqref{eq4}, then the two Weyl nodes locate at $\left(k_x,k_y,k_z\right)=\left(\pm 0.5,0,0\right)$. Similar sphere ${\bf S}: \left(k_x\pm a_0\right)^2+k_{y}^2+k_{z}^2=k_{0}^{2}$ enclosing the node is introduced. With the parameters above, we pick $a_0=0.5$. Choose suitable eigenvectors and calculate the Chern number numerically near the Weyl node A that locates at $\left(k_x,k_y,k_z\right)=\left(0.5,0,0\right)$ we find that when the sphere encloses A the Chern number is 1. When the sphere doesn't enclose A the Chern number becomes zero. Perform a similar calculation near Weyl node B that locates at $\left(k_x,k_y,k_z\right)=\left(-0.5,0,0\right)$ we find that when the sphere encloses B the Chern number is -1. When the sphere does not enclose B the Chern number becomes zero. From different values of the Chern number, nontrivial topological states of the Weyl semimetal phase are confirmed.

\vspace{.25cm}
\noindent {\bf $\bullet$ \em critical and gapped states}

Finally, we calculate the topological invariant for the critical and gapped states. We find that both of Chern number and Berry phase become zero. These null results indicate the trivial topological property of the gapped state. It also indicates that the nodal state realized here is the one with the topological charge $\zeta_2=0$.

\subsubsection{The eight-component effective field theoretic model}

The calculations of the topological invariant in model Eq.\eqref{eq:1Lagrangian} are similar. Take the Weyl-nodal phase((a) in Fig.\ref{fig:eight}) as an example. The corresponding bands that form the nodal ring are $E_{1\pm}=\pm\sqrt{\left(\sqrt{k_{x}^2+k_{y}^2+M_{1}^2}-2b_{xy}\right)^2}$ while the bands to form the Weyl nodes are $E_{2\pm}=\pm\sqrt{\left(\sqrt{k_{x}^2+M_{2}^2}- b_{x}\right)^2+k_{y}^2}$. Obtaining the corresponding eigenvectors numerically, we find that the total Berry phase for the system is $\pi$ and a non-zero Chern number exists when the sphere encloses one of the Weyl nodes. The other eight phases have different combinations of the topological invariant. For example, in the Weyl-critical and Weyl-gap phases we have a non-zero Chern number but a zero Berry phase. While in the nodal line-critical and nodal line-gap phases, the system has a zero Chern number but with its Berry phase equal to $\pi$. These results strongly suggest the distinction of different topological phases and the occurrence of topological phase transitions.

\section{Conclusion and discussion}
\label{sec:cd}
We have studied the topological phase transitions between the Weyl semimetal and the nodal semimetal states through effective field theory models. Two classes of effective field theory models have been built, which are distinguished by different spinors. For the four-component spinor model, we have studied how the Weyl nodes (corresponding to the one-form field) influences the nodal line (corresponding to the two-form field) and vice versa. We find the existence of novel critical states, including the three Weyl nodal state and the triple degenerate nodal point. These new critical topological states are realized for the first time in the effective field theory model. We would also like to mention that two different types of nodal line states, distinguished by the topological charge, are also realized by choosing different non-zero components of the one-form field in the system. While for the eight-component spinor model, we realized the Weyl semimetal and the topological nodal line semimetal states simultaneously by introducing two sets of fermions. The phase structures are intuitively presented and nine phases exist in such a system. Novel topological states also arise during the phase transitions. The occurrence of phase transitions has been checked by calculating the topological invariants for different (topological) phases. By studying the phase transitions between different topological states, we may get further theoretical guidance to the potential applications of topological systems, leading to far-reaching results in physics.

There are still some open questions that we would like to leave for future work. First, what would be the phase transition structures between the $\boldmath{Z}_2$-Weyl semimetal and nodal line semimetal? Could there be more interesting physics? Second, higher dimensional spacetime could be considered to extend the Hilbert space, so what would happen if we consider the phase diagrams in a higher dimension? Third, what would be the behavior in the strong coupling regime? Could we build a holographic model\cite{future} to study this? It would also be interesting to generalize our results to study the phase transitions between other topological states.
\subsection*{Acknowledgments}
We thank K. Landsteiner and Y. Liu for discussions.
This work was supported by the National Key R\&D Program of China (Grant No. 2018FYA0305800),
National Natural Science Foundation of China (Grant Nos. 11875083, 12005255, 12035016),
the Strategic Priority Research Program of Chinese Academy of Sciences (Grant No. XDB2800000),
the Key Research Program of Chinese Academy of Sciences (Grant Nos. XDPB08-1, XDPB15).
The work of Y.W.S. has also been partly supported by starting grants from University of Chinese Academy of Sciences and Chinese Academy of Sciences.
\appendix
\section{the effect of two-form field on the Weyl semimetal state}
\label{app:a}
\vspace{.25cm}
\noindent {\bf  $\bullet$ \em Turning on $b_{xy}$ in the Weyl semimetal phase.}

\begin{figure}[h!]
  \centering
  \begin{subfigure}[b]{0.33\textwidth}
\includegraphics[width=\textwidth]{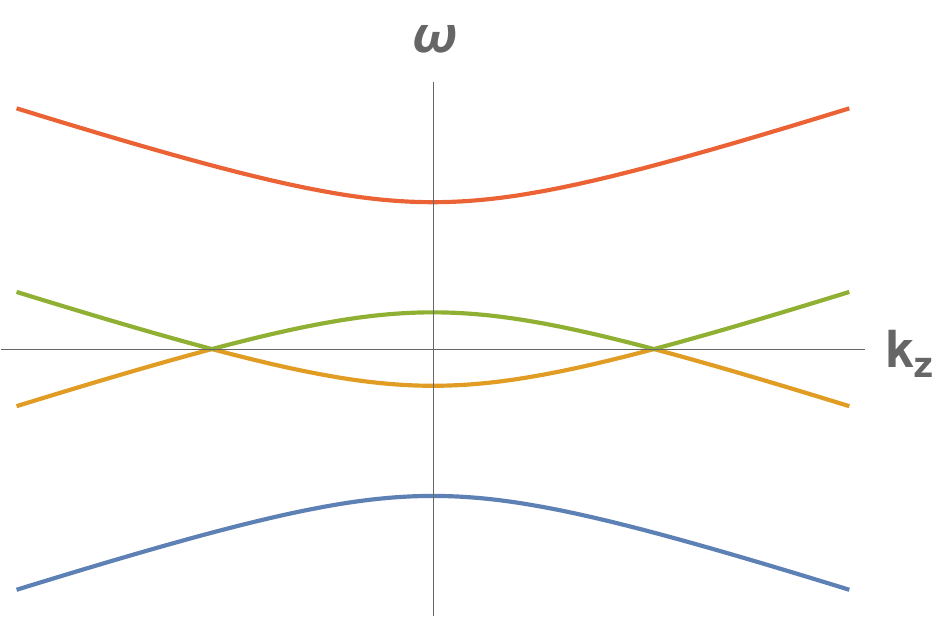}
\caption{\small two Weyl nodes}
\end{subfigure}
\begin{subfigure}[b]{0.32\textwidth}
\includegraphics[width=\textwidth]{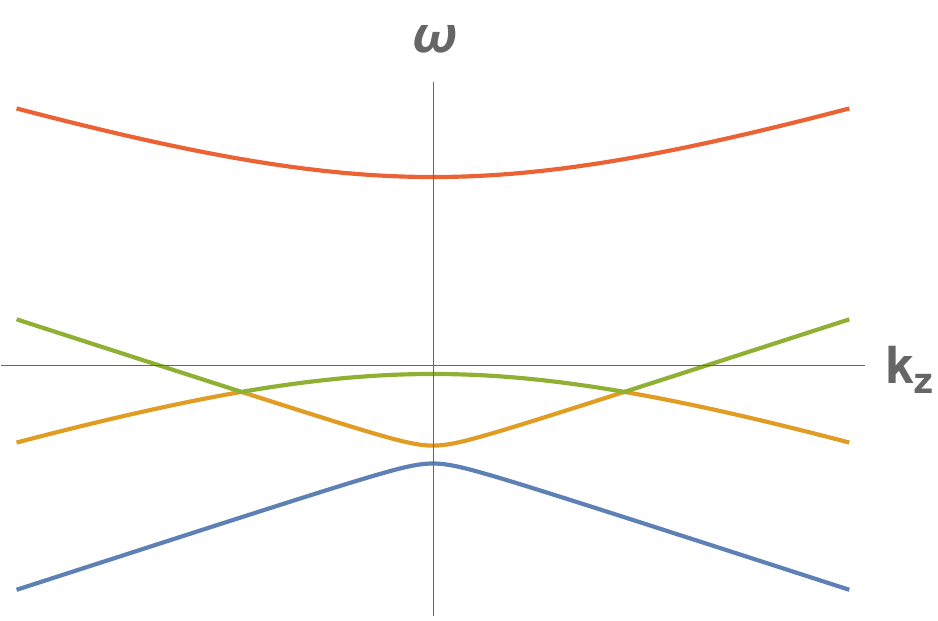}
\caption{\small two shifted Weyl nodes}
\end{subfigure}
\begin{subfigure}[b]{0.33\textwidth}
\includegraphics[width=\textwidth]{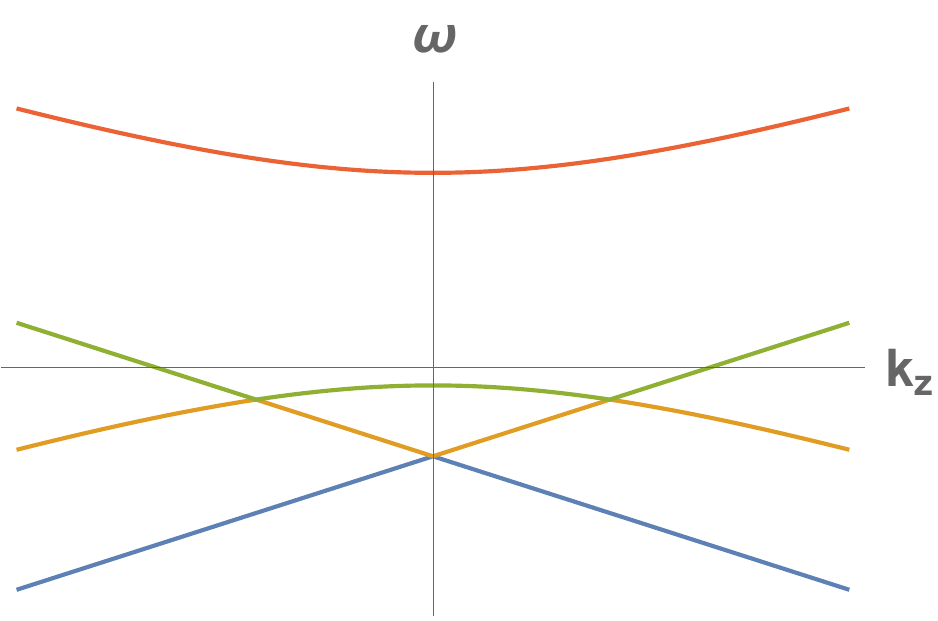}
\caption{\small three nodes}
\end{subfigure}
\begin{subfigure}[b]{0.33\textwidth}
\includegraphics[width=\textwidth]{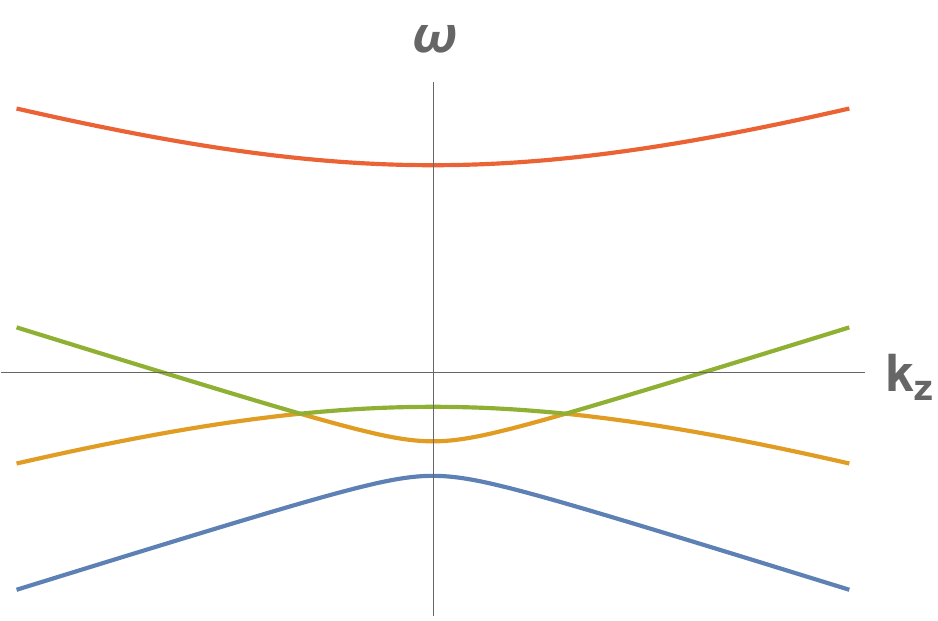}
\caption{\small another two shifted Weyl nodes}
\end{subfigure}
\begin{subfigure}[b]{0.32\textwidth}
\includegraphics[width=\textwidth]{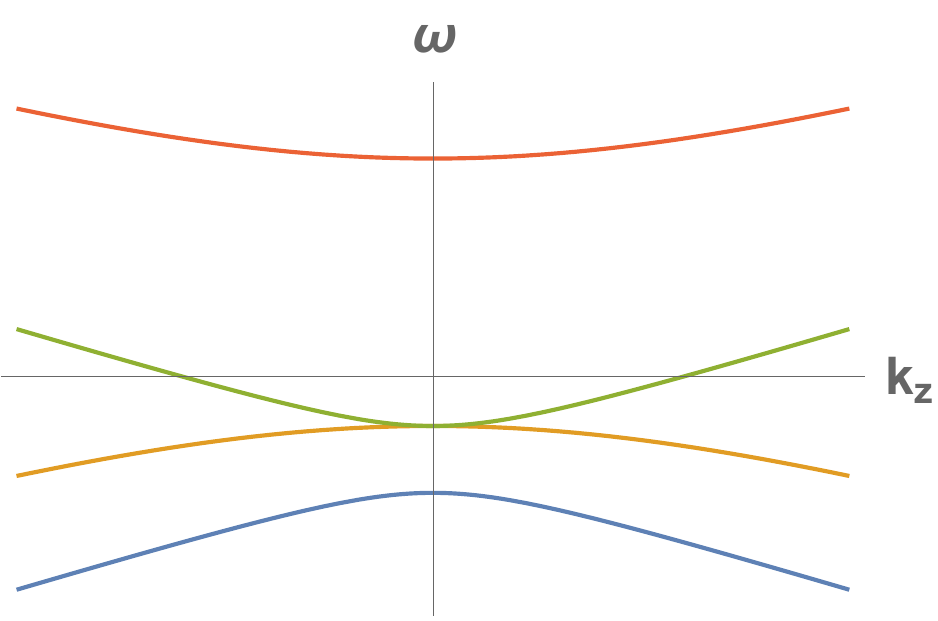}
\caption{\small critical Dirac node}
\end{subfigure}
\begin{subfigure}[b]{0.33\textwidth}
\includegraphics[width=\textwidth]{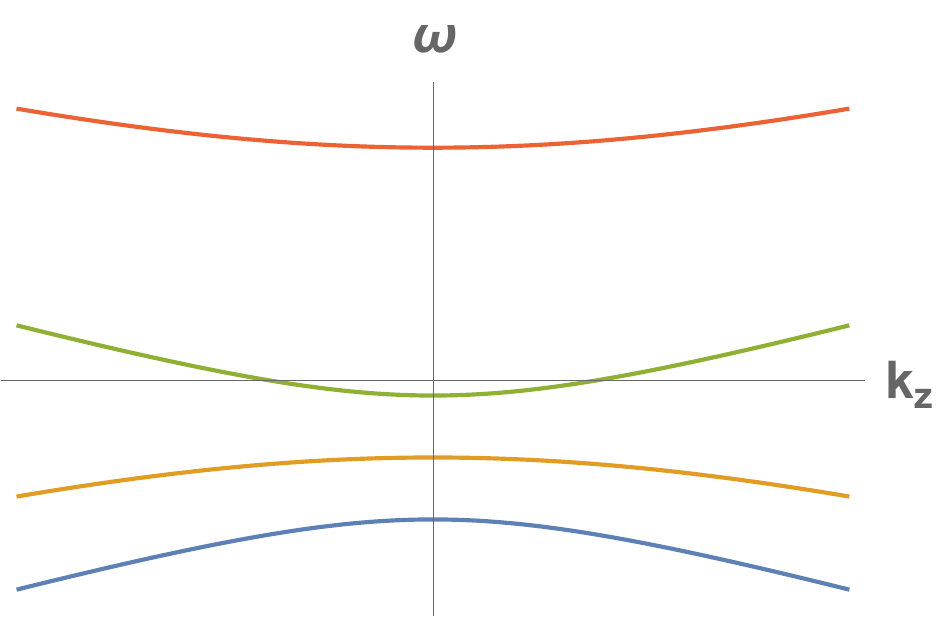}
\caption{\small gap}
\end{subfigure}
 \caption{\small Part of the energy spectrum of \eqref{eq4} as a function of $k_z$ with $k_x=k_y=0$. From (a) to (f): (a): the system has two Weyl nodes when $b>M$ and $b_{xy}=0$, (b): two Weyl nodes move down when $b>M>2b_{xy}$ and $b_{xy}\neq 0$, (c): three nodes at $b>M=2b_{xy}$, (d): another two Weyl nodes when $b>2b_{xy}>M$, (e): a critical Dirac node when $b=2b_{xy}>M$, and (f): the fully gapped state when $2b_{xy}>b>M$. Note that in (b) the system has a nodal ring formed by the lowest and the second lowest energy band (nodal ring A), while in (d),(e) and (f), the system has a nodal ring formed by the second lowest and the third lowest energy band (nodal ring B).)}
  \label{fig3}
\end{figure}

The Weyl semimetal phase has $|b|>|M|$, and the system has two Weyl nodes, located at $\left(k_x,k_y,k_z\right)=\left(0,0,\pm\sqrt{b^2-M^2}\right)$ with the distance between them equal to $2\sqrt{b^2-M^2}$, corresponding to case (a) in Fig.\ref{fig3}. Now keeping the values $b$ and $M$ of (a) in Fig.\ref{fig3} and increasing the value of $b_{xy}$ from 0, we could obtain the following five types of energy spectrums.

I. For $2b_{xy}<M<b$, two Weyl nodes move down along the energy axis\footnote{This can also be seen from the perspective of the nodal ring state. As $b_{xy}$ increases, the third lowest energy band of the system moves down. Since the Weyl nodes are formed by the second and third lowest energy bands, a downward move is observed.}, as shown in (b) in Fig.\ref{fig3}. The distance between these two Weyl nodes begins to decrease, reflecting the effect of the nodal line semimetal phase. 

II. At $2b_{xy}=M<b$, a novel phase with three nodes ``emerges" ((c) in Fig.\ref{fig3}). The newly appearing node actually serves as a critical point. As shown in Fig.\ref{fig11}, when $b_{xy}=M/2$, the radius of nodal ring become zero and a node forms at $\left(k_x,k_y,k_z\right)=\left(0,0,0\right)$, so it can be seen from the perspective of Fig.\ref{fig3}. Note that the third node is a topologically nontrivial one: this is verified by calculating the topological invariant in section \ref{sec:invariant} where we find that this node has a non-zero topological charge. As we have explained above, this nontrivial topological charge reflects the fact that the nodal ring will not disappear but will appear again when tuning the parameter further along this direction. To the best of our knowledge, this is the first time that this critical case is obtained from effective field theory. It is different from the case in \eqref{eq3} where the critical point is a topologically trivial Dirac node. 

III. For $M<2b_{xy}<b$, the third node disappears from the perspective of Fig.\ref{fig3}. This is because the third node grows again into a nodal ring, which cannot be seen in the picture. Besides the nodal line semimetal state, the system still has two Weyl nodes as shown in (d) in Fig.\ref{fig3}. The distance between the two Weyl nodes decreases faster with increasing $b_{xy}$ than in case (b) in Fig.\ref{fig3}. This effect should also be associated with the difference between two types of nodal rings.

\begin{figure}[h!]
  \centering
  \begin{subfigure}[b]{0.35\textwidth}
\includegraphics[width=\textwidth]{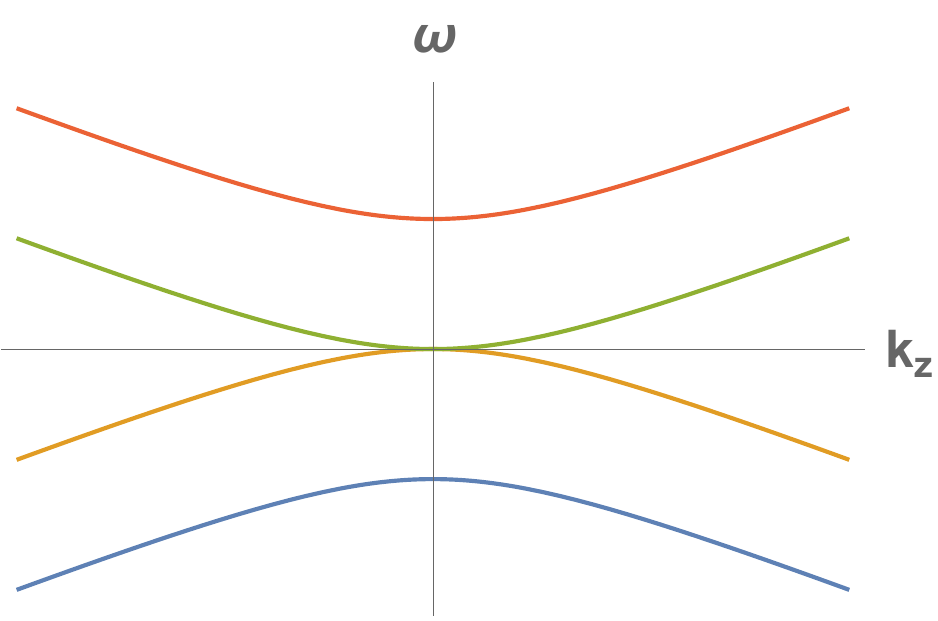}
\caption{\small Dirac point(Critical)}
\end{subfigure}
\begin{subfigure}[b]{0.35\textwidth}
\includegraphics[width=\textwidth]{Diracdown.pdf}
\caption{\small shifted Dirac point}
 \end{subfigure}
 \begin{subfigure}[b]{0.35\textwidth}
\includegraphics[width=\textwidth]{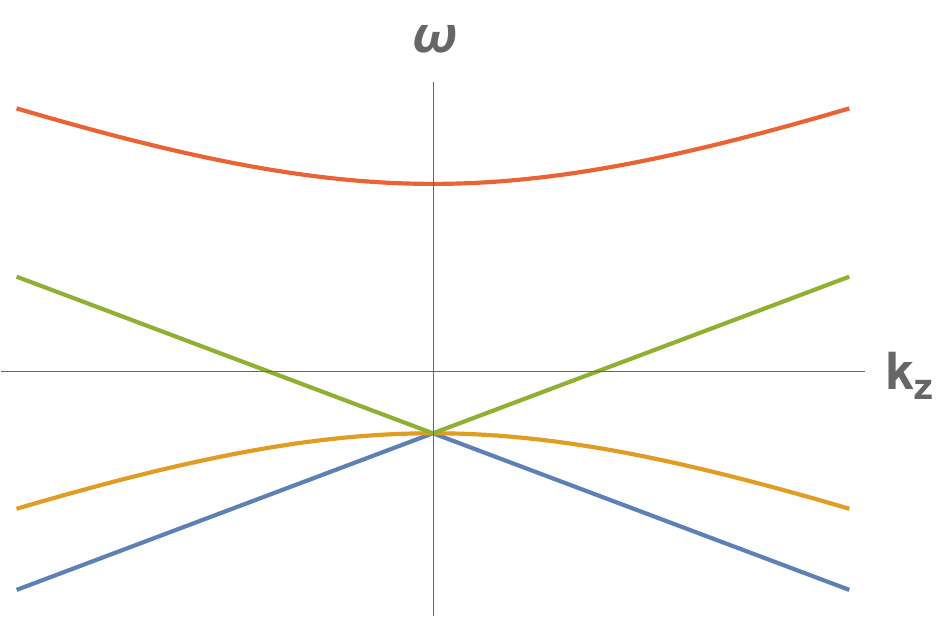}
\caption{\small triple degenerate nodal point}
 \end{subfigure}
  \begin{subfigure}[b]{0.35\textwidth}
\includegraphics[width=\textwidth]{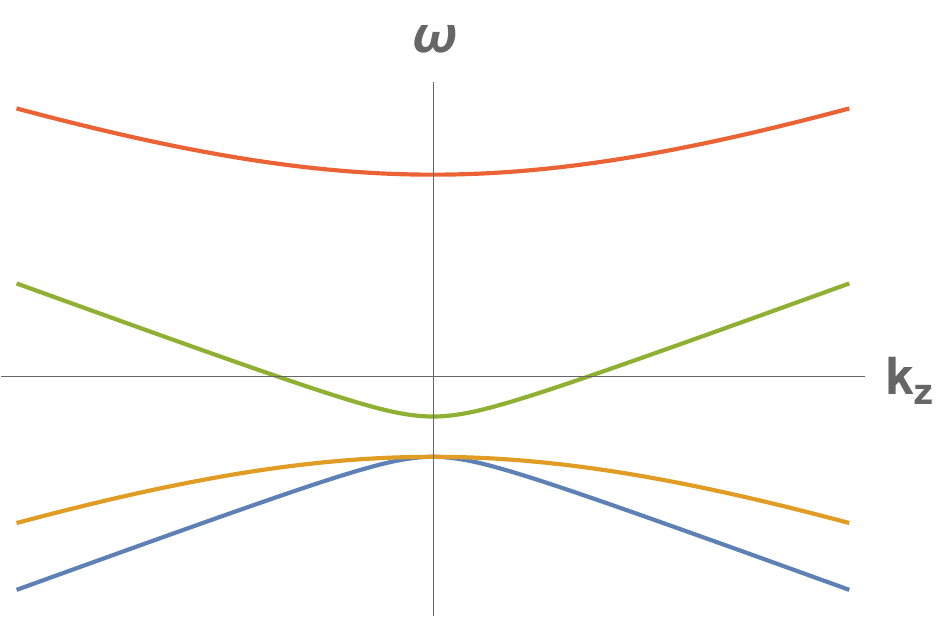}
\caption{\small another shifted Dirac point}
 \end{subfigure}
  \caption{\small Part of the energy spectrum of \eqref{eq4} as a function of $k_z$ with $k_x=k_y=0$. From (a) to (d): (a): the system has one Dirac point when $b=M$ and $b_{xy}=0$, (b): Dirac point moves down for $2b_{xy}<b=M$, (c): triple degenerate nodal point when $2b_{xy}=b=M$, and (d): another Dirac point when $2b_{xy}>b=M$. Note that in (b) the system has a nodal ring formed by the lowest and the second lowest energy band (nodal ring A), while in (d) the system has a nodal ring formed by the second lowest and the third lowest energy band (nodal ring B).}
  \label{fig4}
\end{figure}

IV. At $M<2b_{xy}=b$, a Dirac point is obtained ((e) in Fig.\ref{fig3}). This Dirac point is different from the critical Dirac point represented in Eq.\eqref{eq1} where $b=M$, and we have shown this case in (a) in Fig.\ref{fig4}. Comparing these two Dirac semimetal states, we find that a non-zero $b_{xy}$ pulls down the position of the Dirac node in the frequency axis, which is the same as what happens in the Weyl node case in (b) of Fig.\ref{fig3}. Note that in this case, the nodal ring still exists as in case III.

V. Continue to increase $b_{xy}$ till $M<b<2b_{xy}$, and the system becomes fully gapped ((f) in Fig.\ref{fig3}). In this case, the nodal ring also exists as in cases III and IV.
\vspace{0cm}

\vspace{.25cm}
\noindent {\bf  $\bullet$ \em Turning on $b_{xy}$ in the critical phase}

The critical phase of Eq.\eqref{eq1} is a Dirac point where $b_{xy}=0$ and $b=M$((a) in Fig.\ref{fig4}). The Dirac point is located at $\left(k_x,k_y,k_z\right)=\left(0,0,0\right)$. Fix the values $b$ and $M$ as in (a) in Fig.\ref{fig4} and increase the value of $b_{xy}$ from 0, then we could obtain the following three types of energy spectrums with the presence of the nodal ring A or B (depends on the relative value of $2b_{xy}$ and $M$).

I. For $2b_{xy}<b=M$, the location of the Dirac point moves down as shown in (b) in Fig.\ref{fig4}. The non-zero $b_{xy}$ shifts the position of the Dirac node in the same way as in the case of the Weyl semimetal phase.

II. When $2b_{xy}=b=M$, a new triple degenerate critical point appears ((c) in Fig.\ref{fig4}), which is formed by a band crossing of three bands at one point. Similar to case (c) in Fig.\ref{fig3}, this critical point also corresponds to the critical case where the radius of the nodal ring becomes zero. It is a new topological state called triple degenerate nodal point with nontrivial topological charge\cite{fang2,fang3,weng2}, as will be shown later. The nontrivial topological charge is reflected in the fact that the node will not disappear when we tune the parameter slightly in both the larger and smaller directions: it either becomes a node formed by the upper two bands of the three bands, or a node formed by the lower two of the three bands. 

III. If we fix $b$ and $M$ in (c) in Fig.\ref{fig4}  and further increase $b_{xy}$ to have $2b_{xy}>b=M$, we get another Dirac point ((d) in Fig.\ref{fig4}). Here ``another" means that the crossing energy bands (orange and blue in Fig.\ref{fig4}) are different from the previous case (b) (orange and green in Fig.\ref{fig4})\footnote{This change can also be viewed from the perspective of the nodal ring state. As shown in Fig.\ref{fig11}, the crossing energy bands that form the nodal ring change after the point $2b_{xy}=M$.}, and these two different Dirac points cannot appear without the contributions from the two-form field. (b) in Fig.\ref{fig4} can be obtained by slightly decreasing $M$ or $b_{xy}$ from (c) in Fig.\ref{fig4}, while (d) in Fig.\ref{fig4} can be obtained by slightly increasing $M$ or $b_{xy}$ from (c) in Fig.\ref{fig4}. As we have already mentioned, these behaviors indicate that (c) in Fig.\ref{fig4} is not an accidental degeneracy and is protected by topology. The change of the crossing energy bands indicates the occurrence of a topological phase transition, which we will show by calculating the topological charge in section \ref{sec:invariant}.
\vspace{0cm}

\vspace{.25cm}
\noindent {\bf  $\bullet$ \em Turning on $b_{xy}$ in the gapped phase} 

\begin{figure}[h!]
  \centering
  \begin{subfigure}[b]{0.33\textwidth}
\includegraphics[width=\textwidth]{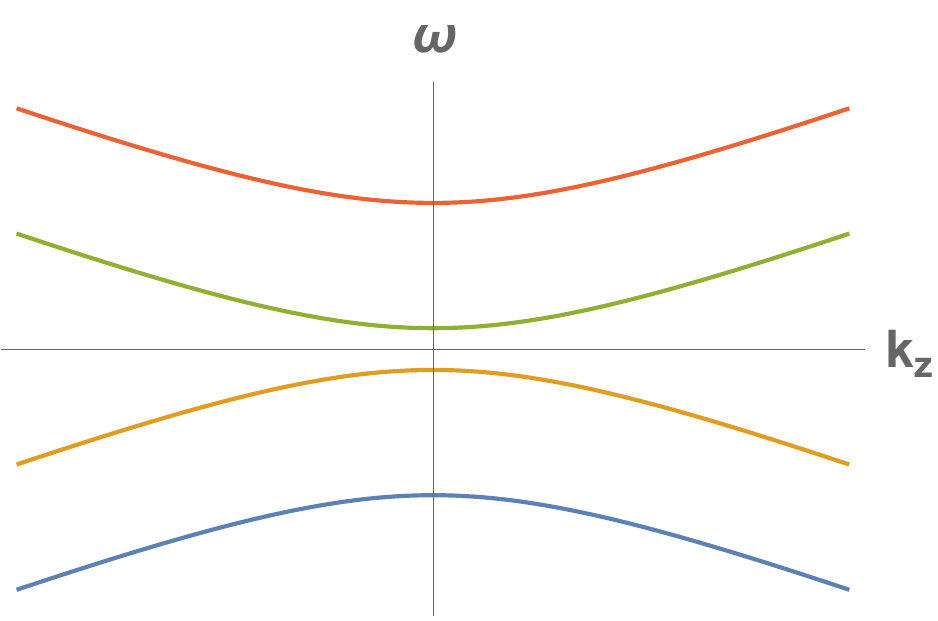}
\caption{\small gapped state}
\end{subfigure}
\begin{subfigure}[b]{0.32\textwidth}
\includegraphics[width=\textwidth]{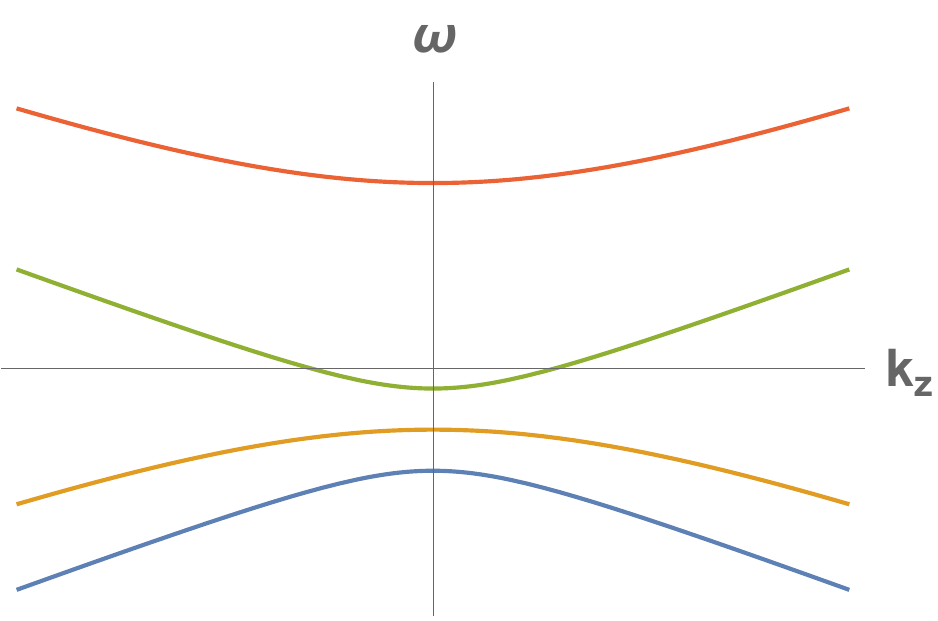}
\caption{\small shifted gapped state}
\end{subfigure}
\begin{subfigure}[b]{0.33\textwidth}
\includegraphics[width=\textwidth]{dd.pdf}
\caption{\small critical Dirac node}
\end{subfigure}
\begin{subfigure}[b]{0.33\textwidth}
\includegraphics[width=\textwidth]{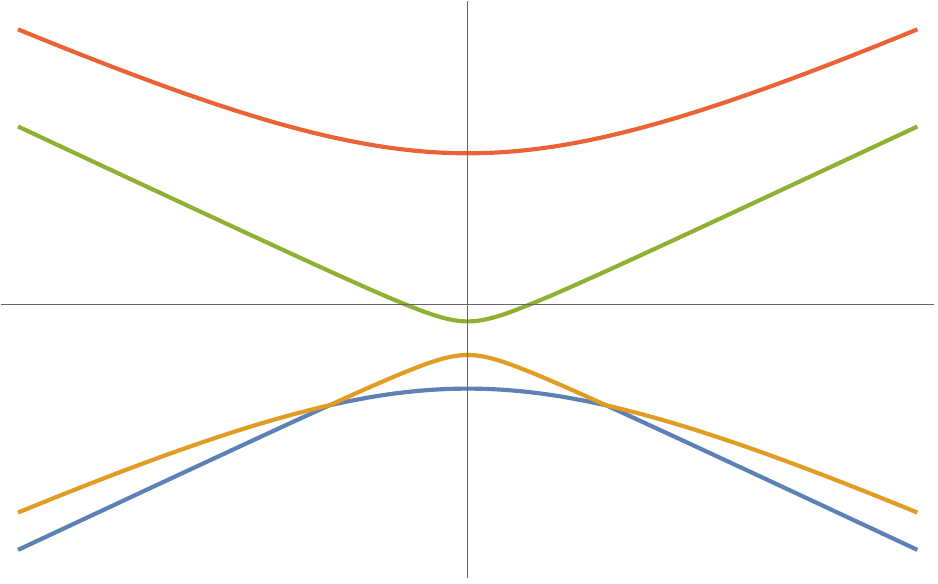}
\caption{\small shifted Weyl nodes}
\end{subfigure}
\begin{subfigure}[b]{0.32\textwidth}
\includegraphics[width=\textwidth]{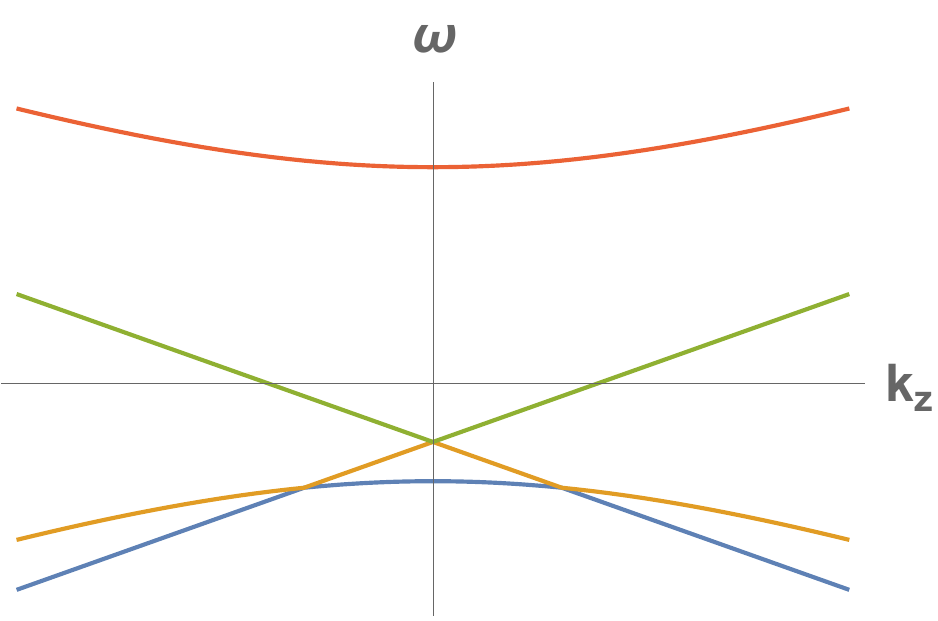}
\caption{\small three nodes}
\end{subfigure}
\begin{subfigure}[b]{0.33\textwidth}
\includegraphics[width=\textwidth]{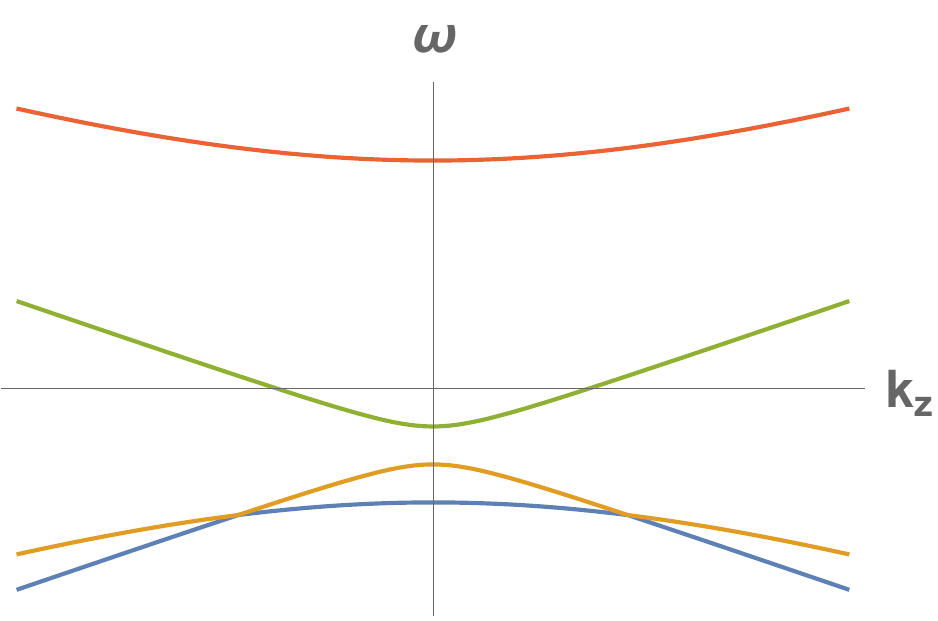}
\caption{\small shift Weyl nodes}
\end{subfigure}
 \caption{\small Part of the energy spectrum of \eqref{eq4} as a function of $k_z$ with $k_x=k_y=0$. From (a) to (e): the system is (a):a gapped state when $b<M$ and $b_{xy}=0$, (b): gap moves down when $2b_{xy}<b<M$, (c): a critical Dirac node when $2b_{xy}=b<M$, (d): two Weyl nodes move down when $b<2b_{xy}<M$, (e): three nodes when $b<2b_{xy}=M$, and (f) two Weyl nodes move down when $b<M<2b_{xy}$. Note that in (b),(c) and (d), the system has a nodal ring formed by the lowest and the second lowest energy band (nodal ring A), while in (f), the system has a nodal ring formed by the second lowest and the third lowest energy band (nodal ring B).}
  \label{fig5}
\end{figure}

The gapped state satisfies $b<M$ and $b_{xy}=0$((a) in Fig.\ref{fig5}). If we continue to increase $b_{xy}$ with the fixed $b$ and $M$ in (a) in Fig.\ref{fig5}, we could obtain the following five types of energy spectrums with the presence of the nodal ring A or B(depends on the realitive value of $2b_{xy}$ and $M$).

I. Increase the value of $b_{xy}$ from 0 to $2b_{xy}<b<M$, and the position of the gap moves down, as shown in (b) in Fig.\ref{fig5}. Note that in this case, a nodal ring also exists in the $k_x-k_y$ plane which cannot be seen in the figure.

II. When $2b_{xy}=b<M$, a critical Dirac point appears ((c) in Fig.\ref{fig5}). A nodal ring also exists as in case I.

III. If we continue to increase $b_{xy}$, while keep $b<2b_{xy}<M$, the Dirac point separates into two Weyl nodes, which are shown in (d) in Fig.\ref{fig5}. It is quite different from the original Lagrangian Eq.\eqref{eq1}, where the gapped state could never again become Weyl nodes with an increasing mass. The $k_x-k_y$ plane nodal ring still exists in this case. 

IV. When $b<2b_{xy}=M$, we obtain three nodes similar to (c) in Fig.\ref{fig3} ((e) in Fig.\ref{fig5}) as a critical point. The reason for the existence of the third node is also the same as in Fig.\ref{fig3}, i.e. the nodal ring shrinks to a point at this point. The same as in that case, this is also a nontrivial node, which can be seen both from the nontrivial topological invariant and the existence of finite radius nodal rings when the parameter is tuned slightly in both directions.

V. In the region $b<M<2b_{xy}$, the third node vanishes\footnote{same reason as (d) in Fig.\ref{fig3}} and the system becomes (f) in Fig.\ref{fig5}. A new nodal ring in the $k_x-k_y$ plane is formed by the second lowest and the third lowest energy band in this case. 

\vspace{0cm}

\section{the calculating process of the surface states}
\label{app:b}
\subsection{the nodal line semimetal case}
\label{app:b1}
Substituting the form of $\Upsilon_{nl}\left(x,y,z\right)$ and $b_{xy}=-b_{yx}=b_{xy}\theta\left({z}\right)$ into $H_{nl}\Upsilon_{nl}\left(x,y,z\right)=E_{nl}\Upsilon_{nl}\left(x,y,z\right)$, we can obtain four differential equations. With the value $E_{nl}=k_y$, these equations can be simplified as follows:
\begin{eqnarray}
&\left(\partial_{z}-k_{x}\right)u_{1}\left(z\right)+i\left(m+2b_{xy}\theta\left(z\right)\right)u_{2}\left(z\right)=0,\nonumber\\
&\left(\partial_{z}-k_{x}\right)u_{1}\left(z\right)+i\left(m-2b_{xy}\theta\left(z\right)\right)u_{2}\left(z\right)=0,\nonumber\\
&-i\left(m+2b_{xy}\theta\left(z\right)\right)u_{1}\left(x\right)+\left(\partial_{z}+k_{x}\right)u_{2}\left(z\right)=0,\nonumber\\
&-i\left(m-2b_{xy}\theta\left(z\right)\right)u_{1}\left(z\right)+\left(\partial_{z}+k_{x}\right)u_{2}\left(z\right)=0.\nonumber
\end{eqnarray}
It is easy to obtain the following solutions as
$u_{2}\left(z\right)\left(z<0\right)=A e^{\sqrt{k_{x}^{2}+m^{2}}z}$,
$u_{2}\left(z\right)\left(z>0\right)=A e^{-\sqrt{k_{x}^{2}+m^{2}-4b_{xy}^{2}}z}$,
$u_{1}\left(z\right)=B u_{2}\left(z\right)$. The normal condition of $u_i\left(z\right)$ gives the parameter region where the surface states survive, i.e. $k_{x}^{2}+m^{2}-4b_{xy}^{2}\geqslant0$. Thus we have  $k_x\leqslant-\sqrt{4b_{xy}^{2}-m^2}$ and $k_x\geqslant\sqrt{4b_{xy}^{2}-m^2}$. The $u_i\left(z\right),i=1,2$ correspond to the surface states of the nodal line semimetal that can be observed from the ARPES.

Here we have to mention that $u_i\left(z\right),i=1,2$ cannot be real functions at the same time according to the eigenenergy equations of the Hamiltonian Eq.\eqref{hnl}. This might be solved by introducing a pure imaginary field $b_{tz}$ as mentioned in \cite{Liu:2020ymx} and we leave this for a future work.

\subsection{The coexistence case}
\label{app:b2}
An eigenvalue with $E=k_y$ can be obtained and we obtain the following four differential equations:
\begin{eqnarray}
&\left(\partial_{x}+k_{z}+b\theta\left(x\right)\right)u_{1}\left(x\right)+\left(m+2b_{xy}\right)u_{2}\left(x\right)=0,\nonumber\\
&\left(\partial_{x}+k_{z}+b\theta\left(x\right)\right)u_{1}\left(x\right)+\left(m-2b_{xy}\right)u_{2}\left(x\right)=0,\nonumber\\
&\left(m+2b_{xy}\right)u_{1}\left(x\right)+\left(\partial_{x}-k_{z}+b\theta\left(x\right)\right)u_{2}\left(x\right)=0,\nonumber\\
&\left(m-2b_{xy}\right)u_{1}\left(x\right)+\left(\partial_{x}-k_{z}+b\theta\left(x\right)\right)u_{2}\left(x\right)=0.\nonumber
\end{eqnarray}

From these equations, we obtain:
\begin{eqnarray}
&u_{1}\left(x\right)\left(x<0\right)=A e^{\sqrt{k_{z}^{2}+m^{2}-4b_{xy}^{2}}x},\nonumber\\
&u_{2}\left(x\right)\left(x<0\right)=\frac{A}{2b_{xy}-m}\left(\sqrt{k_{z}^{2}+m^{2}-4b_{xy}^{2}}+k_{z}\right)e^{\sqrt{k_{z}^{2}+m^{2}-4b_{xy}^{2}}x},\nonumber\\
&u_{1}\left(x\right)\left(x>0\right)=B e^{\left(-b+\sqrt{k_{z}^{2}+m^{2}-4b_{xy}^{2}}\right)x}+C e^{-\left(b+\sqrt{k_{z}^{2}+m^{2}-4b_{xy}^{2}}\right)x},\nonumber\\
&u_{2}\left(x\right)\left(x>0\right)=\frac{A\left(\sqrt{k_{z}^{2}+m^{2}-4b_{xy}^{2}}+k_{z}\right)}{2b_{xy}-m}\left[B e^{\left(-b+\sqrt{k_{z}^{2}+m^{2}-4b_{xy}^{2}}\right)x}+C^{\prime} e^{-\left(b+\sqrt{k_{z}^{2}+m^{2}-4b_{xy}^{2}}\right)x}\right].\nonumber
\end{eqnarray}

According to the continuity of $u_{1}\left(x\right)$ and $u_{2}\left(x\right)$ at $x=0$, we have $A=B+C$ and $A=B+C^{\prime}=B+\frac{k_{z}-\sqrt{k_{z}^{2}+m^{2}-4b_{xy}^{2}}}{\sqrt{k_{z}^{2}+m^{2}-4b_{xy}^{2}}+k_{z}}C$.

With $C=0,A=B$, we obtain
\begin{eqnarray}
&u_{1}\left(x\right)\left(x<0\right)=A e^{\sqrt{k_{z}^{2}+m^{2}-4b_{xy}^{2}}x},\nonumber\\
&u_{1}\left(x\right)\left(x>0\right)=A e^{\left(-b+\sqrt{k_{z}^{2}+m^{2}-4b_{xy}^{2}}\right)x},\nonumber\\
&u_{2}\left(x\right)\left(x>0\right)=\frac{A\left(\sqrt{k_{z}^{2}+m^{2}-4b_{xy}^{2}}+k_{z}\right)}{2b_{xy}-m}u_{1}\left(x\right).\nonumber 
\end{eqnarray}

Since the solution is normalisable only with $-b+\sqrt{k_{z}^{2}+m^{2}-4b_{xy}^{2}}<0$ when $x>0$, thus we obtain the surface state with eigenvalue $E=+k_{y}$ in the region $-\sqrt{b^{2}-m^{2}+4b_{xy}^{2}}<k_z<\sqrt{b^{2}-m^{2}+4b_{xy}^{2}}$. This survive region of surface states is consist with the Hamiltonian of the Weyl semimetal eq.\eqref{eq1} with $b_{xy}=0$. We note that the existence of the two-form field will enlarge this region.

Next we will consider the case with $b_{xy}=-b_{yx}=b_{xy}\theta\left({x}\right)$. Similar as above, we obtain the another four differential equations:
\begin{eqnarray}
&\left(\partial_{x}+k_{z}+b\right)u_{1}\left(x\right)+\left(m+2b_{xy}\theta\left(x\right)\right)u_{2}\left(x\right)=0,\nonumber\\
&\left(\partial_{x}+k_{z}+b\right)u_{1}\left(x\right)+\left(m-2b_{xy}\theta\left(x\right)\right)u_{2}\left(x\right)=0,\nonumber\\
&\left(m+2b_{xy}\theta\left(x\right)\right)u_{1}\left(x\right)+\left(\partial_{x}-k_{z}+b\right)u_{2}\left(x\right)=0,\nonumber\\
&\left(m-2b_{xy}\theta\left(x\right)\right)u_{1}\left(x\right)+\left(\partial_{x}-k_{z}+b\right)u_{2}\left(x\right)=0.\nonumber
\end{eqnarray}

The form of $u_{1}\left(x\right)$ and $u_{2}\left(x\right)$ are:
\begin{eqnarray}
&u_{1}\left(x\right)\left(x<0\right)=B e^{\left(-b+\sqrt{k_{z}^{2}+m^{2}}\right)x}+C e^{-\left(b+\sqrt{k_{z}^{2}+m^{2}}\right)x},\nonumber\\
&u_{2}\left(x\right)\left(x<0\right)=-\frac{1}{m}\left(\sqrt{k_{z}^{2}+m^{2}}+k_{z}\right)\left[B e^{\left(-b+\sqrt{k_{z}^{2}+m^{2}}\right)x}+\frac{k_{z}-\sqrt{k_{z}^{2}+m^{2}}}{k_{z}+\sqrt{k_{z}^{2}+m^{2}}} C e^{-\left(b+\sqrt{k_{z}^{2}+m^{2}}\right)x}\right],\nonumber\\
&u_{1}\left(x\right)\left(x>0\right)=B^{\prime} e^{\left(-b+\sqrt{k_{z}^{2}+m^{2}-4b_{xy}^{2}}\right)x}+C^{\prime} e^{-\left(b+\sqrt{k_{z}^{2}+m^{2}-4b_{xy}^{2}}\right)x},\nonumber\\
&u_{2}\left(x\right)\left(x>0\right)=\frac{A\left(\sqrt{k_{z}^{2}+m^{2}-4b_{xy}^{2}}+k_{z}\right)}{2b_{xy}-m}\left[B^{\prime} e^{\left(-b+\sqrt{k_{z}^{2}+m^{2}-4b_{xy}^{2}}\right)x}+C^{\prime\prime} e^{-\left(b+\sqrt{k_{z}^{2}+m^{2}-4b_{xy}^{2}}\right)x}\right].\nonumber
\end{eqnarray}
Consider the normalisation condition at $x<0$, we have $C=0$. Then from the continuity of $u_{1}\left(x\right)$ and $u_{2}\left(x\right)$ at $x=0$ we can get the relation between $B,C,B^{\prime}$ and $C^{\prime}$, we will not show the explicit form here because it is complicated.

The normalized region of $k_z$ need to obtain from the formula of $u_i\left(x\right),i=1,2$ both with $x<0$ and $x>0$, and this is $-\sqrt{b^{2}-m^{2}+4b_{xy}^{2}}<k_z<-\sqrt{b^{2}-m^{2}}$ or $\sqrt{b^{2}-m^{2}}<k_z<\sqrt{b^{2}-m^{2}+4b_{xy}^{2}}$. Contrary to the case with $b_z=b \theta\left({x}\right)$ where the surface states survive in a continues region. The surface states with $b_{xy}=-b_{yx}=b_{xy}\theta\left({x}\right)$ survive in two discrete region.

\section{Review of \texorpdfstring{$\boldmath{Z}_2$}. Weyl Semimetal and the eight component spinor}\label{appendixz2}
Ideal Weyl semimetal has been realized in cold atoms\cite{pan}. This system retains the basic topological properties of the Weyl semimetal and could serve as a good candidate for studying chiral anomaly. However, most Weyl semimetals have more than one pair of Weyl nodes. Multiple Weyl nodes bring novel topological properties that need to be carefully studied. In another aspect, an electron system also includes degrees of freedom that describe the particle-hole and spin. In addition, a special degree of freedom called valley also exists in the Weyl semimetal. If additional degrees of freedom are considered simultaneously, more pairs of Weyl nodes could be introduced.

One minimum extension of the ideal Weyl semimetal is a system which has two pairs of Weyl nodes. $\boldmath{Z}_2$-Weyl semimetal is such a system where four Weyl nodes exist. The existence of additional nodes introduces another topological charge called $Z_2$ charge to the system\cite{kimb,Morimoto}. $\boldmath{Z}_2$ charge is a spin analog of the chiral charge, which measures the difference between the spin up and spin down fermions with the same chiral charges. Thus, $\boldmath{Z}_2$-Weyl semimetal could study the chiral and spin degree of freedom simultaneously and bring more significant topological properties. An effective field theory model for the $\boldmath{Z}_2$-Weyl semimetal could be written as
 \begin{align}
\label{eq2}
\mathcal{L}_{\boldmath{Z}_2}=\Psi^{\dagger}\left[\Gamma^{0}\left(i\Gamma^{\mu}\partial_{\mu}-e\Gamma^{\mu}A_{\mu}-\Gamma^{\mu} \Gamma^{5}b_{\mu} {\mI}_{1}+ {M}_{1} {\mI}_{1}+ {M}_{2} {\mI}_{2} \right)+\hat{\Gamma}^{0}\left(e\hat{\Gamma}^{\mu}\hat{A}_{\mu}- \hat{\Gamma}^{\mu}\hat{\Gamma}^{5}c_{\mu} {\mI}_{2}\right)\right]\Psi\,,
\end{align}
where $\Psi$ is an eight-component spinor that contains degrees of freedom that describe the chirality, particle-hole, and spin
\begin{eqnarray}
\label{psi}
\Psi=\left(\Psi_{p,+,\uparrow},\Psi_{p,+,\downarrow},\Psi_{h,+,\uparrow},\Psi_{h,+,\downarrow},\Psi_{p,-,\uparrow},\Psi_{p,-,\downarrow},\Psi_{h,-,\uparrow},\Psi_{h,-,\downarrow}\right)^{T},
\end{eqnarray}
and here $p,h$ in the index refers to particle-hole, $\pm$ refers to the chirality and the arrow means spin up and down.

Original $4\times 4$ Gamma matrices also need to be generalized into $8\times8$  matrices for this eight-component spinor system. New `Gamma' matrices are defined as follows:
\be
\label{eq:88marix}
\Gamma^{\mu}\equiv\gamma^{\mu}\otimes \mathbb{I}_2\,,~~\hat{\Gamma}^{\mu}\equiv\gamma^{\mu}\otimes \mathbb{Z}_2\,,~~\Gamma^{5}\equiv\gamma^{5}\otimes \mathbb{I}_2\,,~~\hat{\Gamma}^{5}\equiv\gamma^{5}\otimes \mathbb{Z}_2\,,\ee
where $\mu=0,1,2,3$, $\gamma^\mu$ is the $4\times 4$ Dirac Gamma matrix and
\begin{align}
\label{eq:matrix}
\mathbb{I}_2 &=\left(\begin{array}{cc}1 & 0 \\0 & 1\end{array}\right) ,\qquad
\mathbb{Z}_2=\left(\begin{array}{cc}1 & 0 \\0 & -1\end{array}\right)\,.
\end{align}
Here, $\pm1$ in $\mathbb{Z}_{2}$ means spin up and down, respectively. It could e checked directly that these newly defined Gamma matrices satisfy the Clifford algebra. With this definition, the information of chiral and $\boldmath{Z}_2$ charge have been taken into consideration in the eight-component spinor\cite{Ji:2021aan}.

\begin{figure}[h!]
  \centering
  \begin{subfigure}[b]{0.33\textwidth}
\includegraphics[width=\textwidth]{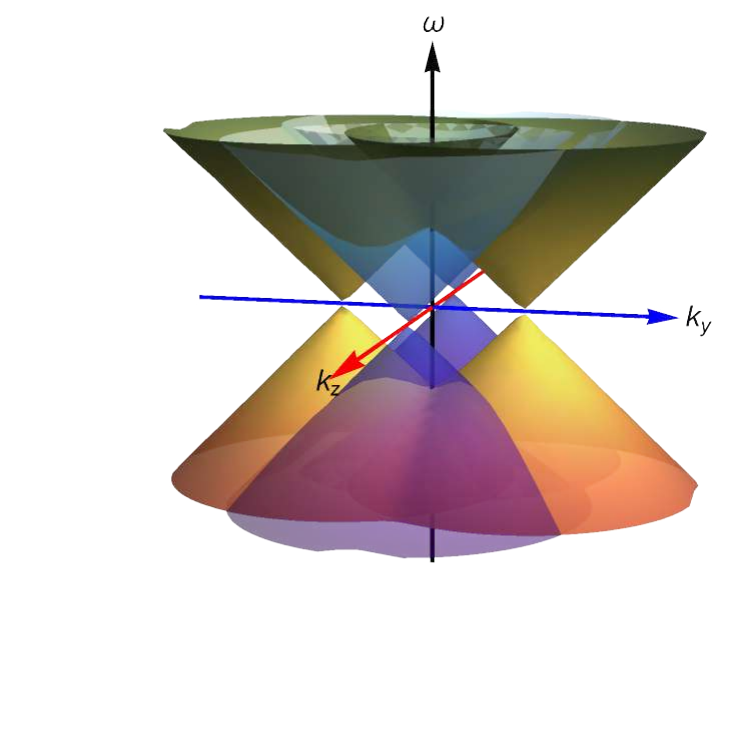}
\vspace{-1.7cm}
\caption{\small Weyl-$\boldmath{Z}_2$}
\end{subfigure}
\begin{subfigure}[b]{0.32\textwidth}
\includegraphics[width=\textwidth]{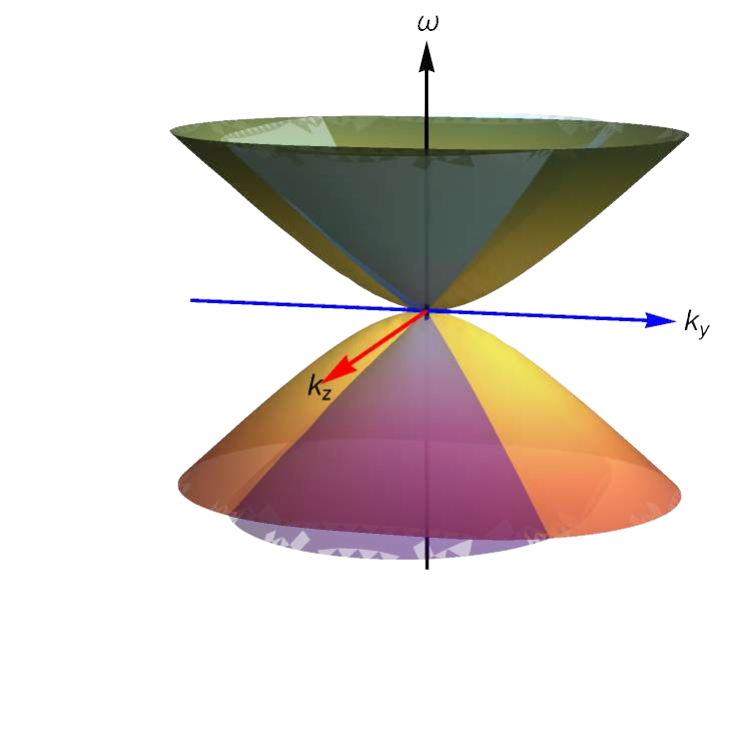}
\vspace{-1.7cm}
\caption{\small double critical}
\end{subfigure}
\begin{subfigure}[b]{0.33\textwidth}
\includegraphics[width=\textwidth]{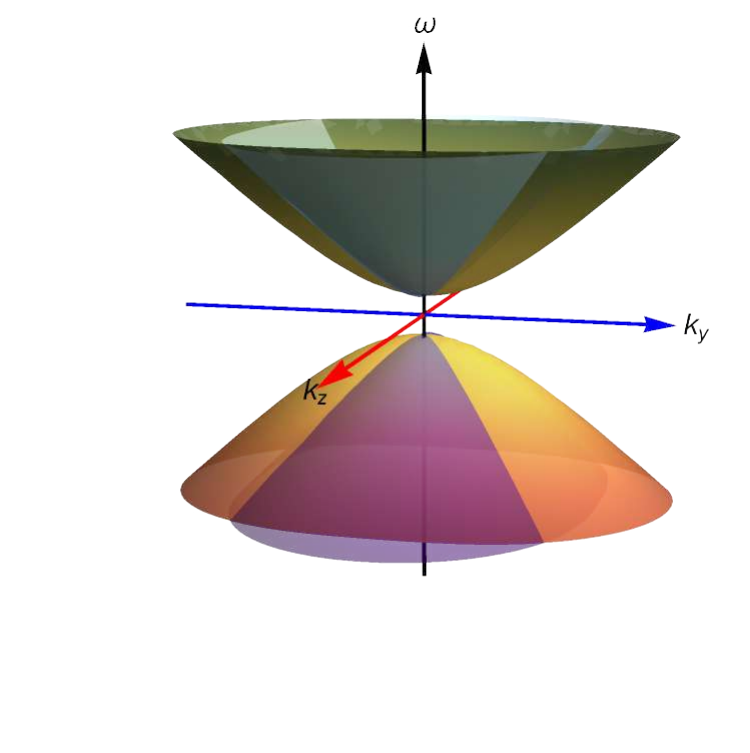}
\vspace{-1.7cm}
\caption{\small gap-gap}
\end{subfigure}
\begin{subfigure}[b]{0.32\textwidth}
\includegraphics[width=\textwidth]{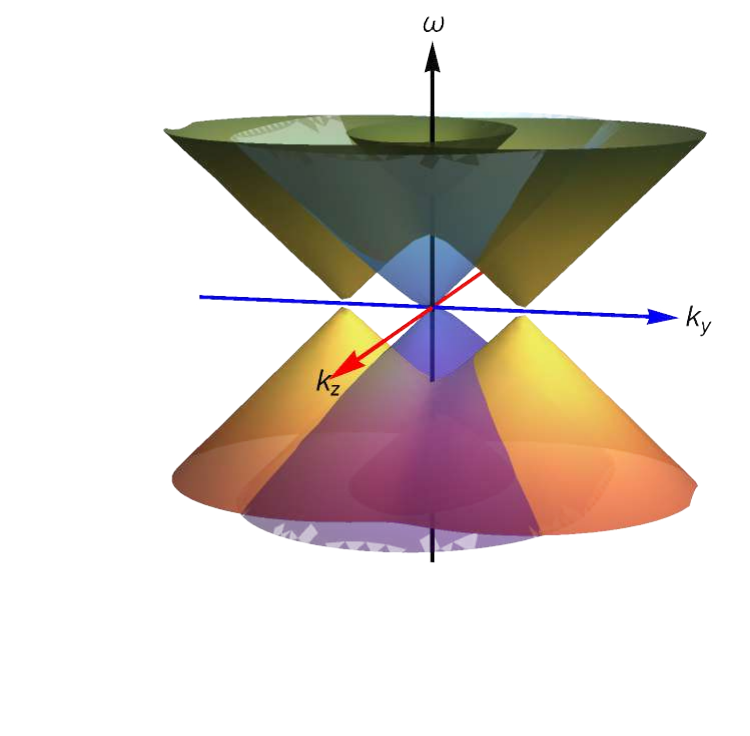}
\vspace{-1.7cm}
\caption{\small Weyl/$\boldmath{Z}_2$-critical}
 \end{subfigure}
\begin{subfigure}[b]{0.32\textwidth}
\includegraphics[width=\textwidth]{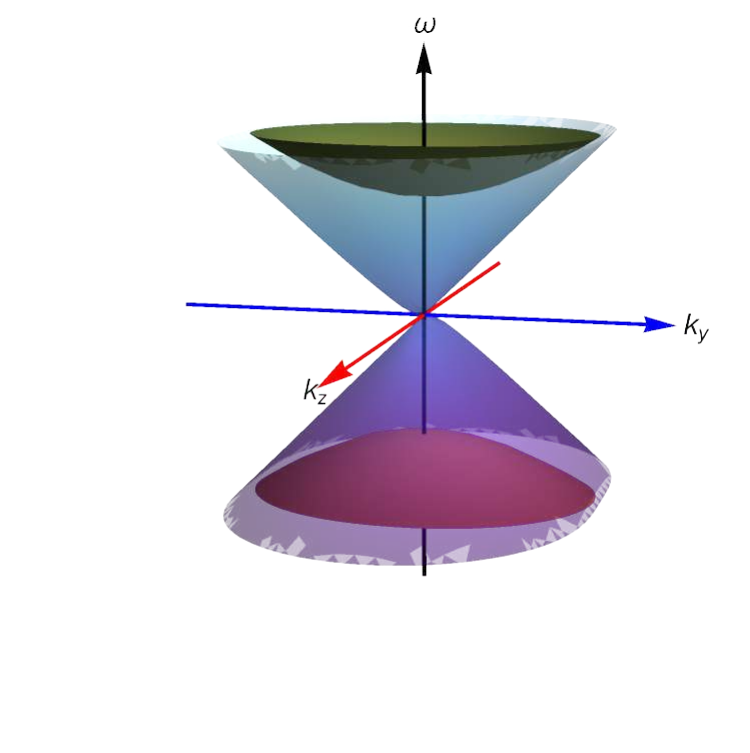}
\vspace{-1.7cm}
\caption{\small critical-gap or gap-critical}
\end{subfigure}
\begin{subfigure}[b]{0.32\textwidth}
\includegraphics[width=\textwidth]{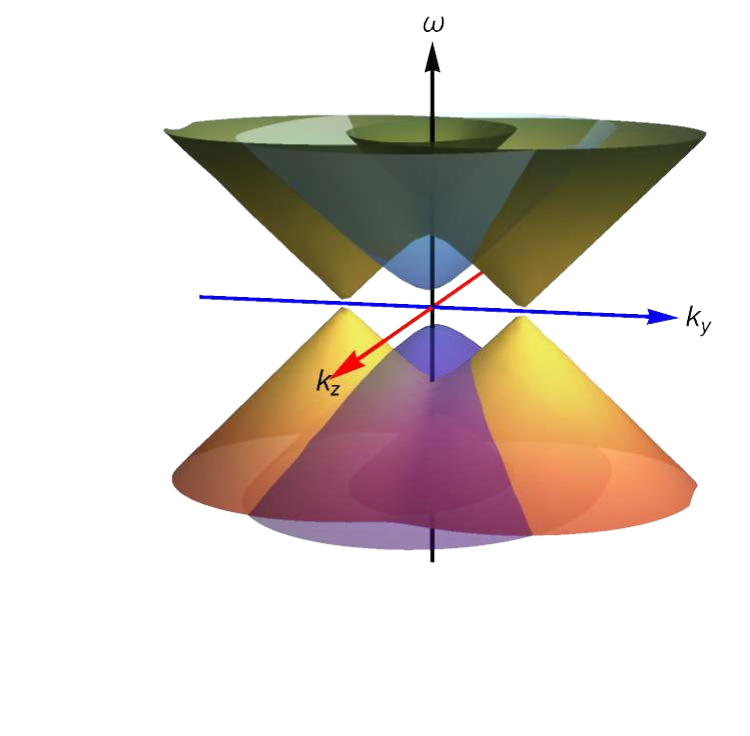}
\vspace{-1.7cm}
\caption{\small Weyl/$\boldmath{Z}_2$-gap
\label{fig:phasef}}
\end{subfigure}
  \caption{\small The energy spectrum \ref{ospectrum} as a function of $k_y$ and $k_z$ with $k_x=0$.  {From (a) to (f): the system has two pairs of Weyl/$\boldmath{Z}_2$ nodes (a), two critical Dirac nodes (b), fully gapped (c), one pair of Weyl/$\boldmath{Z}_2$ nodes and a critical Dirac node (d), a critical Dirac node and a gapped phase (e) and the case of one pair of Weyl/$\boldmath{Z}_2$ nodes with two gapped bands (f)\cite{Ji:2021aan}}.
  }
  \label{fig:phase}
\end{figure}

Choose $b_{\mu}=b\delta^{x}_{\mu}$ and $c_{\mu}=c\delta^{y}_{\mu}$ in the Eq.\eqref{eq2} without loss of generality, we obtain the eight eigenvalues:
\begin{eqnarray}
\label{spectrum}
E_{1}=\pm\sqrt{\left(b_{z}\pm\sqrt{k_{z}^{2}+M_1^{2}}\right)^{2}+k_{y}^{2}}\,,\quad~~~~
 E_{2}=\pm\sqrt{\left(c_{y}\pm\sqrt{k_{y}^{2}+M_2^{2}}\right)^{2}+k_{z}^{2}}\,.
\end{eqnarray}
Note that $E_1$ only depends on $b$ and $M_1$ while $E_2$ only depends on $c$ and $M_2$. $E_1$ can be viewed as the energy spectrum of \eqref{eq1}, with $M$ replaced by $M_1$, and $E_2$ can be viewed as the energy spectrum of \eqref{eq1} with $\vec{b}$ and $M$ replaced by $\vec{c}$ and $M_2$. Nine phases exist in this system(shown in Fig.\ref{fig:phase})  and the topological phase transitions have been studied in\cite{Ji:2021aan}, which we do not repeat here.
\vspace{0cm}

We would like to mention that each pair of the nodes carry topological charges $(1,0), (-1, 0)$ or $(0,1), (0, -1)$. One could identify the pair of nodes in one direction as carrying the charge of chiral $U(1)$, while the nodes in the other direction as carrying the charge of the analog spin $Z_2$ charge. Coexistence of two topological charges are present directly. This reminds us that if we want to study the phase transition between different topological states, we could enlarge the Hilbert space of the system. The enlarged Hilbert space contains more symmetry than the one where a single topological state lives. We will employ this approach to study the phase transition between Weyl semimetal and topological nodal line semimetal in the following.

\subsection*{Data Availability Statement}
The authors confirm that there are no associated data in the paper.


\begin{thebibliography}{99}

\bibitem{zkliu}
Z. K. Liu, B. Zhou, Y. Zhang, {\em et al.},
{\em Discovery of a Three-Dimensional Topological Dirac Semimetal, $Na_{3}Bi$},
\href{https://www.science.org/doi/abs/10.1126/science.1245085}{ Science, {\bf 343}, 6173 (2014)}.

\bibitem{Ylchen}
Z. K. Liu, J. Jiang, B. Zhou, {\em et al.},
{\em A stable three-dimensional topological Dirac semimetal $Cd_{3}As_{2}$},
\href{https://doi.org/10.1038/nmat3990}{Nature.\ Mater, {\bf 13}, 3990 (2014)}.

\bibitem{kane}
S. M. Young, S. Zaheer, J. C. Y. Teo, {\em et al.},
{\em Dirac Semimetal in Three Dimensions},
\href{https://journals.aps.org/prl/abstract/10.1103/PhysRevLett.108.140405}{Phys.\ Rev.\ Lett. {\bf 108}, 140405 (2012)}.

\bibitem{Wan}
X.~Wan, A. M. Turner, A. Vishwanath, and S. Y. Savrasov,
{\em Topological semimetal and Fermi-arc surface states in the electronic structure of pyrochlore iridates},
\href{http://dx.doi.org/10.1103/PhysRevB.83.205101}{\ Phys.\ Rev.\ B {\bf 83}, 205101 (2011)} [\arXiv{1007.0016}{cond-mat.str-el}].

\bibitem{rmb}
N. P.~Armitage, E. J. Mele, A.~Vishwanath,
{\em Weyl and Dirac semimetals in three-dimensional solids},
\href{https://journals.aps.org/rmp/abstract/10.1103/RevModPhys.90.015001}{\ Rev. \ Mod. \ Phys. {\bf 90}, 15001 (2018)}
[\arXiv{1705.01111}{cond-mat.str-el}].

\bibitem{burkov1}
A.~A.~Burkov, M.~D.~Hook and L.~ Balents,
{\em Topological nodal semimetals},
\doi{10.1103/PhysRevB.84.235126}{Phys.\ Rev.\ B {\bf 84}, 235126 (2011)},
[\arXiv{1110.1089}{cond-mat.mes-hall}].

\bibitem{Gorgar1}
E. V.~Gorbar, V. A. Miransky, I. A. Shovkovy, and P. O. Sukhachov,
{\em Dirac semimetals $A_3Bi\left(A=Na,K,Rb\right)$ as Z$_{2}$  Weyl semimetals},
\href{https://journals.aps.org/prb/abstract/10.1103/PhysRevB.91.121101}{\ Phys.\ Rev.\ B {\bf 91}, 121101(2015)}.

\bibitem{Gorgar2}
E. V.~Gorbar, V. A. Miransky, I. A. Shovkovy, and P. O. Sukhachov,
{\em Surface Fermi arcs in Z$_{2}$ Weyl semimetals $A_3Bi\left(A=Na,K,Rb\right)$},
\href{https://journals.aps.org/prb/abstract/10.1103/PhysRevB.91.235138}{\ Phys.\ Rev.\ B {\bf 91}, 235138(2015)}.

\bibitem{liang}
Qi-Feng Liang, Jian Zhou, Rui Yu, Zhi Wang, and Hongming Weng, 
{\em Node-surface and node-line fermions from nonsymmorphic lattice symmetries},
\href{https://journals.aps.org/prb/abstract/10.1103/PhysRevB.93.085427}{\ Phys.\ Rev.\ B {\bf 93}, 085427 (2016)}.

\bibitem{zhong}
Chengyong Zhong, Yuanping Chen, Yuee Xie, {\em et al.}, {\em Towards three-dimensional Weyl-surface semimetals in graphene networks},
\href{https://pubs.rsc.org/en/content/articlelanding/2016/NR/C6NR00882H}{\ Nanoscale {\bf 8}, 7232 (2016)}.

\bibitem{Ninomiya}
H.~B.~Nielsen and M.~Ninomiya,
{\em Adler-Bell-Jackiw anomaly and Weyl fermions in crystal,}{}
\doi{10.1016/0370-2693(83)91529-0}{Phys. Lett. B \textbf{130}, 389-396 (1983)}.

%
\bibitem{Son:2012wh}
D.~T.~Son and N.~Yamamoto,
{ \em Berry Curvature, Triangle Anomalies, and the Chiral Magnetic Effect in Fermi Liquids,}
\doi{10.1103/PhysRevLett.109.181602}{Phys.\ Rev.\ Lett.\  {\bf 109}, 181602 (2012)} [\arXiv{1203.2697}{cond-mat.mes-hall}].

\bibitem{pan}
Z. ~Y. ~Wang, X. ~C. ~Cheng, B. ~Z. ~Wang, {\em et al.},
{\em Realization of an ideal Weyl semimetal band in a quantum gas with 3D spin-orbit coupling},
\href{https://www.science.org/doi/full/10.1126/science.abc0105}{ Science, {\bf 372}, 271 (2021)}[\arXiv{2004.02413}{cond-mat.quant-gas}]. 

\bibitem{Fleury}
Farzad Zangeneh-Nejad and Romain Fleury,
{\em Experimental observation of the acoustic Z$_{2}$ Weyl semimetallic phase in synthetic dimensions},
\href{https://journals.aps.org/prb/abstract/10.1103/PhysRevB.102.064309}{\ Phys.\ Rev.\ B {\bf 102}, 064309(2020)}.

\bibitem{Ji:2021aan}
X.~Ji, Y.~Liu, Y.~W.~Sun and Y.~L.~Zhang,
{\em A Weyl-Z$_{2}$ semimetal from holography},
\doi{10.1007/JHEP12(2021)066}{JHEP \textbf{12}, 066 (2021)}
[\arXiv{2109.05993}{hep-th}].


\bibitem{weng}
H. Weng, C. Fang, Z. Fang, B. A. Bernevig, and X. Dai, {\em Weyl Semimetal Phase in Noncentrosymmetric Transition-Metal Monophosphides} \href{https://journals.aps.org/prx/abstract/10.1103/PhysRevX.5.011029}{\ Phys.\ Rev.\ X {\bf 5}, 011029 (2015)}.

\bibitem{B.Q.Lv}
B. Q. Lv, H. M. Weng, B. B. Fu, X. P. Wang, {\em et al.}, 
{\em Experimental discovery of Weyl semimetal TaAs}, \href{https://journals.aps.org/prx/abstract/10.1103/PhysRevX.5.031013}{\ Phys.\ Rev.\ X {\bf 5}, 031013 (2015)} [\arXiv{1502.04684}{cond-mat.str-el}].

\bibitem{shuang}
S.-M. Huang, S.-Y. Xu, I. Belopolski, C.-C. Lee, {\em et al.}, 
{\em A Weyl Fermion semimetal with surface Fermi arcs in the transition metal monopnictide TaAs class}, \href{https://www.nature.com/articles/ncomms8373}{Nat.\ Commun.\ {\bf 6}, 7373 (2015)}.

\bibitem{SXu}
S.-Y. Xu, I. Belopolski, N. Alidoust, M. Neupane, {\em et al.},
{\em Discovery of a Weyl fermion semimetal and topological Fermi arcs}, \href{https://science.sciencemag.org/content/349/6248/613?ijkey=2bc9168bb187cf6c6774b10ca592d39dd356bf67&keytype2=tf_ipsecsha}{ Science {\bf 349}, 613 (2015)}.

\bibitem{Borisenko}
S. Borisenko, D. Evtushinsky, Q. Gibson, A. Yaresko, {\em et al.},
{\em Time-Reversal Symmetry Breaking Type-II Weyl State in $YbMnBi_2$}, \href{https://arxiv.org/abs/1507.04847}[\arXiv{1507.04847}{cond-mat.str-el}].

\bibitem{Soluyanov}
A. A. Soluyanov, D. Gresch, Z. Wang, Q. Wu,
{\em et al.}, 
{\em Type-II Weyl semimetals},
\href{https://www.nature.com/articles/nature15768}{ Nat. \textbf{527}, 495 (2015)} [\arXiv{1507.01603}{cond-mat.str-el}].

\bibitem{Deng} K. Deng, G. Wan, P. Deng, K. Zhang,
{\em et al.}, 
{\em Experimental observation of topological Fermi arcs in type-II Weyl semimetal $MoTe_2$},\href{https://www.nature.com/articles/nphys3871}{ \ Nat.\ Phys. \textbf{12}, 1105 (2016)} [\arXiv{1603.08508}{cond-mat.str-el}].

\bibitem{HaoZheng} 
H. Zheng, G. Bian, G. Chang, H. Lu,
{\em et al.}, 
{\em Atomic-Scale Visualization of Quasiparticle Interference on a Type-II Weyl Semimetal Surface},\href{https://journals.aps.org/prl/abstract/10.1103/PhysRevLett.117.266804}{\ Phys.\ Rev.\ Lett. {\bf 117}, 266804 (2016)} [\arXiv{1612.05208}{cond-mat.str-el}].

\bibitem{yao}
Xiao-Ping Li, Ke Deng, Botao Fu,{\em et al.},
{\em Type-III Weyl semimetals:$(TaSe_{4})_{2}$I},
\href{https://journals.aps.org/prb/abstract/10.1103/PhysRevB.103.L081402}{\ Phys.\ Rev.\ B {\bf 103}, L081402 (2021)} [\arXiv{1909.12178}{cond-mat.mes-hall}].


\bibitem{Gooth:2017mbd}
J.~Gooth, A.~C.~Niemann, T.~Meng, A.~G.~Grushin, K.~Landsteiner, B.~Gotsmann, F.~Menges, M.~Schmidt, C.~Shekhar and V.~Sue\ss{}, \textit{et al.}
{\em Experimental signatures of the mixed axial-gravitational anomaly in the Weyl semimetal NbP},
\doi{10.1038/nature23005}{Nature {\bf 547}, 324-327 (2017)}
[\arXiv{1703.10682}{cond-mat.mtrl-sci}].

\bibitem{Gooth2}
J. Gooth, F. Menges, N. Kumar, V. Süβ, C. Shekhar,\ss{}, \textit{et al.}
{\em Thermal and electrical signatures of a hydrodynamic electron fluid in tungsten diphosphide},\href{https://www.nature.com/articles/s41467-018-06688-y}{ \ Nat.\ Comm. \textbf{9}, 4093 (2018)}.



\bibitem{Hartnoll:2016apf}
S.~A.~Hartnoll, A.~Lucas and S.~Sachdev,
{\em Holographic quantum matter},
[\arXiv{1612.07324}{hep-th}].

\bibitem{Landsteiner:2019kxb}
K.~Landsteiner, Y.~Liu and Y.~W.~Sun,
{\em Holographic Topological Semimetals,}
\doi{10.1007/s11433-019-1477-7}{Sci. China Phys. Mech. Astron. \textbf{63} (2020) no.5, 250001}
[\arXiv{1911.07978}{hep-th}].


\bibitem{Erdmenger:2008rm}
  J.~Erdmenger, M.~Haack, M.~Kaminski and A.~Yarom,
{\em Fluid dynamics of R-charged black holes,}
 \doi{10.1088/1126-6708/2009/01/055}{JHEP {\bf 0901}, 055 (2009)},
  [\arXiv{0809.2488}{hep-th}].

\bibitem{Banerjee:2008th}
  N.~Banerjee, J.~Bhattacharya, S.~Bhattacharyya, S.~Dutta, R.~Loganayagam and P.~Surowka,
{\em Hydrodynamics from charged black branes,}
 \doi{10.1007/JHEP01(2011)094}{JHEP {\bf 1101}, 094 (2011)},
  [\arXiv{0809.2596}{hep-th}].

\bibitem{Landsteiner:2011cp}
  K.~Landsteiner, E.~Megias, F.~Pena-Benitez,
{\em Gravitational Anomaly and Transport,}
 \doi{10.1103/PhysRevLett.107.021601}{Phys.\ Rev.\ Lett.\  {\bf 107} (2011) 021601},
  [\arXiv{1103.5006}{hep-ph}].

\bibitem{Landsteiner:2011iq}
  K.~Landsteiner, E.~Megias, L.~Melgar, F.~Pena-Benitez,
{\em Holographic Gravitational Anomaly and Chiral Vortical Effect,}
 \doi{10.1007/JHEP09(2011)121}{JHEP {\bf 1109} (2011) 121},
  [\arXiv{1107.0368}{hep-th}].
  
\bibitem{Ji:2019pxx}
X.~Ji, Y.~Liu and X.~M.~Wu,
{\em Chiral vortical conductivity across a topological phase transition from holography,}
\doi{10.1103/PhysRevD.100.126013}{Phys. Rev. D \textbf{100}, no.12, 126013 (2019)}
[\arXiv{1904.08058}{hep-th}]. 

\bibitem{Gao:2023zbd}
L.~L.~Gao, Y.~Liu and H.~D.~Lyu,
{\em Black hole interiors in holographic topological semimetals,}
\doi{10.1007/JHEP03(2023)034}{JHEP {\bf 2303}, 034 (2023)},
[\arXiv{2301.01468}{hep-th}].  


\bibitem{Grushin:2012mt}
  A.~G.~Grushin,
 {\em Consequences of a condensed matter realization of Lorentz violating QED in Weyl semi-metals,}
  \doi{10.1103/PhysRevD.86.045001}{Phys.\ Rev.\ D {\bf 86}, 045001 (2012)},
  [\arXiv{1205.3722}{hep-th}].

\bibitem{Grushin:2019uuu}
  A.~G.~Grushin,
{\em Common and not so common high-energy theory methods for condensed matter physics,}
 \arXiv{1909.02983}{cond-mat.mes-hall}.
 
\bibitem{Kostelecky:2021bsb}
V.~A.~Kosteleck\'y, R.~Lehnert, N.~McGinnis, M.~Schreck and B.~Seradjeh,
{\em Lorentz violation in Dirac and Weyl semimetals,}
\doi{10.1103/PhysRevResearch.4.023106}{Phys.\ Rev.\ Res. {\bf 4}, 023106 (2022)},
Phys. Rev. Res. \textbf{4}, no.2, 023106 (2022)
[arXiv:2112.14293 [cond-mat.mes-hall]].

\bibitem{Kitaev1}
Alexei Kitaev,
{\em Anyons in an exactly solved model and beyond},
\href{https://www.sciencedirect.com/science/article/pii/S0003491605002381}{Annals of Physics {\bf 321}, 2-111(2006)}.


\bibitem{Farjami}
Ashk Farjami, Matthew D. Horner, Chris N. Self, Zlatko Papić, and Jiannis K. Pachos,
{\em Geometric description of the Kitaev honeycomb lattice model},
\href{https://journals.aps.org/prb/abstract/10.1103/PhysRevB.101.245116}{\ Phys.\ Rev.\ B {\bf 101}, 245116(2020)}.

\bibitem{Maiellaro1}
Alfonso Maiellaro, Francesco Romeo and Roberta Citro,
{\em Topological phases of a Kitaev tie},
\href{https://link.springer.com/article/10.1140/epjst/e2019-900180-x}{\ Eur.\ Phys.\ J.\ Special Topics {\bf 229}, 637(2020)}.

\bibitem{Maiellaro2}
Alfonso Maiellaro, Francesco Romeo and Roberta Citro,
{\em Effects of geometric frustration in Kitaev chains},
\href{https://link.springer.com/article/10.1140/epjp/s13360-021-01592-9}{\ Eur.\ Phys.\ J.\ Plus {\bf 136}, 627(2021)}.


\bibitem{read}
N. Read and Dmitry Green,
{\em Paired states of fermions in two dimensions with breaking of parity and time-reversal symmetries and the fractional quantum Hall effect},
\href{https://journals.aps.org/prb/abstract/10.1103/PhysRevB.61.10267}{\ Phys.\ Rev.\ B {\bf 61}, 10267(2000)}.

\bibitem{Golan}
Omri Golan and Ady Stern,
{\em Probing topological superconductors with emergent gravity},
\href{https://journals.aps.org/prb/abstract/10.1103/PhysRevB.98.064503}{\ Phys.\ Rev.\ B {\bf 98}, 064503(2018)}.

\bibitem{Maiellaro3}
Alfonso Maiellaro and Roberta Citro,
{\em Topological Edge States of a Majorana BBH Model},
\href{https://www.mdpi.com/2410-3896/6/2/15}{\ Condens.\ Matter {\bf 6}, 2(2021)}.

\bibitem{fang1}
C.~Fang, H.~Weng, X.~Dai and Z.~Fang,
{\em Topological nodal line semimetals},
\doi{10.1088/1674-1056/25/11/117106}{Chin. Phys. B 25, 117106 (2016)}
[\arXiv{1609.05414}{cond-mat.mes-hall}].


\bibitem{fang2}
Hongming Weng, Chen Fang, Zhong Fang, and Xi Dai, 
{\em Topological semimetals with triply degenerate nodal points in $θ$
-phase tantalum nitride},
\href{https://journals.aps.org/prb/abstract/10.1103/PhysRevB.93.241202}{\ Phys.\ Rev.\ B. {\bf 93}, 241202(R) (2016)}.

\bibitem{fang3}
Hongming Weng, Chen Fang, Zhong Fang, and Xi Dai, 
{\em Coexistence of Weyl fermion and massless triply degenerate nodal points},
\href{https://journals.aps.org/prb/abstract/10.1103/PhysRevB.94.165201}{\ Phys.\ Rev.\ B. {\bf 94}, 165201 (2016)}.

\bibitem{weng2}
Rui Yu, Quansheng Wu, Zhong Fang, and Hongming Weng, 
{\em From Nodal Chain Semimetal to Weyl Semimetal in HfC},
\href{https://journals.aps.org/prl/abstract/10.1103/PhysRevLett.119.036401}{\ Phys.\ Rev.\ Lett. {\bf 119}, 036401 (2017)}.

\bibitem{Ma} J. Z. Ma, Q. S. Wu, Song. M,
{\em et al.}, 
{\em Observation of a singular Weyl point surrounded by charged nodal walls in PtGa},\href{https://doi.org/10.1038/s41467-021-24289-0}{ \ Nat.\ Comm. \textbf{12}, 3994 (2021)}.



\bibitem{Colladay:1998fq}
D.~Colladay and V.~Kostelecky,
{\em Lorentz violating extension of the standard model,}
\doi{10.1103/PhysRevD.58.116002}{Phys. Rev. D {\bf{58}} (1998), 116002}
[\arXiv{hep-ph/9809521}{hep-ph}].

\bibitem{Landsteiner:2015pdh}
  K.~Landsteiner, Y.~Liu and Y.~W.~Sun,
{\em Quantum phase transition between a topological and a trivial semimetal from holography,}{}
\href{https://doi.org/10.1103/PhysRevLett.116.081602}{Phys.\ Rev.\ Lett.\  {\bf 116}, no. 8, 081602 (2016)}
[\arXiv{1511.05505}{hep-th}].


\bibitem{Landsteiner:2015lsa}
K.~Landsteiner and Y.~Liu,
{\em The holographic Weyl semi-metal,}
\doi{10.1016/j.physletb.2015.12.052}{Phys. Lett. B \textbf{753} (2016), 453-457}
[\arXiv{1505.04772}{hep-th}].


\bibitem{Liu:2018bye}
  Y.~Liu and Y.~W.~Sun,
{\em Topological nodal line semimetals in holography,}
  \doi{10.1007/JHEP12(2018)072}{JHEP {\bf 12}, 072 (2018)},
 \arXiv{1801.09357}{hep-th}.
 
\bibitem{Liu:2020ymx}
Y.~Liu and X.~M.~Wu,
{\em An improved holographic nodal line semimetal,}
\doi{10.1007/JHEP05(2021)141}{JHEP {\bf 05}, 141 (2021)},
[\arXiv{2012.12602}{hep-th}].
 
 
\bibitem{Goswami}
Pallab Goswami1 and Sumanta Tewari,
{\em Axionic field theory of (3 + 1)-dimensional Weyl semimetals},
\doi {https://journals.aps.org/prb/abstract/10.1103/PhysRevB.88.245107}{Phys.\ Rev.\ B. {\bf 88},245107 (2013)}.


\bibitem{Ammon:2016mwa}
M. Ammon, M. Heinrich, A. Jim\'enez-Alba and S. Moeckel,
{\em Surface States in Holographic Weyl Semimetals},
\doi{10.1103/PhysRevLett.118.201601}{Phys.\ Rev.\ Lett. {\bf 118},201601 (2017)},
 \arXiv{1612.00836}{hep-th}.



\bibitem{Witten:2015aoa}
  E.~Witten,
{\em Three Lectures On Topological Phases Of Matter,}
  \doi{10.1393/ncr/i2016-10125-3}{Riv.\ Nuovo Cim.\  {\bf 39}, no. 7, 313 (2016)},
 [\arXiv{1510.07698}{cond-mat.mes-hall}].

\bibitem{berry}
M. V. Berry,
{\em Quantal phase factors accompanying adiabatic changes,}
\href{http://rspa.royalsocietypublishing.org/content/392/1802/45}{Proc. Roy. Soc. Lond. A 392, 45 (1984)}.



\bibitem{Liu:2018djq}
Y.~Liu and Y.~W.~Sun,
{\em Topological invariants for holographic semimetals,}
\doi{10.1007/JHEP10(2018)189}{JHEP {\bf 10}, 189 (2018)},
[arXiv:1809.00513 [hep-th]].

\bibitem{ayang}
Xiaolong Feng, Weikang Wu, Yuexin Huang, Zhi-Ming Yu, and Shengyuan A. Yang, 
{\em Triply degenerate point in three-dimensional spinless systems},
\href{https://journals.aps.org/prb/abstract/10.1103/PhysRevB.104.115116}{\ Phys.\ Rev.\ B. {\bf 104}, 115116 (2021)},[\arXiv{2105.07340}{[cond-mat.mes-hall}]. 


\bibitem{future}
X.~Ji and Y.~W.~Sun, working in progress. 

\bibitem{kimb}
A. A. Burkov and Y. B. Kim,
{\em Z$_2$ and Chiral Anomalies in Topological Dirac Semimetals},
\doi{10.1103/PhysRevLett.117.136602}{Phys.\ Rev.\ Lett. {\bf 117},136602 (2016)}
[\arXiv{1606.08446}{cond-mat.mes-hall}]

\bibitem{Morimoto}
T.~Morimoto, A.~ Furusaki,
{\em Weyl and Dirac semimetals with $Z_2$ topological charge},
\doi{10.1103/PhysRevB.89.235127}{Phys.\ Rev.\ B {\bf 89},235127 (2014)}
[\arXiv{1603.7962}{cond-mat.mes-hall}].
\end{thebibliography}
\end{document}